\documentclass[useAMS,usenatbib]{mn2e}
\usepackage{cancel}	
\usepackage{color}
\usepackage{graphicx, subfigure}
\usepackage[authoryear]{natbib}
\usepackage{placeins}
\usepackage{amsmath}
\usepackage{tabulary}

\def\lesssim{\mathrel{\hbox{\rlap{\hbox{\lower4pt\hbox{$\sim$}}}\hbox{$<$}}}}
\def\gtrsim{\mathrel{\hbox{\rlap{\hbox{\lower4pt\hbox{$\sim$}}}\hbox{$>$}}}}
\newcommand{\gps}{\ensuremath{g_{\rm P1}}}
\newcommand{\rps}{\ensuremath{r_{\rm P1}}}
\newcommand{\ips}{\ensuremath{i_{\rm P1}}}
\newcommand{\zps}{\ensuremath{z_{\rm P1}}}
\newcommand{\yps}{\ensuremath{y_{\rm P1}}}
\newcommand{\wps}{\ensuremath{w_{\rm P1}}}
\newcommand{\grizy}{\ensuremath{grizy_{\rm P1}}}
\newcommand{\rizy}{\ensuremath{rizy_{\rm P1}}}

\newcommand{\PS}{\protect \hbox {Pan-STARRS1}}

\newcommand{\msun}{\mbox{M$_{\odot}$}}

\newcommand{\kms}{\mbox{$\rm{\,km\,s^{-1}}$}}

\newcommand{\mpcy}{\mbox{Mpc$^{-3}$\,yr$^{-1}$}}

\DeclareMathAlphabet{\mathsc}{OT1}{a}{m}{sc}
\def\testbx{bx}%
\DeclareRobustCommand{\ion}[2]{%
\relax\ifmmode
\ifx\testbx\f@series
{\mathbf{#1\,\mathsc{#2}}}\else
{\Mathrm{#1\,\mathsc{#2}}}\fi
\else\textup{#1\,{\mdseries\textsc{#2}}}%
\fi}

\title[GW150914]{Pan-STARRS and PESSTO search for an optical counterpart to the LIGO gravitational wave source GW150914}

\author[ Smartt et al.]{S.~J. ~Smartt$^{1}$\thanks{Corresponding author Email : s.smartt@qub.ac.uk},
K. C. Chambers$^{2}$,
K. W. Smith$^{1}$,
M. E. Huber$^{2}$,
D. R. Young$^{1}$,
\newauthor{E. Cappellaro$^{15}$,
D. E. Wright$^{1}$, 
M. Coughlin$^{3}$,
A. S. B. Schultz$^{2}$,
L. Denneau$^{2}$,}
\newauthor{H. Flewelling$^{2}$,
A. Heinze$^{2}$,
E. A. Magnier$^{2}$,
N. Primak$^{2}$,
A. Rest$^{4}$,
A. Sherstyuk$^{2}$,}
\newauthor{B. Stalder$^{2}$,
C. W. Stubbs$^{3}$,
J. Tonry$^{2}$,
C. Waters$^{2}$, 
M. Willman$^{2}$,
J. P.  Anderson$^{10}$}
\newauthor{C. Baltay$^{24}$,M. T. Botticella$^{19}$,
H. Campbell$^{16}$, 
M. Dennefeld$^{5}$,
T.-W. Chen$^{9}$,}
\newauthor{M. Della Valle$^{19}$, N. Elias-Rosa$^{15}$,
M. Fraser$^{16}$,
C. Inserra$^{1}$, 
E. Kankare$^{1}$,
R. Kotak$^{1}$,}
\newauthor{T. Kupfer$^{20,24}$,J. Harmanen$^{14}$, L. Galbany$^{12,13}$, A. Gal-Yam$^{26}$,
L. Le Guillou$^{6,7}$,}
\newauthor{J. D. Lyman$^{8}$, K. Maguire$^{1}$, A. Mitra$^{7}$, M. Nicholl$^{11}$,
F. Olivares E$^{12,23}$
D. Rabinowitz$^{24}$}
\newauthor{A. Razza$^{12,13}$, J. Sollerman$^{21}$, M. Smith$^{22}$,G. Terreran$^{15,1}$,
S. Valenti$^{17,18}$, B. Gibson$^{2}$}
\newauthor{T. Goggia$^{2}$}
\\
$^1$Astrophysics Research Centre, School of Mathematics and Physics, Queens University Belfast, Belfast BT7 1NN, UK \\
$^2$Institute for Astronomy, University of Hawaii at Manoa, Honolulu, HI 96822, USA\\
$^3$Department of Physics, Harvard University, Cambridge, MA 02138, USA\\
$^4$Space Telescope Science Institute, 3700 San Martin Drive, Baltimore, MD 21218, USA\\
$^5$ Institut d'Astrophysique de Paris, CNRS, and Universite Pierre et Marie Curie, 98 bis Boulevard Arago, 75014, Paris, France\\
$^6$  Sorbonne Universites, UPMC Univ. Paris 06, UMR 7585, LPNHE, F-75005, Paris, France \\
$^7$ CNRS, UMR 7585, Laboratoire de Physique Nucleaire et des Hautes Energies, 4 place Jussieu, 75005 Paris, France\\
$^8$ Department of Physics, University of Warwick, Coventry CV4 7AL, UK\\
$^9$ Max-Planck-Institut f{\"u}r Extraterrestrische Physik, Giessenbachstra\ss e 1, 85748, Garching, Germany\\
$^{10}$European Southern Observatory, Alonso de Cordova 3107, Vitacura, Santiago, Chile\\
$^{11}$ Harvard-Smithsonian Center for Astrophysics, 60 Garden St, Cambridge, MA 02138, United States\\
$^{12}$ Millennium Institute of Astrophysics, Casilla 36–D, Santiago, Chile\\
$^{13}$Departamento de Astronomia, Universidad de Chile, Camino El Observatorio 1515, Las Condes, Santiago, Chile\\
$^{14}$Tuorla Observatory, Department of Physics and Astronomy, University of Turku, V{\" a}i{\"a}l{\" a}ntie 20, FI-21500 Piikki{\" o}, Finland\\
$^{15}$ INAF - Osservatorio Astronomico di Padova, Vicolo del l'Osservatorio 5, 35122 Padova, Italy\\
$^{16}$ Institute of Astronomy, University of Cambridge, Madingley Road, Cambridge CB3 0HA, UK\\
$^{17}$ Las Cumbres Observatory Global Telescope Network, 6740 Cortona Dr.,
Suite 102, Goleta, California 93117, USA \\
$^{18}$ Department of Physics, University of California Santa Barbara, Santa Barbara, CA 93106, USA\\
$^{19}$INAF, Osservatorio Astronomico di Capodimonte, Salita Moiariello 16, 80131, Napoli, Italy\\
$^{20}$California Institute of Technology, MC 290-17 Pasadena, CA, 91125, USA\\
$^{21}$ Department of Astronomy and the Oskar Klein Centre, Stockholm University, AlbaNova, SE-106 91 Stockholm, Sweden\\
$^{22}$ School of Physics and Astronomy, University of Southampton,  Southampton, SO17 1BJ, UK\\
$^{23}$ Departamento de Ciencias Fisicas, Universidad Andres Bello, Avda. Republica 252, Santiago, Chile\\
$^{24}$Division of Physics, Mathematics, and Astronomy, California Institute of Technology, Pasadena, CA 91125, USA\\
$^{25}$Physics Department, Yale University, New Haven, CT 06520, USA\\
$^{26}$ Benoziyo Center for Astrophysics, Weizmann Institute of Science, 76100 Rehovot, Israel\\
}
\begin{document}

\maketitle

\clearpage


\begin{abstract}
 We searched for an optical counterpart to the first gravitational wave source discovered by  LIGO (GW150914),
using a combination of the Pan-STARRS1 wide-field telescope and the PESSTO spectroscopic follow-up programme. 
As the final LIGO sky maps changed during analysis, the total probability of the 
source being spatially coincident with our fields was finally only 4.2 per cent. Therefore we discuss our
results primarily as a demonstration of the survey capability of Pan-STARRS and spectroscopic capability 
of PESSTO. We 
 mapped out 442 square degrees of the northern sky region of the initial map. We discovered 56 astrophysical
transients over a period of  41 days from the discovery of the source. Of these, 19 were spectroscopically 
classified and a further 13 have host galaxy redshifts. All transients appear to be fairly normal supernovae
and AGN variability and none is obviously linked with GW150914. 
We illustrate the sensitivity of our survey  by defining parameterised lightcurves with 
timescales of 4, 20 and 40 days and use the sensitivity of the Pan-STARRS1 images to set limits on the 
luminosities of possible sources. The Pan-STARRS1 images reach limiting magnitudes of 
$\ips = 19.2, 20.0$ and $20.8$ respectively for the three timescales.  
For long timescale parameterised lightcurves (with FWHM$\simeq$40d) we set upper limits 
of
$M_i \leq-17.2^{-0.9}_{+1.4}$  if the distance to  GW150914 is 
$D_L = 400\pm200$\,Mpc. 
The number of type Ia SN we find in the survey is similar to that expected from the cosmic SN rate, indicating
a reasonably complete efficiency  in recovering supernova like transients out to 
$D_L = 400\pm200$\,Mpc.  
\end{abstract}


\section{Introduction}
\label{sec:intro}

The first observations of Advanced LIGO  \citep[aLIGO: ][]{2015CQGra..32g4001T},  consisting of two 4 km gravitational-wave interferometers, one at Hanford, WA and the other at Livingston, LA began in September 2015 and ended January 2016.
Advanced Virgo \citep{2015CQGra..32b4001A} is due to come online in 2016.
Due to significant instrumental upgrades \citep{2015CQGra..32g4001T}, the sensitivities in the LIGO interferometers have reached strain noise amplitudes below  $10^{-23}$Hz$^{-1/2}$ in the
frequency regime 10$^2$ to 10$^3$\,Hz. 
The improved sensitivity over previous science
runs  \citep{2012arXiv1203.2674T}
is about a factor 3 in strain sensitivity in the most sensitive band,  which corresponds to an increase in survey volume of more than an order of magnitude
\citep[see][]{theprizepaper}.

Sources of gravitational-waves that the detectors are sensitive to are the compact binary coalescences of black holes and neutron stars and potentially asymmetric core-collapse of massive stars  \citep{2013PhRvD..87b2002A,2012PhRvD..85h2002A,2010CQGra..27q3001A}.
The advanced gravitational-wave detectors are currently sensitive to  binary neutron star mergers within about 100\,Mpc (or beyond if black hole mergers are involved).
The rate of such events is extremely uncertain, by several orders of magnitude, but the detectors are expected to be sensitive to a few neutron-star coalescences per year and potentially more for black hole mergers \citep{2010CQGra..27q3001A}. 
Compact binaries are one of the most promising sources for simultaneous detection of gravitational-wave and electromagnetic emission, which can occur on timescales from seconds to months and wavelengths from X-ray to radio \citep{2012ApJ...746...48M,2010MNRAS.406.2650M}.
One potential source of electromagnetic emission from compact binaries containing at least one neutron star 
are kilonovae \citep{2015MNRAS.446.1115M,2013Natur.500..547T,2013ApJ...774L..23B}.
Compact binary mergers are the working model for short GRBs (with gamma-ray emission lasting less than 
$\sim$2 sec) in which beamed high energy emission occurs due to observers being on the 
binary rotation axis 
\citep{1992ApJ...395L..83N}. 
As discussed in 
\cite{2015arXiv150803608G},
short GRBs (sGRBs) have been observed  from $0.2\lesssim z \lesssim 2$ by $Swift$ and indeed the 
kilonova candidate of \cite{2013Natur.500..547T} was at 
$z=0.356$. Another plausible candidate for a kilonova has been recently identified by
\cite{2015ApJ...811L..22J} and \cite{2015NatCo...6E7323Y} in the data  for GRB060614
which likely has a host galaxy at at $z=0.125$. 
\cite{2015arXiv150803608G} also point out that in 10.5 years of $Swift$ operations, there have been no detections of  sGRBs with redshifts $z<0.1$. This corresponds to a distance of about 400\,Mpc which is 
much further (a factor $\sim$4) than the estimated sensitivity range of LIGO for NS-NS mergers. 
However the beaming factors for sGRBs is still quite uncertain and a much higher volumetric rate 
of NS-NS mergers is plausible if they are currently evading detection due to unfavourable 
beaming angles. 
Radiative transfer models of simulated NS-NS  and BH-NS mergers predict a range of
electromagnetic flux \citep[e.g.][]{2013ApJ...775..113T,2013ApJ...775...18B,2014ApJ...780...31T,2015MNRAS.450.1777K}

There are algorithms to localise the gravitational-wave transients on the sky, which vary from unmodeled sources \citep{2015ApJ...800...81E,2015CQGra..32m5012C} to compact binary signals \citep{2014ApJ...795..105S,2015ApJ...804..114B,2015arXiv150900055C}.
These algorithms result in likelihood sky areas typically spanning $\approx 100 - 1000\,\textrm{deg}^2$.
The program we will discuss in this paper uses a sky-map produced by the unmodeled source algorithms.
The initial LIGO and Virgo science runs included an electromagnetic follow-up program \citep{2014ApJS..211....7A}, which has expanded for the recent run to include many partners. 
Efforts have been made to optimise the success of this program, including the use of gravitational-wave catalogs and optimising multiple-telescope pointings \citep{2011CQGra..28h5016W,2014ApJ...789L...5K,2013ApJ...767..124N,2015arXiv150803608G,2015ApJ...814...25C,2014ARAA..52...43B,2014ApJ...784....8H}, with lessons learned from other multi-messenger efforts \citep{2015ApJ...811...52A,2015ApJ...806...52S,2013ApJ...769..130C}.

The first discovery of gravitational waves from a binary black Hole merger has been 
announced 
\citep{theprizepaper}. 
The  signal, known initially as candidate G184098, 
was initially announced on 16 September 2015 to the broad network of follow-up facilities who signed confidential Memoranda of Understanding with the LIGO/Virgo team. 
As discussed in \cite{theprizepaper}, the confirmed source GW150914 
is  a compact binary merger of two black holes with masses $M_1 = 36^{+5}_{-4}$ M$_{\odot}$ and 
$M_2 = 29^{+4}_{-4}$ M$_{\odot}$.
GW150914 was found with a network signal to noise ratio of 24 corresponding to an $\approx 5.1\sigma$ detection.   The distance to the source is estimated at 
$D =  410^{+160}_{-180}$\,Mpc  or $z= 0.09^{+0.03}_{-0.04}$, \citep[see][]{2016arXiv160203840T}. 
From this event, \cite{theprizepaper} 
and \cite{2016arXiv160203842A}
 estimate the  rate of BBH coalescences to be  $0.002-0.4 {\rm \; Mpc^{-3} Myr^{-1}}$ in the local Universe. 
To put this rate in perspective, we list the volumetric rates of various types of  exploding transients in the Local Universe in Table\,\ref{tab:rates}.  The rate of BBH mergers 
are surprisingly comparable other exotic transients we know of. 
The initial LIGO skymaps returned by the unmodeled pipelines returned sky-areas with 90\% confidence levels of $\approx$ 310 square degrees, with the final LIGO analysis which was published corresponding to compact binary coalescences with skymaps covering 90\% confidence levels of $\approx$ 600 square degrees.  In this paper, we provide our first attempts to place upper-limits on optical counterparts to the direct detection of a gravitational wave signal. 

This search for an optical counterpart to GW150914 requires pre-existing images of the available sky as well
as the ability to survey the LIGO error regions efficiently. 
The Pan-STARRS facility and the Pan-STARRS1 Surveys (Chambers et al, 2016, in prep) provide the survey capability and the reference images. 
Analysis of the difference images by the Image Processing Pipeline  or IPP \citep{2015MNRAS.449..451W}
provides candidate detections and their attributes. 
These data are then further analysed by machine learning algorithms 
\citep{2015MNRAS.449..451W}
to provide the down-selected list of candidates for follow-up
spectroscopy by Hawaii and, the Public ESO Spectroscopic Survey of Transient Objects
PESSTO. 
The Pan-STARRS1 telescope began its science surveys under the umbrella of the PS1 Science Consortium in 2010. These surveys ran
until mid 2014, with the whole northern sky above a declination of $\delta\simeq-30^{\circ}$
degrees covered in the filters \grizy\ (called the 3$\pi$ survey). The three other major
surveys were the Medium Deep Field survey, as described in the papers
\cite{2012ApJ...745...42T,2014ApJ...795...44R,2015MNRAS.448.1206M}, the Pandromeda project 
\citep{2014ApJ...797...22L} and PanPlanets 
\citep{2009A&A...494..707K}.  
In addition an NEO focused survey on the sweet spots at opposition at the beginning and end of the nights was carried out \citep[see][for a description of the PS1 moving object processing system]{2013PASP..125..357D}. 
The search for transients in the Medium Deep field has been 
well documented in, for example the recent papers of 
\cite{2014ApJ...780...44C,2014ApJ...794...23D,2015ApJ...799..208S,2015ApJ...804...28G}. 
A number of 
papers have used data from the 3$\pi$ survey for transient science
\citep[e.g.][]{2013ApJ...770..128I,2013Natur.502..346N,2013ApJ...779L...8F}. 
These have been a combination of finding and discovering transients
in the 3$\pi$ data \citep{2015A&A...580L..15P,2013ApJ...770..128I} and retrospectively searching the data for interesting epochs for known supernovae and massive star outbursts
\cite[e.g.][]{2013ApJ...779L...8F,2015A&A...581L...4K}. 

Since the 
creation of an all-sky image stack (which was produced in a PS1 internal 
processing version called PV2) the single epochs in 3$\pi$ have
been differenced with respect to this reference sky and transients
have been catalogued. The survey has transitioned from the PS1SC 
funded operations (2010 to mid-2014) to the Pan-STARRS Near Earth
Object Science Consortium (PSNSC; mid-2014 to present) and the search for transients has continued. 
We currently run the  ``Pan-STARRS Survey for Transients" (PSST),
with first results in 
\cite{2015A&A...580L..15P} and \cite{2015IAUGA..2258303H}. We search the NASA PSNSC survey data for stationary transients and make our discoveries public on timescales of 12-24hrs after first observations. Additionally, we are using the Pan-STARRS1 telescope in pointed, triggered, mode to survey the sky localisation regions of LIGO/Virgo gravitational wave searches. This paper describes the first pointed search from this
new operational mode of the PSNSC surveys. In doing this, we fed the targets discovered to two 
spectroscopic follow-up programmes. One of these was PESSTO, the Public ESO Spectroscopic Survey 
of Transient Objects which has 90 nights allocated per year on the ESO New Technology Telescope and 
has a partnership with Pan-STARRS to classify and follow targets from the PSST discovery stream
\citep[e.g.][]{2015ATel.7102....1G}. The other was an extensive programme on the University of Hawaii 2.2m telescope with the SNIFS spectrometer. 

\begin{table*}
\caption{
They have all been converted to rates per Mpc$^{3}$ per Myr. The rate of GRBs are  estimates  of the true rates, after correction for the (uncertain) beaming factor. The distance columns refers to the distance within which the rates are calculated}
\label{tab:rates}
\begin{tabular}{llll}
\hline
\hline
Type       & Rate                                         &  Distance          & Reference  \\
               &  ${\rm \; Mpc^{-3} Myr^{-1}}$   &    Mpc                                    &    \\
\hline
Binary BH mergers      &   0.002-0.4              &            $\sim$400               & \cite{theprizepaper}    \\ 
Core-collapse SN        &    96$-$140       &  $30- 400$                     &   \cite{2011ApJ...738..154H,2011MNRAS.412.1441L,2009MNRAS.395.1409S}       \\ 
Broad-lined Ic SN         &   1$-$4           &   $30 - 400$ &       \cite{2015arXiv151101466G,2012ApJ...759..107K,2009MNRAS.395.1409S}      \\ 
LGRBs (true)                &  0.1-0.6                                            &    $\lesssim$450      &     \cite{2007ApJ...657L..73G} \\
sGRBs  (true)                 &    1                                             &       $\lesssim$400    &       \cite{2014ARAA..52...43B}           \\
Superluminous SN        &  0.01                                         &    900-2600          &    \cite{2011Natur.474..487Q,2013MNRAS.431..912Q,2015MNRAS.448.1206M} \\ 
\hline
\end{tabular}
\end{table*}

\section{Instrumentation and observational details}
\label{sec:inst}

\subsection{The Pan-STARRS1 telescope, camera and photometric system} 
The \PS\  telescope has a 1.8-m diameter primary mirror with  $f/4.4$ cassegrain focus. It was designed as a 
a high-\'{e}tendue wide-field imaging system, 
and is located near the summit of Haleakala on the island of Maui. A 1.4 Gigapixel camera is mounted at cassegrain consisting of 
sixty Orthogonal Transfer Array devices, each of which has a detector area of 48460 x 48680 microns, of which 3.7 per cent  is not active
due to further division into OTA cells with gaps. 
The 10 micron pixels have a plate scale of 0.26 arcsec. The devices are arranged in the focal plane as an 8x8 pattern minus the four corner chips. 
The nominal focal plane is 418.88 mm in diameter or 3.0 degrees. With the gaps and masked regions, the 7.06 square degree FOV has
an active region of about 5 square degrees.
A   description of the Pan-STARRS1 system is provided by 
\cite{2010SPIE.7733E..12K}.
The \PS\ observations are obtained through a set of five broadband filters, which are designated as \gps, \rps, \ips, \zps, and \yps. A further, broad filter \wps, has been used primarily for near earth asteroid (NEO) searches 
\citep[e.g.][]{2015Icar..261...34V}
and is currently being employed in the NASA funded survey that produces both NEO discoveries and 
stationary transients
\citep{2015IAUGA..2258303H,2015A&A...580L..15P}. Although the $griz$-band 
filter system for \PS\ is  similar to the SDSS 
\citep{2009ApJS..182..543A}, there are some significant differences.  
The  \gps\ filter extends 20~nm redward of the $g_{SDSS}$ filter bandpass, which provides extra sensitivity but does include the 
 5577\AA\ night sky emission line. The  \zps\ filter has a red cut off at 930~nm, whereas  $z_{SDSS}$  had a red response
which was effectively defined by the detector efficiency. \PS\ does not have a $u-$band, but extends redder to include a \yps band, 
and has produced the first all sky image in this waveband.  The \PS\ photometric system in discussed in detail in \cite{2012ApJ...750...99T}.  

Images obtained by the \PS\ system are processed with the Image Processing Pipeline (IPP) \citep[see details in][]{2013ApJS..205...20M}, originally on a computer cluster at the Maui High Performance Computer Center (MHPCC) but now located in Maui Research and Technology Center, Kihei, Maui.  The pipeline runs the images through a succession of stages including device ``de-trending'', a flux-conserving warping to a sky-based image plane, masking and artefact location (Waters et al. 2016, in prep.). 
De-trending involves bias and dark correction and flatfielding using white light flatfield images from a dome screen, in combination with an illumination correction obtained by rastering sources across the field of view.   After determining 
an initial astrometric solution the flat-fielded images were then warped onto the tangent plane of the sky using a flux conserving 
algorithm.  The warping to a sky-based image plane involves mapping the camera pixels to a defined set of skycells, of which there
are approximately 51 per PS1 pointing. 
The plate scale for the warped images is 0.25 arcsec/pixel. 
Photometry from \PS\ is in the ``AB system''  \citep[see][for a discussion]{2012ApJ...750...99T}   where the  monochromatic AB magnitude is the logarithm of 
flux density 

\begin{align}
m_{AB}(\nu) &= -2.5\log(f_\nu/3631~\hbox{Jy})\\
 &= -48.600 - 2.5\log(f_\nu[\hbox{erg/sec/cm$^2$/Hz])}
\end{align}

where $1~\hbox{Jy} = 10^{-23}\hbox{erg/sec/cm$^2$/Hz}$. The calibration of the \cite{2012ApJ...750...99T} PS1 photometric system through a reference star catalogue is described in detail in \cite{2012ApJ...756..158S} and \cite{2013ApJS..205...20M}.  For the nightly processing, the zeropoints of the full-field camera chips are set  from a catalogue of photometric reference stars from the ``ubercal" analysis during the first reprocessing of all of the PS1 3$\pi$ data as described in \cite{2012ApJ...756..158S,2013ApJS..205...20M}. The 
internal calibration of this reference catalogue has a relative precision of around 1\%.  However the automated 
zeropoint applied in the difference imaging is currently an average full-field zeropoint calculated and this 
can lead to variations across skycells up to  $\pm0.15$.

The search for transients is greatly aided by having the pre-existing sky images from the Pan-STARRS1 Sky Surveys (Chambers et al 2016, in prep.)
carried out by the PS1 Science Consortium.  
Transient sources are identified by the Image Processing Pipeline through analysis of difference images, created by subtracting the stacked 
image from the PS1SC 3$\pi$ survey as a template from the observed image taken as part of the search for the counterpart.  As described in \cite{2015IAUGA..2258303H} the \gps, \rps, \ips, \zps, and \yps
images employ the PS1 Science Consortium 3$\pi$ reference stack (currently PV3 in the internal processing
labelling) for difference imaging. The  \wps
static sky stack is not available over the whole sky, 
but is employed where it exists on a nightly basis. 
Difference imaging by IPP has been further described in \cite{2014MNRAS.437..656M,2015MNRAS.448.1206M,2015A&A...580L..15P,2015MNRAS.449..451W}. Similar processing was implemented on the
PS1 images discussed here and the imaging survey is further described in Section\,\ref{sec:search}.

%
%
%
%

\begin{figure*}
 \centering
\caption{Skymaps showing the probability contours produced by LIGO (solid contours), the PS1 pointings (green circles with diameters
 of 2.7$^{\circ}$ degrees and the positions of transients.
The panels are separated into time intervals from discovery epoch of GW150914. A transient is plotted on the relevant panel if it was discovered in the epochs shown. All time is in the observer frame.}
  \begin{tabular}{cc}
    \includegraphics[width=70mm]{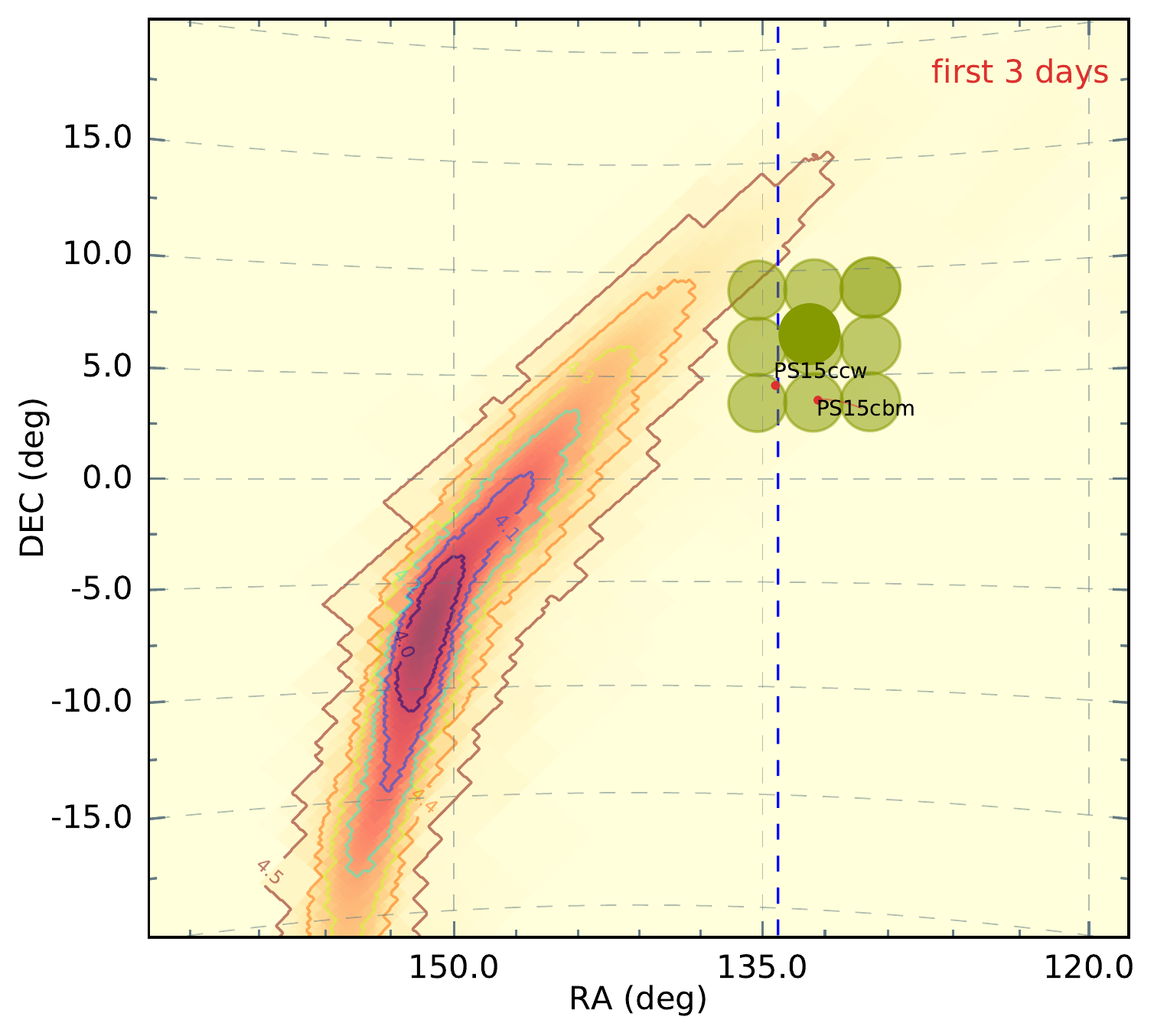}&
    \includegraphics[width=70mm]{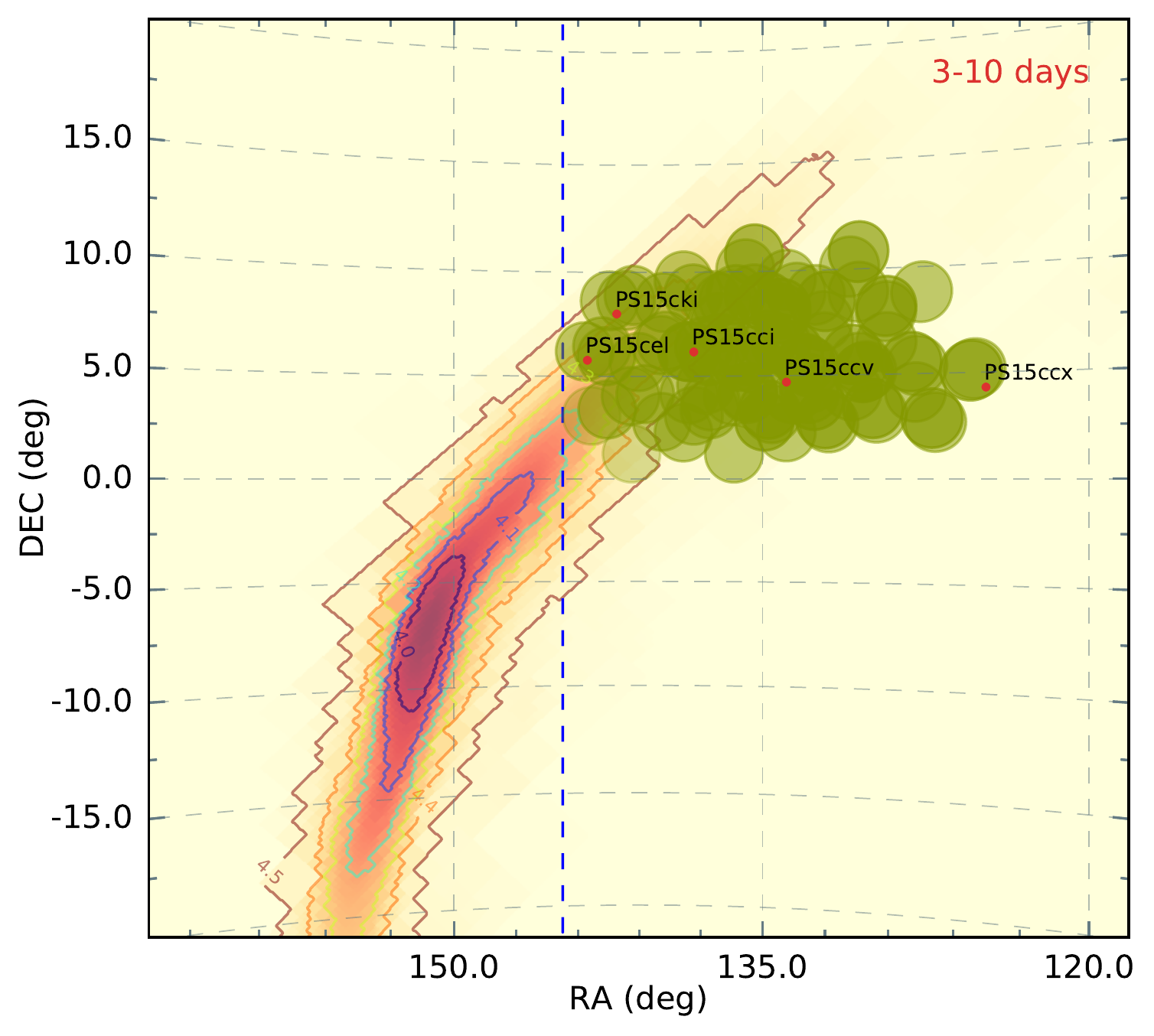}
\\
    \includegraphics[width=70mm]{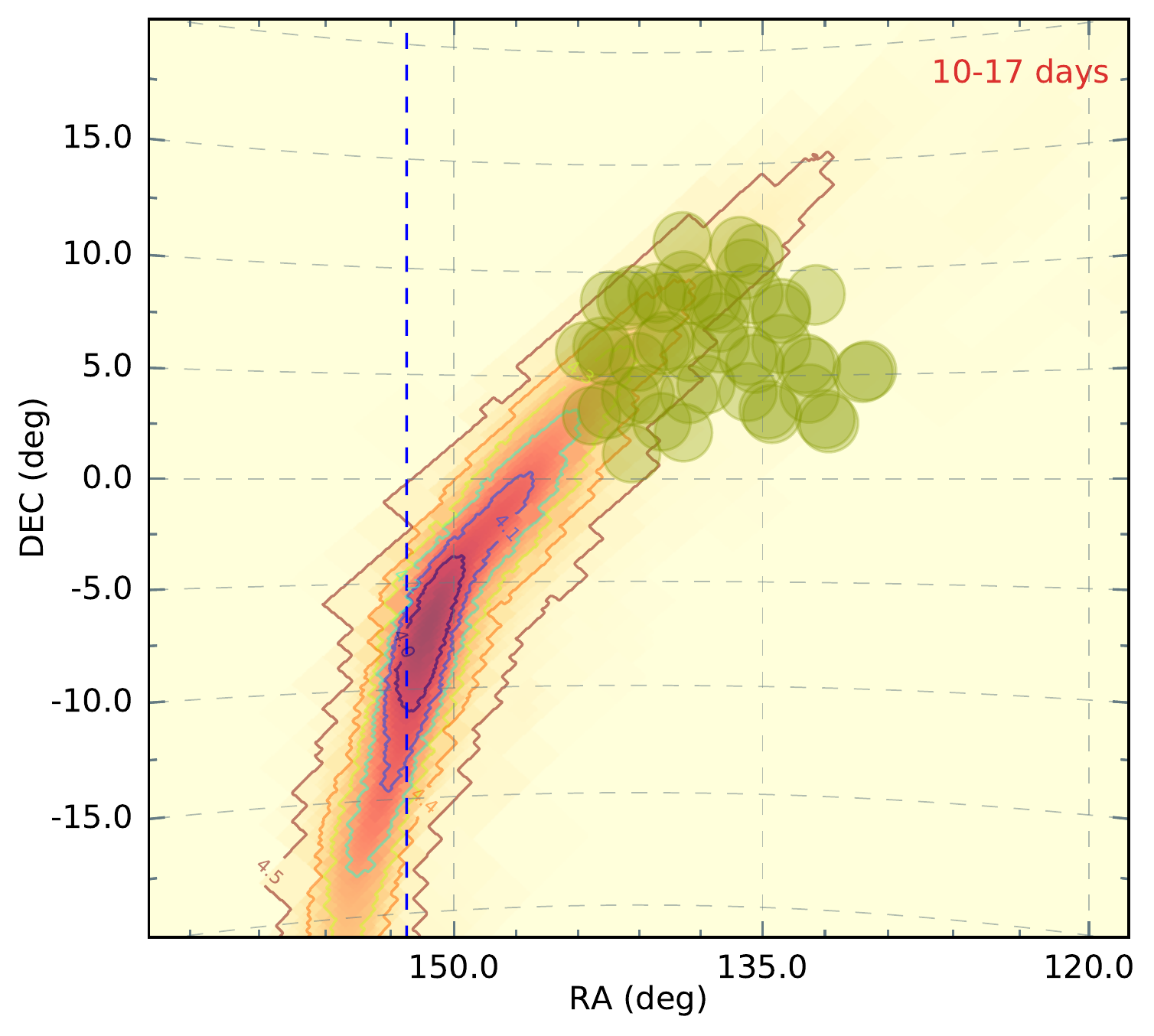}
&
    \includegraphics[width=70mm]{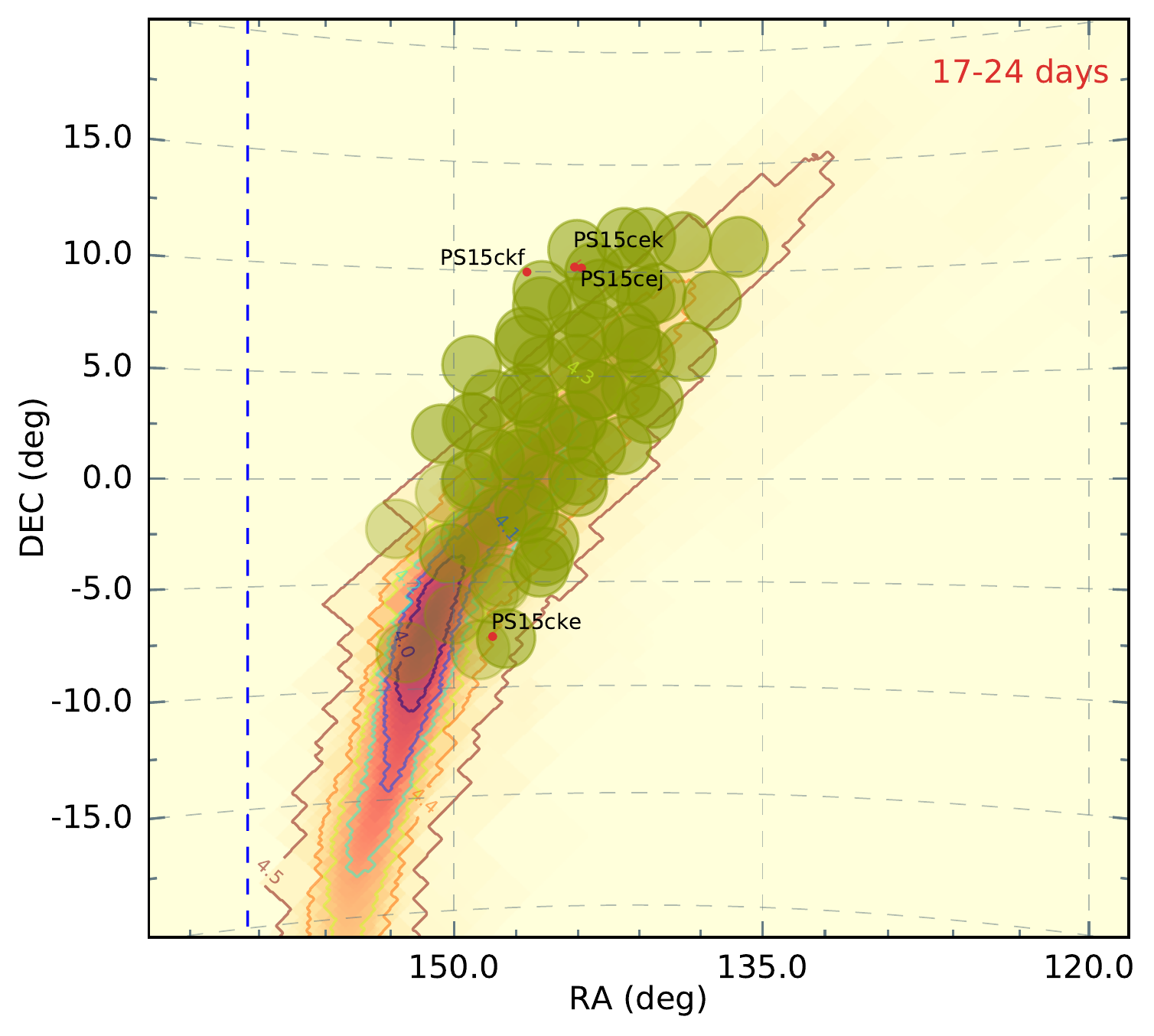}
\\
    \includegraphics[width=70mm]{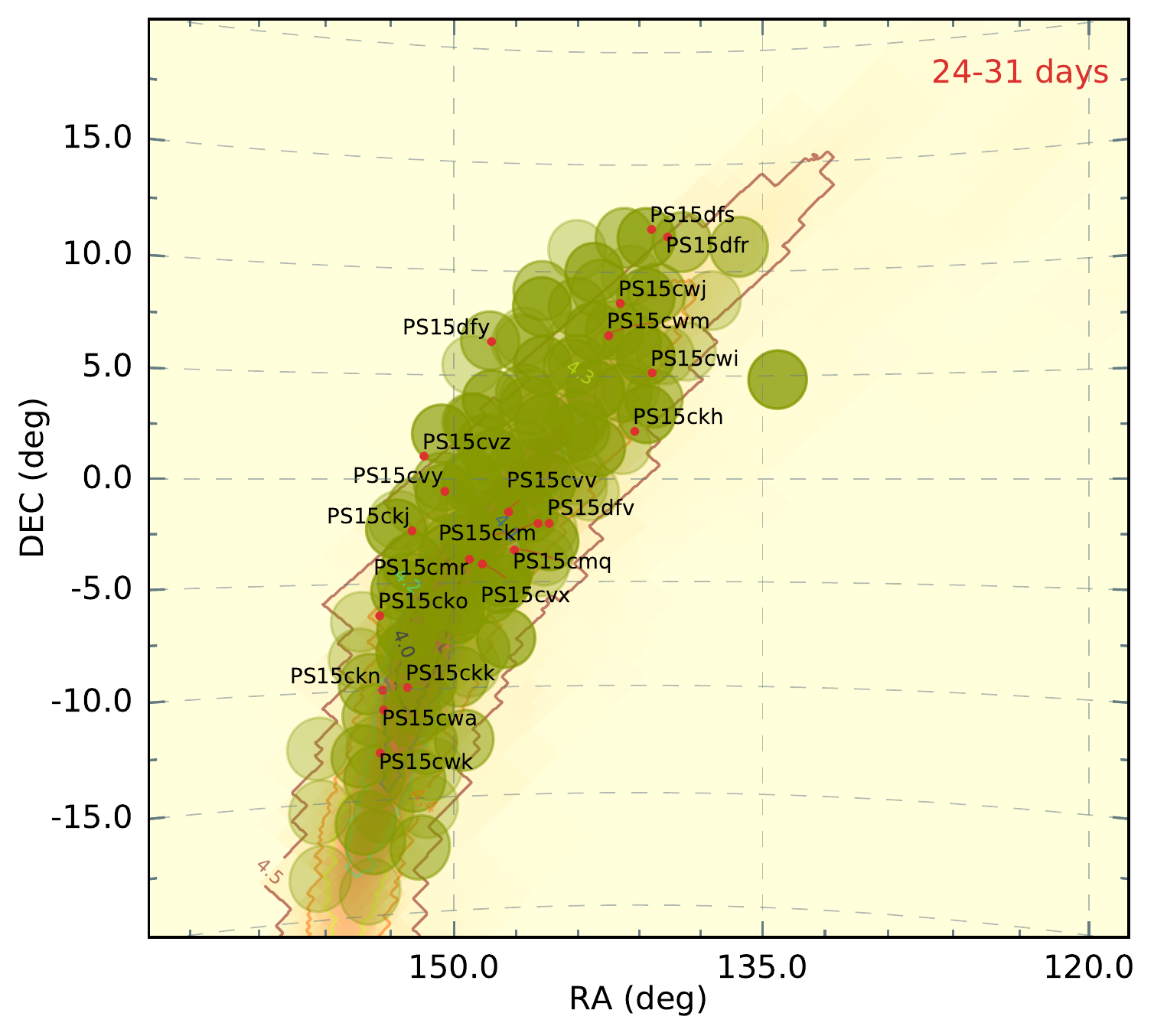}
&
    \includegraphics[width=70mm]{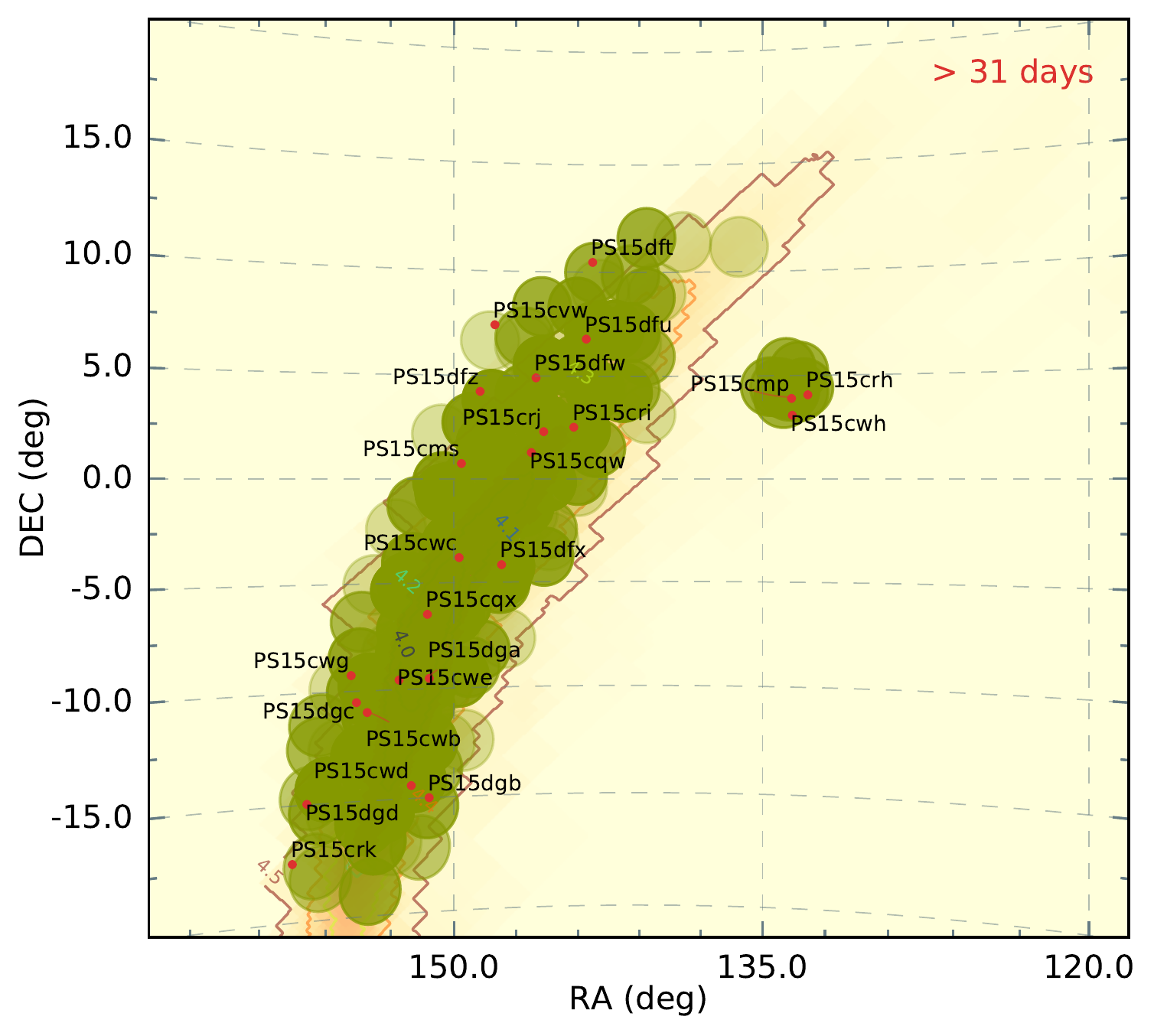}
\\
\end{tabular}
\label{fig:skymaps}
\end{figure*}

\subsection{Spectroscopic instruments}
Spectroscopic follow-up was carried out with two main instruments, namely the SuperNovae Integral Field Spectrograph (SNIFS) on the University of Hawaii 2.2m Telescope (UH 2.2m) and the EFOSC2 spectrometer on the 
ESO New Technology Telescope (NTT) as part of the Public ESO Spectroscopic Survey of Transient Objects (PESSTO\footnote{www.pessto.org}).
One further spectrum was taken with the 200 inch Hale telescope at Palomar with the Double Spectrograph (DBSP). 

\subsubsection{SNIFS on the UH 2.2m telescope}
SNIFS  provides integral field spectra and was designed for supernovae spectroscopy
\citep{2004SPIE.5249..146L}. 
It provides coverage over two arms : the blue covers  
3200$-$5600\AA\ at  2.4\AA\ per pixel sampling and $R\simeq1000$ at 4300\AA; the red arm 
covers 5200-10000\AA\ at 2.9\AA\ per pixel sampling and  $R\simeq1300$ at 7600\AA. 
The dichroic mirror which reflects light to the blue arm and transmits to the red  has a 
cross-over region of 5000-5250\AA\ within which it is difficult to recover reliable flux calibrated spectra 
during standard reductions. We hence chose to remove this region in the spectra presented here. 
Data reductions were performed using the SNIFS SuperNova Factory (SNF) summit pipeline \citep{2001MNRAS.326...23B,2006ApJ...650..510A}. Standard CCD processing included bias and flat-field corrections with images then converted into blue and red channel data cubes for bad pixel and cosmic-ray cleaning. Wavelength calibration used arc lamp exposures taken at position of the target to minimise errors from telescope flexure and a rough flux calibration was applied using a historical response function for each data cube. A one-dimensional spectrum was extracted from each data cube using a point-spread function model with sky subtraction.

\subsubsection{EFOSC2 on the NTT with PESSTO}

The PESSTO survey is described in \cite{2015A&A...579A..40S}. The EFOSC2 spectrometer is 
employed on the ESO NTTT at the nasmyth focus and Gr\#13 is used to provide 
spectra between 3650-9250\AA\ at a dispersion of 5.5\AA\,pix$^{-1}$. 
A slit width of 1\farcs0 is nominally used (unless poor seeing 
forces a change to 1\farcs5) which results in a resolution of 18.2\AA. 
All data presented in this paper were taken with a 1\farcs0 slit. 
PESSTO employs a dedicated pipeline as described in \cite{2015A&A...579A..40S}, and 
all spectra were reduced using these standard procedures. The reduced spectra will be 
made public through the ESO Science Archive Facility during the summer of 2016, when 
the PESSTO project plans to release years 3 and 4 of the public survey in Spectroscopic 
Survey Data Release 3. Before then, all classification
spectra will be placed on WISeREP \footnote{http://wiserep.weizmann.ac.il/home}. 

\subsubsection{DPSP on the Hale}
One spectrum was taken by T. Kupfer at the Palomar 200-inch Telescope (the Hale) using the 
Double Spectrograph (DBSP). This is a low to medium resolution spectrometer at 
cassegrain focus that uses a dichroic at 5500\AA\ to split the beam and direct it to 
two separate red and blue channels. A 1\farcs5 slit was used on the night of 
25 Sep 2015 to observe one candidate (PS15cci, see Section\,\ref{sec:trans}) at airmass=2.627.  The blue arm used the 
grating 600/4000 and the red arm had the 316/7500 grating installed, 
resulting in resolutions of approximately 4\AA\ and 8\AA\  FWHM respectively. 
Standard reduction procedures were employed to reduce the red and blue arms, including 
bias frame subtraction, extraction and sky subtraction, arc wavelength calibration and flux calibration
(using the standard G191-B2B). A single good signal-to-noise spectrum (S/N$\sim30$ at 6000\AA) resulted from the 1500s exposure time and the red and blue arms were merged for object classification.

\section{The Pan-STARRS observing campaign of LIGO source GW150914}
\label{sec:obs}

GW150914 was announced to the collaborating groups who had offered dedicated observing facilities during 
the first  observing run (O1) of LIGO on 16 September 2015 (06:39 UT) as source G\,184098.
The source was detected  by the LIGO Hanford Observatory  and LIGO Livingston Observatory
at 14 Sep 2015 09:50:45 UT \citep{GCN18330,theprizepaper}. 
Two sky maps 
were produced, one from the rapid localisation 
\citep[the cWB map;][]{2016PhRvD..93d2004K}
and the other from the  refined localisation 
\citep[LALInference Burst : LIB, with date and time stamp 14  Sep 2015  19:02:54;][]{2015arXiv151105955L}

A northern and southern sky banana shaped region was produced
for each, with the 90\%  credible regions spanning about 700-800 square degrees. The
night of the 16 Sep 2015 on Haleakala (05:00 - 15:00 UT) was wiped out due to poor weather and 
no Pan-STARRS observations took place. We started observing the G150914 northern banana 
on the night of 17 Sep 2015 (14:50 - 15:40 UT). 
Due to the RA location, only the north western
tip of the localisation banana was observable in the 
first 10 days, and at low airmass close to twilight. 
Figure\,\ref{fig:skymaps} shows all the PS1 footprints
(2.9 degrees diameter) in a binned time sequence 
and the probability sky contours from the final LIGO/Virgo
LIB analysis. 

During the first few days, we observed the cWB map 
region. This was due to us misinterpreting the time sequence
of release and assuming the cWB was the more likely
due to having a later date stamp from the LIGO/Virgo analysis. However this choice had little real effect as
the more easterly, higher probability regions 
of the LIB analysis were inaccessible due their RA and 
indeed the first 3 day PS1 pointings shown in Figure\,\ref{fig:skymaps} represent the eastern limits of
the telescope's pointing capability at that time. 
As further illustrated in Figure\,\ref{fig:skymaps}, 
we stepped through the region in time to cover the 
bulk of the northern LIB localisation banana within 
30-40 days.  Pan-STARRS continued to observe this field 
until 26 October 2015 (MJD 57321).  A combination of the 
\rps, \ips, \zps and \yps\ filters were used, for two reasons. 
One was that expected NS-NS mergers have predicted
intrinsically red transients 
\citep{2013ApJ...774...25K} and more practically, the sky brightness during the 
early observations (in astronomical twilight) forced a 
shift to the redder bands. The most common filter used was 
the \ips-band. Exposure times were typically between 30-80 sec, 
and small dither strategies were employed to mitigate the 
fill factor problems of the GPC1 focal plane. 
A list of telescope pointings are available
on request from the authors. 

During subsequent analysis, the 
LIGO/Virgo team further refined the 
sky probability map, 
producing a final analysis result 
which they released to the collaborating teams as 
GCN 18858 
\citep{GCN18858} and then publicly in 
\citep{2016arXiv160203840T}.
The two maps released were BAYESTAR \citep{2016PhRvD..93b4013S}
and LALInference \citep{2015PhRvD..91d2003V}. 
 The LALInference map with date stamp 
2016 Jan 9, 1:29:46 was referred to 
 ``as the most
 accurate and authoritative localisation for this event" in GCN 18858, and described in detail in
\citep{2016arXiv160203840T}. 
However this 
map  favours the
southern portion of the annulus originally 
released at the time we began follow-up, 
This resulted in a final 
90 per cent credible region of 600 square 
degrees that was mostly in the south and
a severe reduction in the northern 
probability.  Hence our coverage is 
low, but this paper demonstrates the 
methods we have employed and will 
employ in the future to search for 
EM counterparts of GW sources. 

We used  \ips, \zps, \yps band observations progressively to map as much of the GW150914 error region as possible due to the proximity of twilight.  We have measured the boundaries of twilight in each of the PS1 filters
where the end of evening twilight is defined as the time when the the sky brightness at constant altitude becomes constant,  and the start of morning twilight is when the sky brightness ceases to be constant. For the PS1 \grizy\ bands, this is when the sun is at 16, 16, 16, 14, and 10 degrees below the horizon at PS1 respectively (Chambers et al 2016, in prep).  The proximity to twilight of the GW150914 error region limited the accessible RA at an hour angle of 4.5 hours or  an airmass of  approximately 2.6. Figure\,\ref{fig:skymaps}  shows the boundary of the accessible region in the \ips band and Table Appendix\,X lists all the data. 

\begin{table*}
\caption{Transient candidates in the field of 
GW150914 
(56 in total). Discovery dates refer to the date of the 
initial detection by Pan-STARRS. For reference, 
GW150914
 was discovered at   20150914.41 (MJD 57279.41). 
}
\label{tab:alltrans}
\begin{center}
\begin{tabular}{lllllllll}
\hline \hline
Name & RA     & Dec     & RA  & Dec  & Discovery & Discovery & Disc & Disc.  \\ 
     & (J2000) & (J2000)& (J2000) & (J2000)    &  Date       & MJD    & mag. & filt.\\ \hline
PS15cbm & 08 49 19.85 & +03 48 17.8 & 132.33271 & +3.80494  & 20150917.62 & 57282.62 & 18.55 & \ips \\
PS15ccw & 08 57 30.60 & +04 31 56.1 & 134.37750 & +4.53225  & 20150917.63 & 57282.63 & 19.31 & \ips \\
PS15cci & 09 13 22.76 & +06 10 47.3 & 138.34483 & +6.17981  & 20150919.63 & 57284.63 & 18.32 & \ips \\
PS15ccx & 08 18 03.91 & +04 18 04.2 & 124.51629 & +4.30117  & 20150919.63 & 57284.63 & 19.42 & \zps \\
PS15ccv & 08 55 23.05 & +04 41 19.0 & 133.84604 & +4.68861  & 20150922.62 & 57287.62 & 20.03 & \ips \\
PS15cel & 09 34 11.58 & +05 46 45.2 & 143.54825 & +5.77922  & 20150923.63 & 57288.63 & 19.53 & \ips \\
PS15cki & 09 28 27.24 & +08 00 51.5 & 142.11350 & +8.01431  & 20150923.64 & 57288.64 & 19.17 & \zps \\
PS15cej & 09 35 19.41 & +10 11 50.7 & 143.83087 & +10.19742 & 20151002.62 & 57297.62 & 18.13 & \ips \\
PS15cek & 09 36 41.04 & +10 14 16.2 & 144.17100 & +10.23783 & 20151002.63 & 57297.63 & 17.24 & \zps \\
PS15cke & 09 52 35.14 & -07 36 32.0 & 148.14642 & -7.60889  & 20151002.64 & 57297.64 & 16.72 & \zps \\
PS15ckf & 09 45 57.71 & +09 58 31.4 & 146.49046 & +9.97539  & 20151003.65 & 57298.65 & 17.57 & \yps \\
PS15cwj & 09 27 44.89 & +08 31 32.1 & 141.93704 & +8.52558  & 20151013.60 & 57308.60 & 20.02 & \ips \\
PS15cwi & 09 21 31.27 & +05 10 26.8 & 140.38029 & +5.17411  & 20151013.61 & 57308.61 & 20.43 & \ips \\
PS15ckm & 09 43 47.15 & -02 10 13.3 & 145.94646 & -2.17036  & 20151013.61 & 57308.61 & 19.57 & \ips \\
PS15ckj & 10 07 58.59 & -02 29 47.9 & 151.99412 & -2.49664  & 20151013.61 & 57308.61 & 18.31 & \ips \\
PS15cko & 10 14 01.69 & -06 30 46.9 & 153.50704 & -6.51303  & 20151013.62 & 57308.62 & 19.51 & \ips \\
PS15ckh & 09 24 55.83 & +02 19 25.1 & 141.23263 & +2.32364  & 20151013.62 & 57308.62 & 19.40 & \ips \\
PS15cvz & 10 05 41.49 & +01 05 33.2 & 151.42288 & +1.09256  & 20151013.62 & 57308.62 & 19.55 & \ips \\
PS15cvy & 10 01 45.13 & -00 36 06.8 & 150.43804 & -0.60189  & 20151013.63 & 57308.63 & 19.76 & \ips \\
PS15ckn & 10 13 29.31 & -10 00 06.1 & 153.37213 & -10.00169 & 20151014.62 & 57309.62 & 19.44 & \ips \\
PS15ckk & 10 08 48.60 & -09 54 50.7 & 152.20250 & -9.91408  & 20151014.62 & 57309.62 & 16.43 & \ips \\
PS15dfs & 09 21 37.60 & +12 01 38.0 & 140.40667 & +12.02722 & 20151015.60 & 57310.60 & 20.94 & \ips \\
PS15dfr & 09 18 29.04 & +11 40 10.4 & 139.62100 & +11.66956 & 20151015.60 & 57310.60 & 21.31 & \ips \\
PS15cwm & 09 30 01.03 & +06 58 12.6 & 142.50429 & +6.97017  & 20151015.61 & 57310.61 & 20.90 & \ips \\
PS15dfy & 09 52 48.76 & +06 38 04.5 & 148.20317 & +6.63458  & 20151015.61 & 57310.61 & 19.82 & \ips \\
PS15cvv & 09 49 30.25 & -01 36 37.5 & 147.37604 & -1.61042  & 20151015.62 & 57310.62 & 20.14 & \ips \\
PS15cmr & 09 57 03.59 & -03 53 24.3 & 149.26496 & -3.89008  & 20151015.62 & 57310.62 & 19.35 & \ips \\
PS15cmq & 09 48 22.97 & -03 27 41.4 & 147.09571 & -3.46150  & 20151015.62 & 57310.62 & 20.19 & \ips \\
PS15cvx & 09 54 35.48 & -04 07 22.3 & 148.64783 & -4.12286  & 20151015.62 & 57310.62 & 20.32 & \ips \\
PS15dfv & 09 41 38.31 & -02 10 21.8 & 145.40963 & -2.17272  & 20151015.62 & 57310.62 & 20.83 & \ips \\
PS15cwa & 10 13 18.75 & -10 54 43.9 & 153.32812 & -10.91219 & 20151015.63 & 57310.63 & 20.27 & \ips \\
PS15cwk & 10 13 55.42 & -12 52 49.2 & 153.48092 & -12.88033 & 20151015.63 & 57310.63 & 20.11 & \ips \\
PS15cms & 09 58 35.10 & +00 44 34.7 & 149.64625 & +0.74297  & 20151017.62 & 57312.62 & 19.93 & \ips \\
PS15cvw & 09 52 09.25 & +07 26 48.3 & 148.03854 & +7.44675  & 20151018.61 & 57313.61 & 19.86 & \ips \\
PS15cmp & 08 54 24.40 & +03 54 00.5 & 133.60167 & +3.90014  & 20151019.58 & 57314.58 & 21.82 & \rps \\
PS15crh & 08 51 16.19 & +04 03 57.9 & 132.81746 & +4.06608  & 20151019.58 & 57314.58 & 21.39 & \rps \\
PS15cwh & 08 54 15.18 & +03 04 59.0 & 133.56325 & +3.08306  & 20151019.58 & 57314.58 & 22.09 & \rps \\
PS15cri & 09 36 50.66 & +02 31 20.0 & 144.21108 & +2.52222  & 20151021.60 & 57316.60 & 20.67 & \ips \\
PS15cwb & 10 16 21.58 & -11 00 10.5 & 154.08992 & -11.00292 & 20151021.61 & 57316.61 & 20.25 & \ips \\
PS15dgc & 10 18 20.86 & -10 31 28.3 & 154.58692 & -10.52453 & 20151021.61 & 57316.61 & 20.42 & \ips \\
PS15cwe & 10 10 24.74 & -09 33 10.0 & 152.60308 & -9.55278  & 20151021.63 & 57316.63 & 20.47 & \ips \\
PS15crk & 10 30 03.48 & -17 31 38.7 & 157.51450 & -17.52742 & 20151021.63 & 57316.63 & 19.97 & \ips \\
PS15dgb & 10 04 43.54 & -15 00 03.8 & 151.18142 & -15.00106 & 20151021.63 & 57316.63 & 20.71 & \ips \\
PS15dga & 10 04 42.37 & -09 31 14.8 & 151.17654 & -9.52078  & 20151021.63 & 57316.63 & 20.34 & \ips \\
PS15dfx & 09 50 52.07 & -04 09 46.3 & 147.71696 & -4.16286  & 20151023.60 & 57318.60 & 20.80 & \ips \\
PS15cwg & 10 19 19.55 & -09 16 01.2 & 154.83146 & -9.26700  & 20151023.61 & 57318.61 & 20.39 & \ips \\
PS15crj & 09 42 42.16 & +02 18 09.8 & 145.67567 & +2.30272  & 20151023.62 & 57318.62 & 20.88 & \ips \\
PS15dfz & 09 54 59.64 & +04 14 08.1 & 148.74850 & +4.23558  & 20151023.62 & 57318.62 & 20.68 & \ips \\
PS15dfu & 09 34 24.28 & +06 48 01.0 & 143.60117 & +6.80028  & 20151023.62 & 57318.62 & 21.19 & \ips \\
PS15dft & 09 33 09.38 & +10 28 02.2 & 143.28908 & +10.46728 & 20151023.62 & 57318.62 & 19.41 & \ips \\
PS15dfw & 09 44 11.65 & +04 54 52.1 & 146.04854 & +4.91447  & 20151024.60 & 57319.60 & 21.00 & \ips \\
PS15cwc & 09 59 01.22 & -03 48 04.3 & 149.75508 & -3.80119  & 20151024.61 & 57319.61 & 21.11 & \ips \\
PS15cqx & 10 05 03.70 & -06 29 44.7 & 151.26542 & -6.49575  & 20151024.61 & 57319.61 & 20.32 & \ips \\
PS15dgd & 10 27 26.07 & -14 58 20.1 & 156.85862 & -14.97225 & 20151024.61 & 57319.61 & 20.55 & \ips \\
PS15cqw & 09 45 06.43 & +01 17 02.0 & 146.27679 & +1.28389  & 20151025.60 & 57320.60 & 20.99 & \ips \\
PS15cwd & 10 08 06.70 & -14 25 08.5 & 152.02792 & -14.41903 & 20151025.62 & 57320.62 & 20.93 & \ips \\
\hline
\end{tabular}
\end{center}

\end{table*}

\begin{table*}
\caption{Classifications of all transient candidates in the field of 
GW150914. 
Discovery dates refer to the date of the 
first detection by Pan-STARRS. ``Spec. date" refers to the date of 
the classification spectrum. 
The Type designation of (SN)
means that a spectrum was not available, 
but the absolute magnitude and lightcurve data
available are consistent with a normal supernova.  The 
instrument or survey data are listed in ``Notes" and described in the text.  Some objects have no classification spectrum, 
but have 
spectroscopic host redshifts from NED. ``Phase Spec." refers to the estimated phase of the spectrum with respect to 
peak.  ``Exp. Date" refers to the estimated time of the explosion of the supernova (from the spectroscopic typing
information) with respect to the time of 
GW150914 
(the day of 14 Sep 2015) e.g. $-17$d means the SN is estimated to have exploded 17 days before 
GW150914. 
The uncertainties are discussed in the text. 
}
\label{tab:alltrans_spec}
\begin{center}
\begin{tabular}{ l l l l l l l l p{1cm} }
\hline \hline
Name       & Disc. Date & Spec Date & Type & Phase Spec & Spec $z$ &   Exp. Date  & Notes  \\
\hline
PS15cbm  &20150917.62   &  20151003  &  SN Ia  & $>$20d  & 0.059                     &      -17d                  & PESSTO \\
PS15ckm   &20151013.61   & 20151022    &  SN Ia   &   peak                & 0.080                 & +20d                                 & UH2.2m+SNIFS \\     
PS15cwm  &20151015.61    &  ...              &  (SN)        &   ...                     &  (0.22)                    &      ...          & SDSS DR12  photo$-z$, association uncertain  \\
PS15ccw  & 20150917.63   &   ...              &  (SN)     &     ...                    & 0.072                       &      ...           &  host $z$, SDSS DR12  \\
PS15cci   & 20150919.63    &  20150926  &  SN Ia   &  pre-peak        &  0.055                   & -1d                    & DBSP, SN1991T-like \\  
PS15ccx  &  20150919.63   &  20150927  &  SN Ia   &  15-20d        & 0.097                 &   -25d                & GCN18371 , LT+SPRAT    \\
PS15ccv   & 20150922.62   &  ...               &   (SN)     &  ...                       &0.071                       &       ...          &    host $z$, SDSS DR12    \\
PS15cel   & 20150923.63    &   20151012 &  SN II   & $>20$d     &  0.057                   & -2d                & PESSTO \\
PS15cki   & 20150923.64    &   20151020 &  SN II   & $>25$d  & 0.024                    &   +6d               &   UH2.2m+SNIFS      \\  
PS15cej   & 20151002.62    &   20151010  & SN Ia   &  peak                & 0.049      &   +8d                 & PESSTO\\     
PS15cek   & 20151002.63   &   ...              &  AGN     &  ...                     & 0.060                         &       ...           &   2MFGC 7447,  SDSS DR12 spectrum   \\
PS15cke   & 20151002.64   &   20151014 &  SN Ia   & 50d      &  0.030                    & -38d               &  UH2.2m+SNIFS\\
PS15ckf    & 20151003.65   &   20151020 & SN II    &  $>$15d     &  0.019                    & +13d                & UH2.2m+SNIFS \\  
PS15cwj    & 20151013.60   &  20151117   & SN Ia  &  21d                &   0.135           & +25d                    &             PESSTO\\
PS15cwi    & 20151013.61    &  20151118   & SN II   &   +30d     &  0.058                    & +32d                   & PESSTO \\
PS15ckj    & 20151013.61   & 20151022    &  SN II   &     peak                      & 0.020       & +33d                                       & UH2.2m+SNIFS, in IC593 $z=0.02018$\\        
PS15cko    & 20151013.62  &  20151025   &  SN II    &     post-peak                       &    0.217                       &  uncertain                     & SNIFS,PESSTO         \\       
PS15ckh    & 20151013.62   &  ...               &  QSO     &    ...                     & 1.523                        &     ...                   &   Lensed QSO [IBB2003] J0924+0219 \\
PS15cvz    & 20151013.62   & ...                & (SN)       &    ...                       & 0.069                    &     ...                     & host $z$, SDSS DR12\\
PS15cvy    & 20151013.63   &   20151022 & SN II          &  young             &        0.046     &      + 33d                    & UH2.2m+SNIFS \\
PS15ckn    & 20151014.62  &  20151027   &  SN Ic   & +7d                    &               0.08  & +23d                      & UH2.2m+SNIFS \\   
PS15ckk    & 20151014.62  & ...                & AGN        &   ...                     & 0.058                     &                ...                    & NPM1G-09.0361	     \\
PS15dfr &   20151015.60       & ...            & (QSO)      &       ...                &           (0.549)  &                        ...      &      Coincident with MLS140206-091829+114011 \\ 
PS15dfs &  20151015.60        & ...           & (SN)               &       ...                &           (0.424)  &                        ...         & Offset from host galaxy, SDSS DR12 photo$-z$   \\  
PS15dfy &   20151015.61        & ...         &  (QSO)                &       ...                &           (0.582)  &                        ...           &  Likely QSO, MLS130216-095249+063804  \\
PS15cvv   & 20151015.62    & ...               &  (SN)        &   ...                     &  ...                         &                ...                &  hostless\\
PS15cmr   & 20151015.62   &  20151027  & SN Ia &  +7d         & 0.093                            & +18d                   &  UH2.2m+SNIFS , host $z=0.093$ SDSS DR12\\      
PS15cmq  & 20151015.62   &  20151117    & SN II     &   +11d          &   0.065                  & +48d                    & PESSTO \\
PS15cvx   & 20151015.62   &   ...              &  (SN)         &  ...                 &    0.152                  &                          & host spec$-z$ 2dFGRS                 \\
PS15dfv &   20151015.62        & ...         & (QSO)               &       ...                &           ...  &                        ...           & Likely QSO SDSS J094138.31-021021.7 \\ 
PS15cwa   & 20151015.63   &   ...              &  (SN)       &  ...                   &  ...                         &                      ...                   & 3\farcs9 from host galaxy                  \\
PS15cwk   & 20151015.63   &   ...              &  (SN)        &     ...       &       ...                             &                    ...                 & 15\farcs4   from host galaxy    \\
PS15cms   & 20151017.62   &  20151024  &  SN Ia  &  peak               & 0.065                   &  +22d                           &  UH2.2m+SNIFS\\         
PS15cvw   & 20151018.61    &   ...              &  (SN)      &  ...                   &      ...                      &                           ...                 & faint, $r=22.7$ host galaxy        \\
PS15cmp   & 20151019.58   &  ...              &   (SN)  &  ...                         & 0.097                     &     ...             &  host $z$,  SDSS DR12     \\
PS15crh    & 20151019.58    &   ...              &  (SN)           & ...              &   ...                          &                      ...               & faint, $r=21.9$ host galaxy       \\
PS15cwh   & 20151019.58    &  ...              &   (SN)  &     old                      & 0.028                   &          ...         &  host $z$, SDSS DR12,  old        \\
PS15cri     & 20151021.60    &     ...              &  (SN)           &      ...                 &               ...       &                    ...               & hostless    \\
PS15cwb   & 20151021.61    &  20151121 & SN Ia   &  +20d            & 0.145                      &  +30d                          &   PESSTO \\
PS15dgc &  20151021.61        & ...         & (SN)               &       ...                &           0.056  &                        ...           &  Offset 6\farcs6 from host, spec$-z$ in NED \\
PS15cwe   & 20151021.63    &    ...              &  (SN)           &        ...               &        ...         &                      ...             & hostless     \\
PS15crk    & 20151021.63     &  ...              &  (SN)           &     ...                  &         ...        &                         ...          & 6\farcs5 from host galaxy \\   
PS15dga &  20151021.63        & ...         & (SN)               &       ...                &           ...  &                        ...          &  Offset from host galaxy  \\ 
PS15dgb &  20151021.63        & ...         & (SN)               &       ...                &           ...  &                        ...          &  Offset from host galaxy in PS1 \\
PS15dfx &   20151023.60        & ...         & (SN)               &       ...                &           ...  &                        ...           &  Offset from host galaxy in PS1 \\
PS15cwg   & 20151023.61     &  ...              &  (SN)         &    ...                   &         ...            &                     ...              & 0\farcs7 from host galaxy    \\
PS15crj   & 20151023.62       &   ...              &  (SN)           &        ...               & (0.41)              &    ...               & 1\farcs9 from host galaxy, SDSS DR12 photo$-z$   \\
PS15dft &   20151023.62        & ...          & CV               &         ...                &           ...  &                        ...            & Known CV, ASASSN-15se \\  
PS15dfu &  20151023.62        & ...          & (SN)               &       ...                &           0.0866  &                        ...    & 2\farcs7 from host galaxy SDSS DR12 spec$-z$      \\   
PS15dfz &   20151023.62        & ...         &  QSO     &       ...                &           0.844  &                        ...          & QSO, SSDR12 spectrum \\
PS15dfw &  20151024.60        & ...         & (SN)               &       ...                &           (0.113)  &                        ...      &  Offset from host galaxy SDSS, DR12 photo$-z$     \\  
PS15cwc   & 20151024.61     &   ...              &  (SN)           &     ...                  &   ...              &                   ...                  & hostless   \\
PS15cqx   & 20151024.61      &    ...              &  (SN)  &       ...                    & 0.054              &    ...                &  host $z$  from NED  \\
PS15dgd &  20151024.61        & ...         & (SN)               &       ...                &           ...  &                        ...          & Offset 5\farcs6 from host\\  
PS15cqw   & 20151025.60      &  ...              &  (SN)           &   ...                    & (0.22)          &                       ...            & 1\farcs9 from host galaxy, SDSS DR12 photo$-z$      \\
PS15cwd   & 20151025.62      &  ...              &   (SN)              &       ...                &           ...  &                        ...           & 5\farcs5 from host galaxy       \\
\hline
\end{tabular}
\end{center}

\end{table*}

\begin{figure*}
\includegraphics[width=16cm,angle=0]{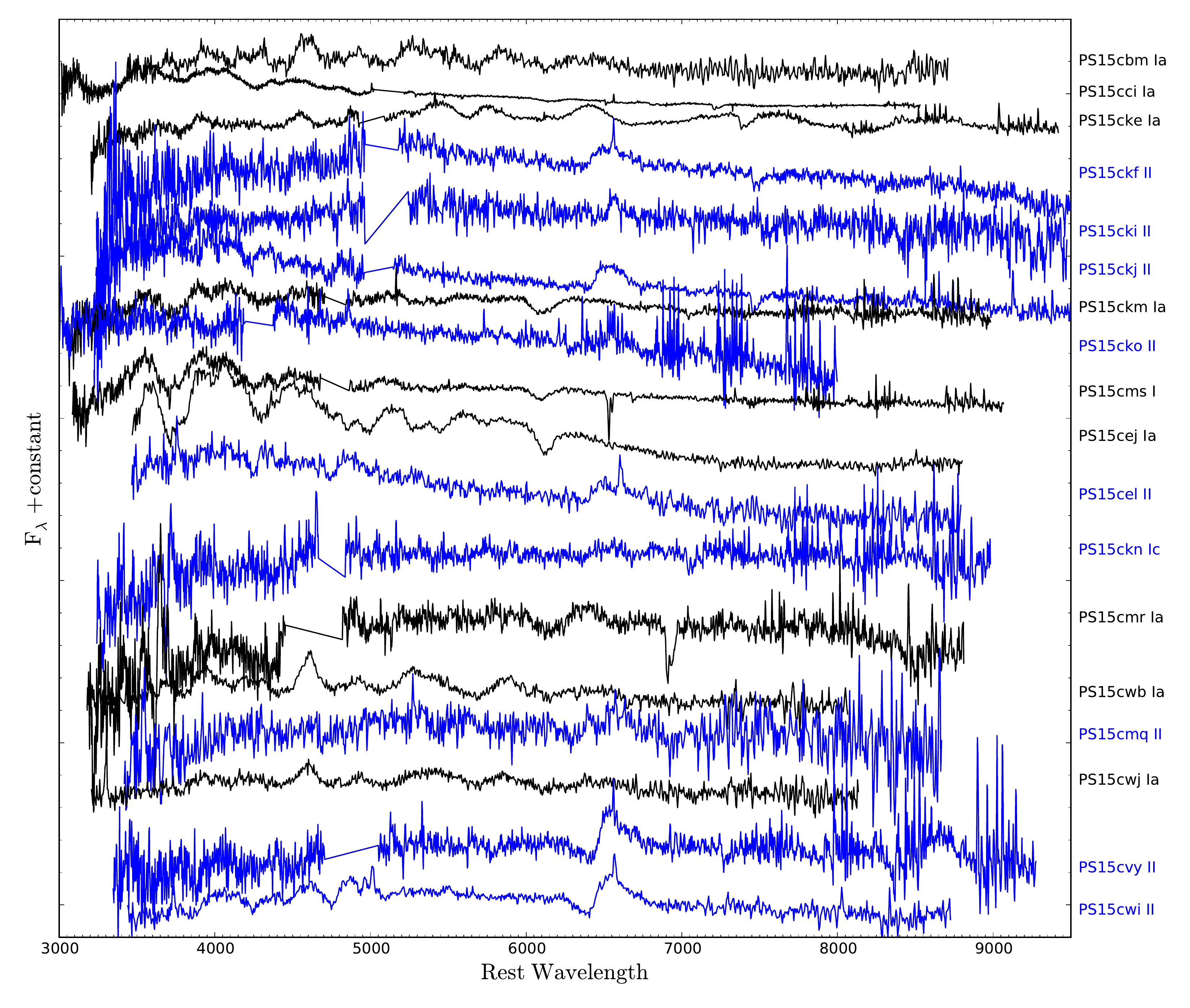}
\caption{ Spectra of all the candidates for which data were taken. The type Ia, and likely type Ia are shown in black, and the core-collapse SNe are shown in blue.}
\label{fig:specall}
\end{figure*}

\subsection{The search for transients in the Pan-STARRS1 fields}
\label{sec:search}

The PS1 3$\pi$ stack in the \rizy\ filters was used as the reference image to perform image subtraction on each of the images taken 
by Pan-STARRS during the campaign described  in Section\,\ref{sec:obs}.  Each of the individual frames were differenced, 
and no stacks of overlapping frames (due to small dithers) were 
used. The difference images were processed by the IPP in Hawaii, 
producing catalogues of potential astrophysical transients. As normal with CCD based difference image surveys, the false positives
introduced by the processing outweigh the real objects by 2-3 orders of magnitude \citep[e.g.][]{2013MNRAS.435.1047B,2015MNRAS.449..451W,2015AJ....150...82G}. The following lists the steps 
taken during our filtering of this data stream. 
In total 10,173,434 difference detections were processed.

\begin{itemize}
\item Pre-database rejection : during the database ingest process objects from the detection catalogues 
are rejected which have any one of 10 poor quality FLAGS set by IPP. These image quality flags are determined by IPP and are a fairly robust way of removing artefacts by detecting saturated and defective pixels, inconsistent sky variance, and objects for which the XY moments can not be determined or which have very poor point-spread-function fits.  All objects without any of these FLAGS set are ingested into the Transient Science Server Database (a MySQL relational database). 
The FLAG check resulted in the rejection of 5,133,123 difference detections.  The remaining 5,040,311 detections were ingested into the MySQL database. Detections within 0.5 arcsec of other detections are aggregated into objects which resulted in the creation of 3,060,121 distinct objects.

\item To further restrict the numbers, we require a minimum of 2 separate detections on separate images
each with a  signal-to-noise greater than 3$\sigma$ and a value for MOMENTS\_XY$< 3.6$ (the second moment in$xy$). The two detections must be within 0\farcs5 separation and the RMS of positional scatter (of all detections) must be $\leq0\farcs25$.  This immediately 
helps remove any uncatalogued minor planets that move between two images, cosmic ray hits and some types of artefacts.

\item Like most CCD mosaic arrays, the giga-pixel camera (GPC1) suffers from cross-talk, such that bright stars cause electronic ghost (PSF-like) sources on other cells and arrays with well known rules. The size of the FOV means that the number of bright stars that cause such cross-talk ghosts is significant and care must be taken to identify them. 
 The IPP checks and sets an \texttt{ON\_GHOST} flag if an object has a parent bright star at a known cross-talk position. If an object has this flag set on any one detection, then it is rejected. 

\end{itemize}

These basic checks, and an insistence of 2 separate detections 
resulted in a total of 308,262 objects which were labelled as candidate  
astrophysical transients in the Transient Science Server database. 
To further refine the search, the following filters were applied. 

\begin{itemize}
\item The ratio of positive, negative and NaN (masked) pixels within an object aperture is used to reject objects with significant numbers of negative or masked pixels. This 
efficiently removes ``dipoles'' which come from convolution artefacts during differencing.  The current thresholds are set as below, where  $nPos$ = number of positive pixels,  $fPos$ = flux from positive pixels  within the aperture (similar definitions for $nNeg$ and $fNeg$) 

  \begin{itemize}
    \item diff\_npos $= nPos >  3$
    \item diff\_fratio $= fPos / (fPos + fNeg) > 0.6$
    \item diff\_nratio\_bad $= nPos / (nPos + nNeg) > 0.4$
    \item diff\_nratio\_mask $= nPos / (nPos + nMask) >\ 0.4$
   \item diff\_nratio\_all = $nPos/(nPos + nMask + nNeg)>0.3$
  \end{itemize}
\item All objects within $1\farcs0$ of a catalogued star in the 2MASS Point source Catalogue, the Guide Star Catalogue and a PS1 stellar catalogue 
\citep[taken from an extended ubercal catalogue based on][]{2012ApJ...756..158S,2015arXiv151201214F} were rejected. 
The 1 arcsec search radius was used for 2MASS, GSC and PS1 Ubercal stars.  For the SDSS star catalog we varied the association radius between 19 (for an $r=13$ star) and 2.5 arcsec (for an $r=18$ star).  

\item All objects within a variable radius of very bright stars are rejected, as likely diffraction spikes, saturation bleeds or other optical effects. The radius of this mask applied is magnitude dependent varies from 15 to 
40 arcsec for stars between 8 to 13 magnitudes.   We typically lose about 0.5 per cent of the sky area to this bright star masking. 
\end{itemize}

These three filters reduced the number of objects to 48,550.  The Transient Science Server then located 300$\times$300 pixel postage stamp image data for all these objects that pass the above filters. We also cross match against the Minor Planet Centre and identify known movers. Although the requirement of a positional scatter of $RMS<0\farcs25$ inherently removes many movers,  there are some slow moving objects that get through this filter and these 
real, slow movers are useful for consistency checks on the image
analysis that follows (in this case there were only 3 known slow movers that made it this far).
A machine learning algorithm was 
run on these pixel stamps that calculates a real-bogus factor between 0 (bogus) and 1 (real) as described in \cite{2015MNRAS.449..451W}.  The machine learning 
algorithm employed for PSST \citep{2015IAUGA..2258303H} is similar in concept to that described 
in \cite{2015MNRAS.449..451W} but has now been revised to be a Convolutional Neural Network, based on the approach of \cite{Ngiam11}  and further details are available in 
\citep{Wright15}.   For completeness, the threshold cut was chosen such that the machine allowed 10 per cent  false positives and 1 per cent missed detections.  The resulting number of objects remaining for human inspection was 5,861. This resulted finally in 
56 transients which had high real-bogus factors which we
confirm as  real astrophysical objects. We found a further 71 faint, more marginal objects that were either 
coincident with known AGN  \citep[in the catalogue of][]{2001A&A...374...92V} or coincident with
candidate AGN/QSOs from the MILLIQUAS catalogue \footnote{http://heasarc.gsfc.nasa.gov/w3browse/all/milliquas.html} \citep{2015PASA...32...10F}, or coincident with faint stars, or had faint detections on only 
one night. These are likely to be a combination of low-level AGN or stellar variability, faint movers, 
or in some cases convolution artefacts occurring when the image quality was sharper than the reference stack. They may also be faint SNe that 
were only detected on one night. 
We consider these as dubious objects for which we could not be certain they were of astrophysical origin and 
a list is available on request.

\section{Results of transients located}
\label{sec:trans}


Of the 
56
transients that we detected, 
a total of 
19 
were observed spectroscopically. 
18  
spectra are presented in this paper, with one classification from a GCN announcement by 
\cite{GCN18371}. 
A further 
13
have archival spectra of the host galaxies, providing host redshifts (with 4 of 
them being known AGN/QSOs.\footnote{These 4 were high significance transients that we kept in our good catalogue in case they showed any sign of being circum-nuclear transients rather than AGN/QSO variability.
We put the other faint and marginal sources which were coincident with candidate AGN/QSO in the 71 possible sources, as we were uncertain if they were real or difference image artefacts. }). 
The spectroscopic classifications and host redshifts are listed in Table\,\ref{tab:alltrans_spec}.
All transients
appear to be normal population supernovae/AGN/QSOs that we would expect to detect in the Pan-STARRS1 
survey fields  (together with 1 CV).  In the following subsections, we discuss the transients with and without direct 
spectroscopic confirmation and classification. 
Throughout the analysis we use $H_{\rm 0} = 72$\kms\,Mpc$^{-1}$, $\Omega_{\rm M} = 0.3, \Omega_{\Lambda} = 0.7$. 

\subsection{Spectroscopically classified transients} 
Table\,\ref{tab:alltrans_spec} lists the spectroscopic classifications of the 19 transients for which spectra 
were taken. We find ten type Ia SNe, eight type II SNe, one type Ic. None of these candidates  could be immediately linked
to a GW signal and none are markedly peculiar. They appear to be typical SNe that  are 
discovered and spectroscopically classified in wide-field surveys. 

The spectra of these supernovae allow estimations of the explosion epochs.  We
estimate the explosion epochs from matching the observed spectra with template spectra in 
both the supernova spectral matching tools SNID \citep{2007ApJ...666.1024B} and Gelato \citep{2008A&A...488..383H}. Uncertainties in these
values vary depending on the spectral type and the availability of good templates. Where we
quote uncertainties below these are the epoch ranges that are reasonably matched to the 
the spectra. \cite{2014ApJ...786...67A} showed that such spectral matching gives consistent results to lightcurve
data with non-detections. 
If any transient or supernova were to be plausibly linked to a GW trigger, 
then one would reasonably expect that the explosion epoch would be close in time to that
of the GW detection time.  From the classification phase of the 
spectrum, we can approximately estimate the time of explosion of the SNe and find that 
all but three or four have estimated explosion epochs more than 2 weeks before or after the time of 
GW150914.  The types and epochs as listed in Table\,\ref{tab:alltrans_spec} are therefore mostly unremarkable.
There are 5 transients worth some further  discussion and we provide specific sections on these below, 
while the rest are effectively discounted from being related and the information is available in Table\,\ref{tab:alltrans_spec}. 

\subsubsection{PS15cci}
PS15cci was discovered as a rising transient in a relatively faint and compact host galaxy 
(SDSS J091322.78+061047.6, $r=19.78\pm0.03$) with a photometric redshift of $z=0.08\pm0.04$. 
It was found in the first few days of our searching (19 Sept 2015, or 4 days after the trigger) and 
the rising nature immediately caught our attention. 
We classified it as a type Ia SN on 26 Sept 2015 (with a high S/N spectrum from the Palomar 200 inch and the DBSP spectrometer), with a spectroscopic
age of roughly 4$\pm4$\,days before maximum light, at $z=0.055$. 
This means that assuming a rise-time of 18$\pm$2 days from explosion to maximum light for 
type Ia SNe 
\citep{2011MNRAS.416.2607G}
our estimated explosion epoch is $-1\pm6$d from the detection of GW150914. The explosion epoch of 
PS15cci is broadly consistent with that of GW150914, but being a normal type Ia, at $z=0.055$
\citep[although resembling the bright SN1991T-like subclass;][]{1992AJ....103.1632P,1997ARA&A..35..309F}
we discount it as being  related. 

\subsubsection{PS15cel}
\begin{figure}
\includegraphics[width=8cm]{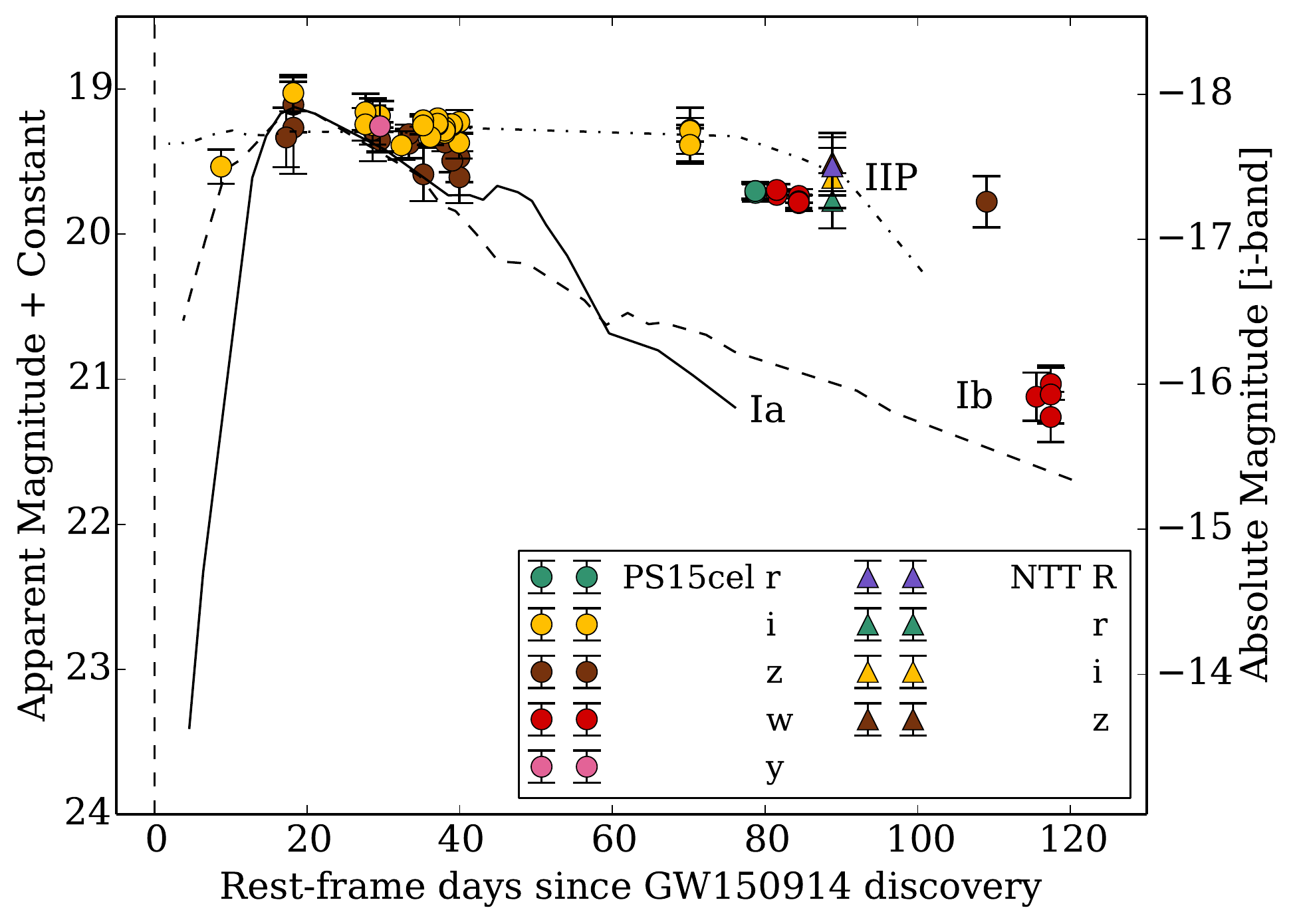}
\caption{The lightcurve of PS15cel with all data points  in the PS1 photometric system. Five points were
 taken in the PS1 $w-$band during routine PSST operations and can be considered to be similar to \rps (as seen in the close agreement
at 80 days).  The PESSTO $riz$ points (at  +75d) were also transformed to the PS1 AB mag system for meaningful comparison. The
right hand axis shows the absolute magnitudes in the \ips-band (the other filters can be considered similar, apart from the small
differences in foreground extinction between the bands). }
\label{fig:ps15cel}
\end{figure}

PS15cel was discovered on 23 Sept 2015 (9 days after the GW source detection), 
with a spectrum taken by PESSTO on 12 Oct 2015. It
is a type II with a broad and narrow component of H$\alpha$. The narrow component is 
very likely from the host galaxy (SDSS J093411.53+054644.7) which we measure at 
$z=0.057$. The lightcurve is plotted in Fig.\,\ref{fig:ps15cel} and is fairly flat  in the first 30 days.  
All photometric points on the lightcurve are from PS1 apart from one set of $BVRiz$ taken with EFOSC2 as part of PESSTO 
\cite[see][for a description of the filters]{2015A&A...579A..40S}. The transient reaches a peak mag of 
 \ips$=-18.0$ mag assuming only foreground Milky Way extinction. The plateau phase is observed to last at least 85 days, which is not unusual for a II-P SN, although the more luminous type II events tend to have faster 
decline rates as shown by the recent work of  \cite{2014ApJ...786...67A} and 
\cite{2016AJ....151...33G}. PS15cel is in the regime of the brightest type II SNe known, as 
defined by the Anderson et al. sample, and also declines relatively slowly (at 0.6 mag per 100d in the \ips band). 
While it sits on the edge of the brightness vs decline rate plot as defined in Fig.\,7 of  \cite{2014ApJ...786...67A}
it is not a far enough from this locus to be an  outlier.  It is also not an outlier in the peak magnitude vs
$\Delta m_{15}$ relation for type II SNe of \cite{2015arXiv151200733R}. 
The last point in the $r-$band shows the
SN is dropping off the plateau onto the nebular, tail phase which is powered by the decay of $^{56}$Co, 
which is  again normal behaviour for  luminous type II SNe \citep[e.g.][]{2013A&A...555A.142I}.

\begin{figure}
\includegraphics[width=8cm]{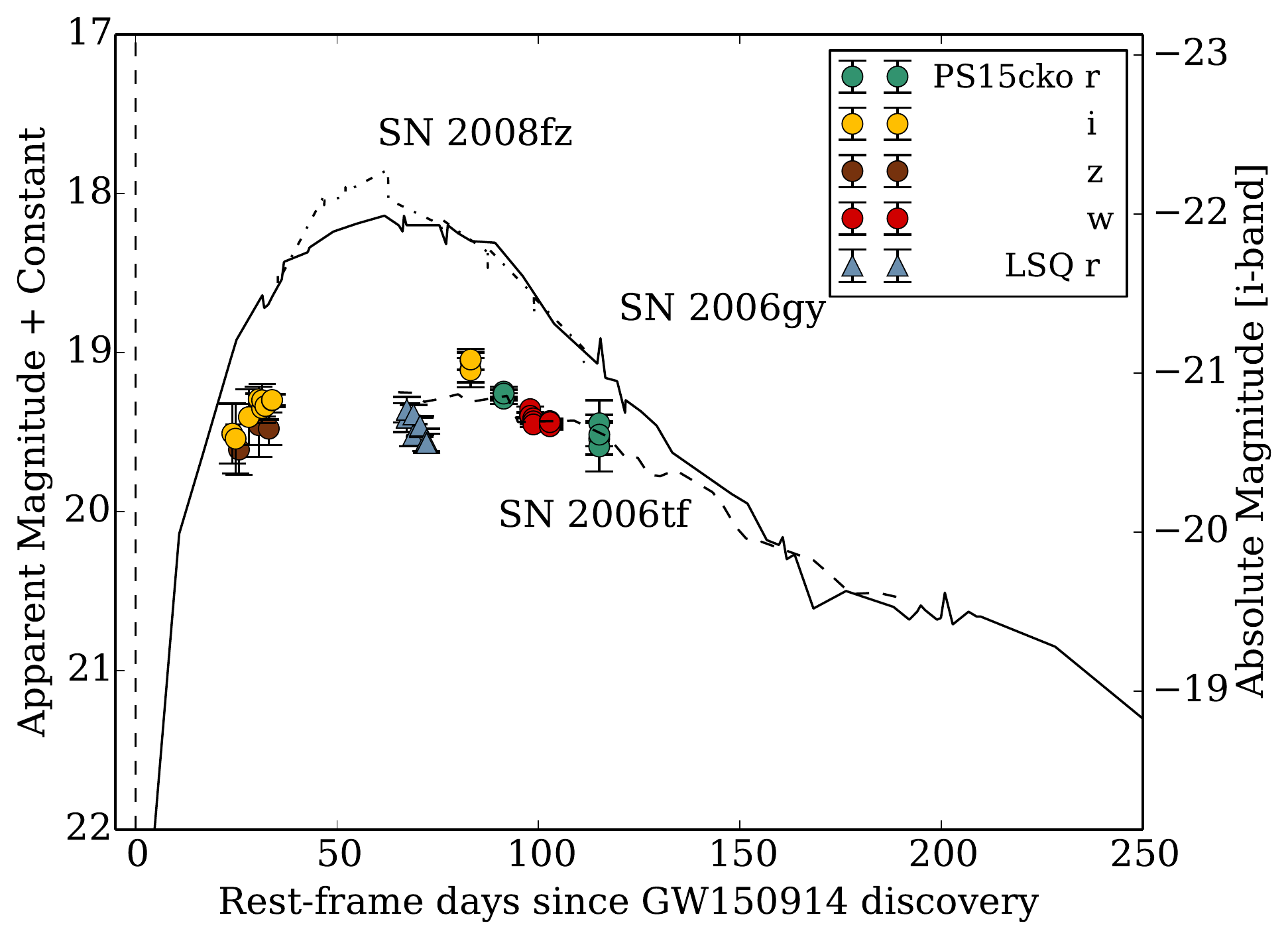}
\includegraphics[width=8cm]{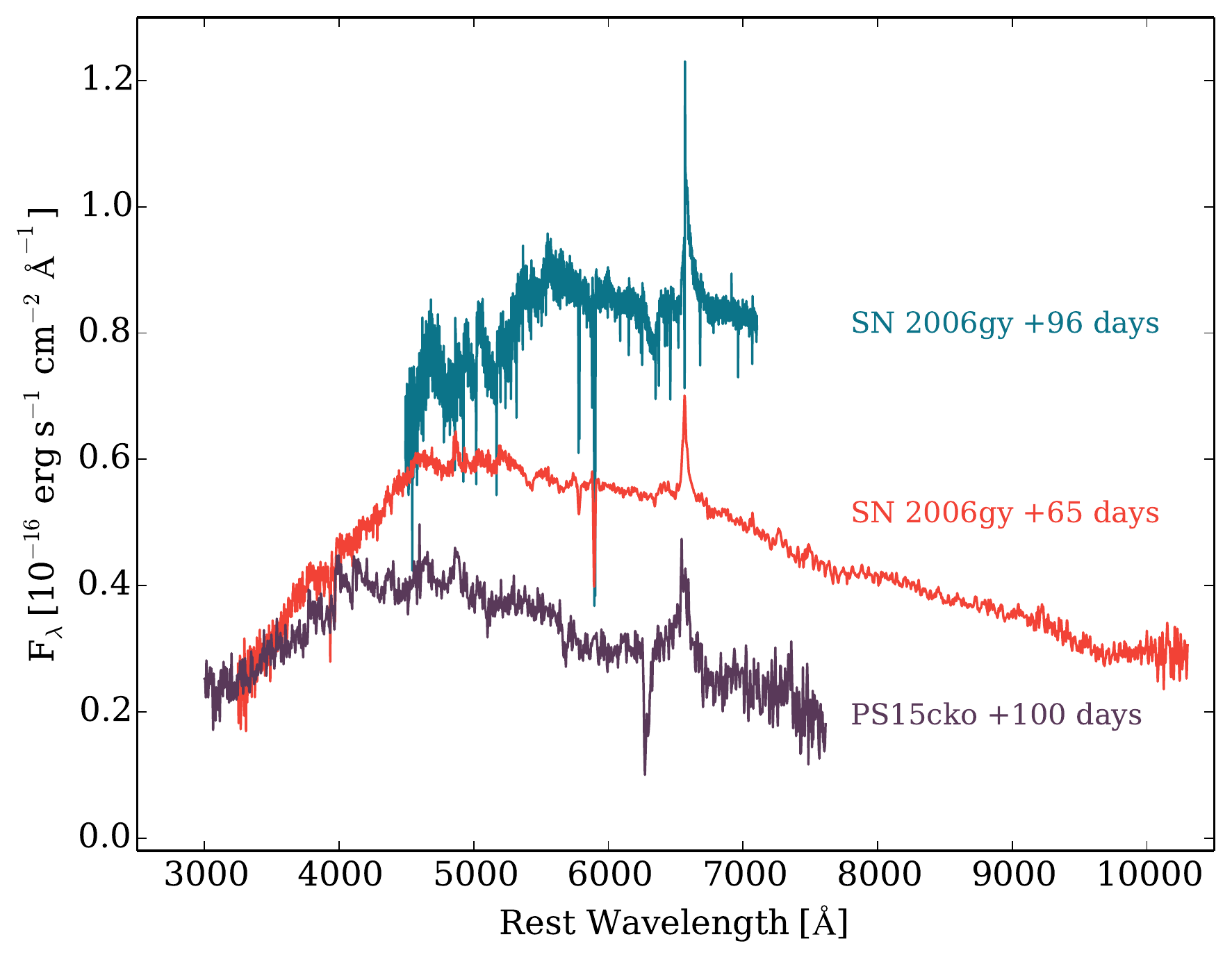}
\caption{{\bf Top :}   The lightcurve of the luminous PS15cko compared to two three other superluminous type IIn SNe. No host reddening has been applied to PS15cko. {\bf Bottom:} Spectral comparison of PS15cko with the two spectra of SN2006gy. Each spectrum has been corrected for foreground extinction only, since the host extinction to PS15cko is uncertain. The lightcurves and spectral comparison indicate this is a luminous type IIn SN.}
\label{fig:ps15cko}
\end{figure}

The PESSTO spectrum taken on 12 Oct 2015 (19 days after discovery) is plotted in Fig.\,\ref{fig:specall}.
 The P-Cygni absorption troughs of the Balmer line series 
are weak, and the normal Fe\,{\sc ii} lines
have not become prominent. The spectrum resembles  type II-P SNe at a few days past 
explosion (SN2004et, SN1999em and SN2006bp), but the Pan-STARRS lightcurve clearly 
dates the SN at least 19\,days after explosion. 
This SN bears some resemblance to the bright and luminous type II SNe SN1992H and SN2009kf
\citep{2010ApJ...717L..52B,1996AJ....111.1286C}
for which \cite{2010ApJ...723L..89U}
have proposed explosion energies of $2\times10^{52}$\,ergs, ejecta masses of $\sim$30\msun\ and 
$^{56}$Ni masses of $\sim$0.4\msun. To produce such an energetic explosion in a  very massive star, 
\cite{2010ApJ...723L..89U}
proposed  a rapid disk accretion onto a central black hole at the time of core-collapse. 
Although the time of discovery  was 23 Sept 2015, 9 days after GW150914, we
do not have data to rule out that the SN exploded before the discovery date and indeed the 
flat plateau would suggest that it did. Type II have rise times ranging from 5-10 days 
\citep{2015A&A...582A...3G, 2015MNRAS.451.2212G,2015arXiv151200733R} and as 
we did not detect that rise time, an explosion date close to the epoch of GW150914 is 
quite possible.  Although  the rapid disk accretion onto a central black hole model of 
\cite{2010ApJ...723L..89U} does involve a massive compact remnant it is certainly not
quantitatively similar to the BH-BH merger that is proposed for GW150914
\citep{theprizepaper}. Therefore  one could not reasonably link the two on the basis of the data in hand and 
the predictions for the GW 
signal from massive asymmetric core collapse 
realistically sets the distance limits at 10-50 Mpc
\citep{2012ApJ...761...63P}.

\subsubsection{PS15cki}
This is offset by 3\farcs3 from a faint diffuse host galaxy visible in the PS1 reference image and 
measured as $r=19.91\pm0.11$ in SDSS DR12 (SDSS J092827.13+080048.6). 
The spectrum is low signal-to-noise, but a broad H$\alpha$ line is visible indicating a type II SN
with an age around 30$\pm10$ days after explosion. This would put the explosion epoch at
about MJD=57285$\pm10$ which broadly covers the GW150914 detection at MJD=57279.41. 
However the absolute magnitude of $M_i=-16.1$ and redshift of $z=0.024$ imply this
is a fairly normal type II SN in a faint host galaxy
and unrelated to the GW burst 

\subsubsection{PS15ckn}
The SNIFS spectrum is of moderate quality with fairly low signal (S/N$\sim10$ after binning
to 10\AA\ per pixel), but the spectrum can be classified as a type Ic SN at about 7$\pm3$\,days after
maximum light. Type Ic SNe have rise times of around 14$\pm5$\,days 
\citep{2011MNRAS.416.3138V,2015A&A...574A..60T,2016arXiv160201736P}
hence the explosion date of this can be estimated at MJD=57302$\pm8$, which is 
around $23\pm8$ days after GW150914.  While some broad lined type Ic SNe are associated with 
GRBs
\citep{2006ARA&A..44..507W}
and black hole formation, 
PS15ckn does not appear to be a broad lined or energetic Ic. At a redshift of $z=0.08$, our 
brightest measured 
magnitude is $M_i = -18.4$ which implies it is not unusually luminous.  The SN is 2\farcs9 
from the core of a spiral galaxy (probably SA-SB type in the PS1 images)  which 
is also a UV source  (GALEXASC J101329.20-100003.6).  
In conclusion, the data we have
suggests PS15ckn unlikely to be unrelated to the GW source.

\subsubsection{PS15cko}
A spectrum was taken  with SNIFS on 2015 October 25, which was 12 days after the discovery. 
The signal to noise is low and the redshift is not securely determined from either broad lines
from the transient or narrow lines from the host galaxy. There is a resolved host galaxy in the 
PS1 image reference stack separated by 3\farcs4 which is likely the host. The spectrum is shown in 
Fig.\ref{fig:specall},  has no strong and distinct features. The lightcurve shows that 
when it was discovered on 2015 October 13, it continued to rise in the $i$ and $z-$bands for about 12 days. 
The discovery date is already 29 days after GW150914, so for it to be related would require a very 
long rise time which would be in the regime of superluminous SNe \citep{2015MNRAS.452.3869N}
or very bright type IIn explosions \citep{2007ApJ...666.1116S}. 

It was recovered in both routine La Silla QUEST  \cite{2013PASP..125..683B} 
and PSST \citep{2015IAUGA..2258303H,2015ATel.7153....1H} operations 
on MJD=57361 to 57419, more than 100 days past first discovery and still relatively bright at 
\ips=20 and $r_{\rm LSQ}=19.4$ (LSQ filter has a broad optical filter and this mag is approximately
an AB $r-$mag equivalent). 
A PESSTO spectrum taken on 4 Feb 2016 conclusively sets it at $z=0.217$, with a 
very strong H$\alpha$ 
line profile in emission. Figure\,\ref{fig:ps15cko} shows that this is likely a luminous SN IIn. 
The initial peak means a restframe $r$-band magnitude of $M_{r} = -20.6$, assuming  the 
observer frame $\ips$ transforms to the $r$-band with a nominal $k$-correction of $2.5\log (1+z)$. 
Figure\,\ref{fig:ps15cko} shows the lightcurve comparisons with  SN2006gy, SN2006tf
and SN2008fz  \citep[data from][]{2007ApJ...666.1116S,2007ApJ...659L..13O,2008ApJ...686..467S,2010ApJ...718L.127D}, 
The explosion epoch is uncertain. The lightcurve implies it was probably after GW150914
but the range in uncertainty is large enough that it could be compatible. We currently discount this 
as related to  GW150914 on three counts.  There are no plausible models linking IIn SNe to GW production, 
it is much further than the estimate for GW150914 and the explosion epoch is not obviously compatible with 
the time of GW150914. 

\subsubsection{PS15ccx}
This was discovered on MJD=57284.63 at
\zps$=19.42\pm0.14$. A resolved host galaxy is 
visible in SDSS and the PS1 stack, 
at 4\farcs8 distant (SDSS J081804.11+041807.9). 
After we reported a preliminary list of transients to the 
Ligo-Virgo EM follow-up GCN circulation list, 
a spectrum of PS15ccx was taken with the 
Liverpool Telescope by 
\cite{GCN18371}. They report it as a SN Ia 
at age 15-23 days post maximum at a
redshift $z\simeq0.097$.

\begin{figure}
\includegraphics[scale=0.42]{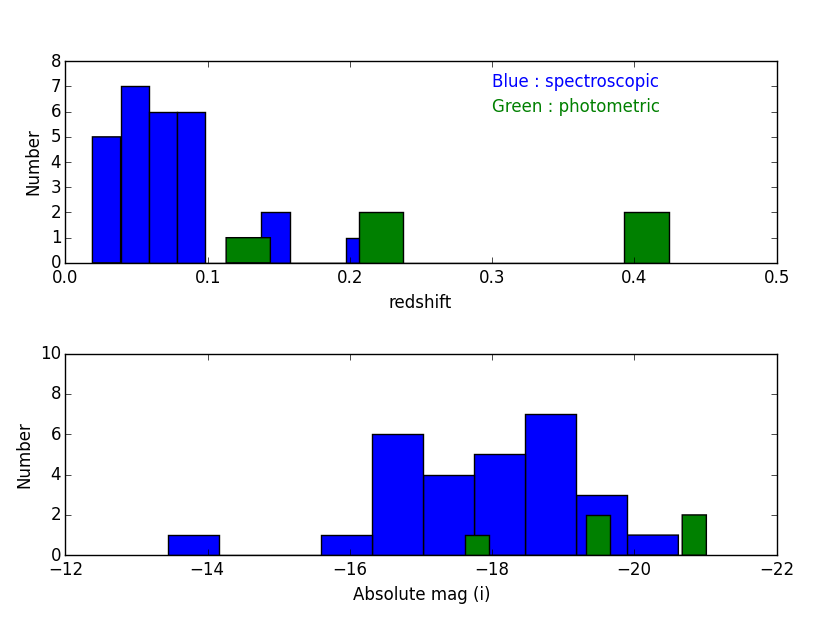}
\caption{Histograms of the redshift distribution 
and absolute magnitudes ($M_{\ips}$) of the confirmed SN or likely SN transients. The objects in blue either have 
a spectroscopic redshift from the transient itself, or host galaxy. The  objects in green have 
 a photometric redshift of the host galaxy from SDSS DR12. The AGN/QSO 
candidates are not included. All magnitudes are in  
\ips apart from one (PS15ckf) which is in \zps.}
\label{fig:zMi_hist}
\end{figure}

\subsection{Transients with no spectroscopic classification} 


For those transients that do not have spectroscopic data, we have assessed their nature
from their proximity to host galaxies and the spectroscopic, or photometric redshifts of those
hosts. This allows an estimate of the absolute magnitude of the transient, which is the 
absolute magnitude of the brightest Pan-STARRS1 point in the lightcurve. In doing this, 
foreground extinctions from 
\cite{2011ApJ...737..103S} 
are applied.  
Where the association with a host galaxy is 
likely (where the transient is within a projected distance of 30\,kpc), this is then a lower limit to the 
absolute peak magnitude since the transient may have been brighter before PS1 discovery and 
internal host galaxy reddening may apply. However it serves to illustrate which transients are consistent  
with the standard, known supernova population. 
In addition to the 
19 spectroscopically classified transients, and 
4 spectroscopically known 
AGN/QSOs, a  further 
9 of the 43  were associated with galaxies with host redshifts. There are 
3 objects which 
are either known CVs or likely QSO/stellar variables. Which leaves 
21 objects with no confirmed spectroscopic redshifts (of these, 5 have photometric
redshifts).  These 21 objects
are discussed in the following 
subsections and we outline the most probable nature.

\subsubsection{PS15ccw}
\begin{figure}
\includegraphics[scale=0.42]{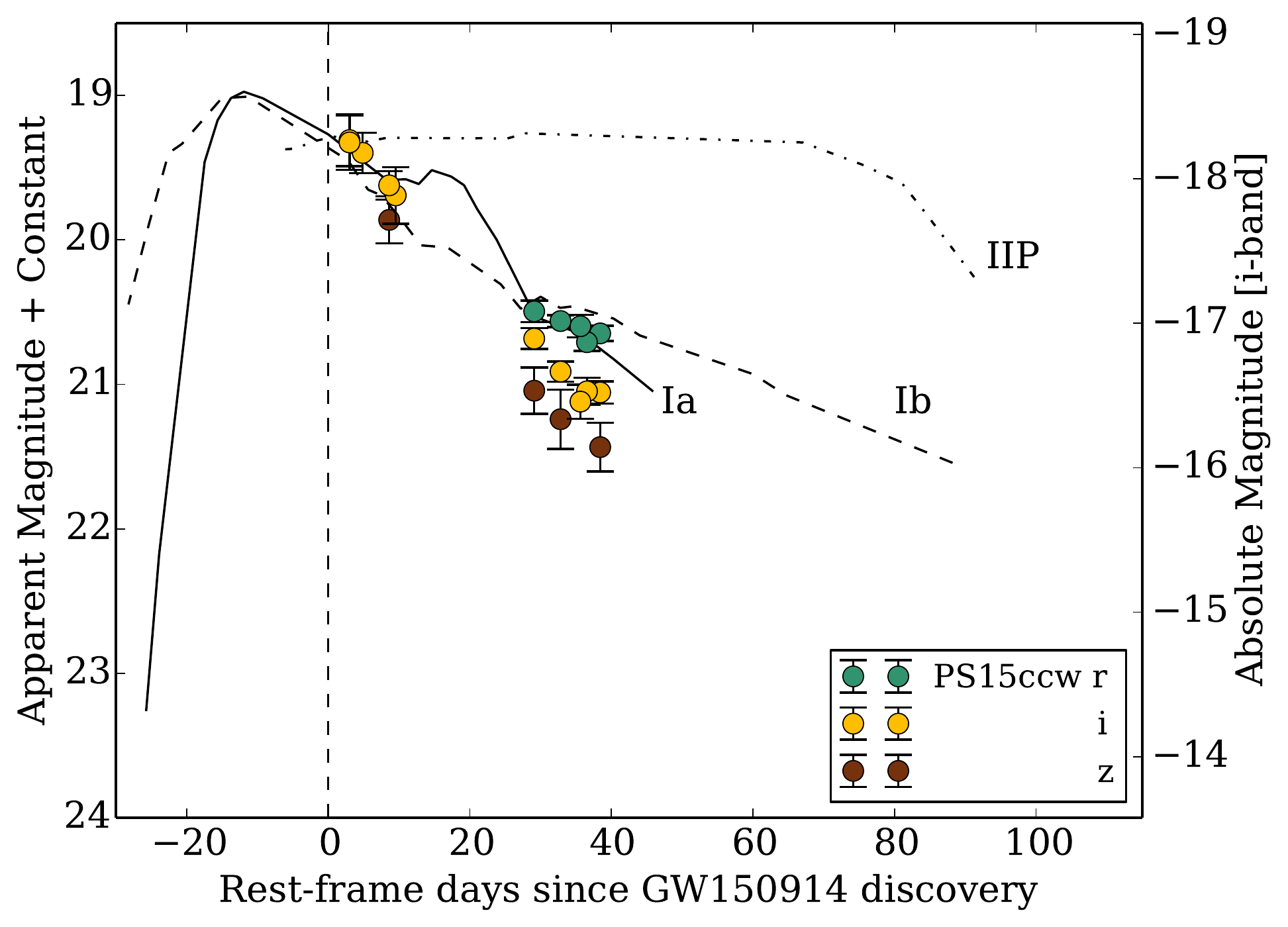}
\includegraphics[scale=0.42]{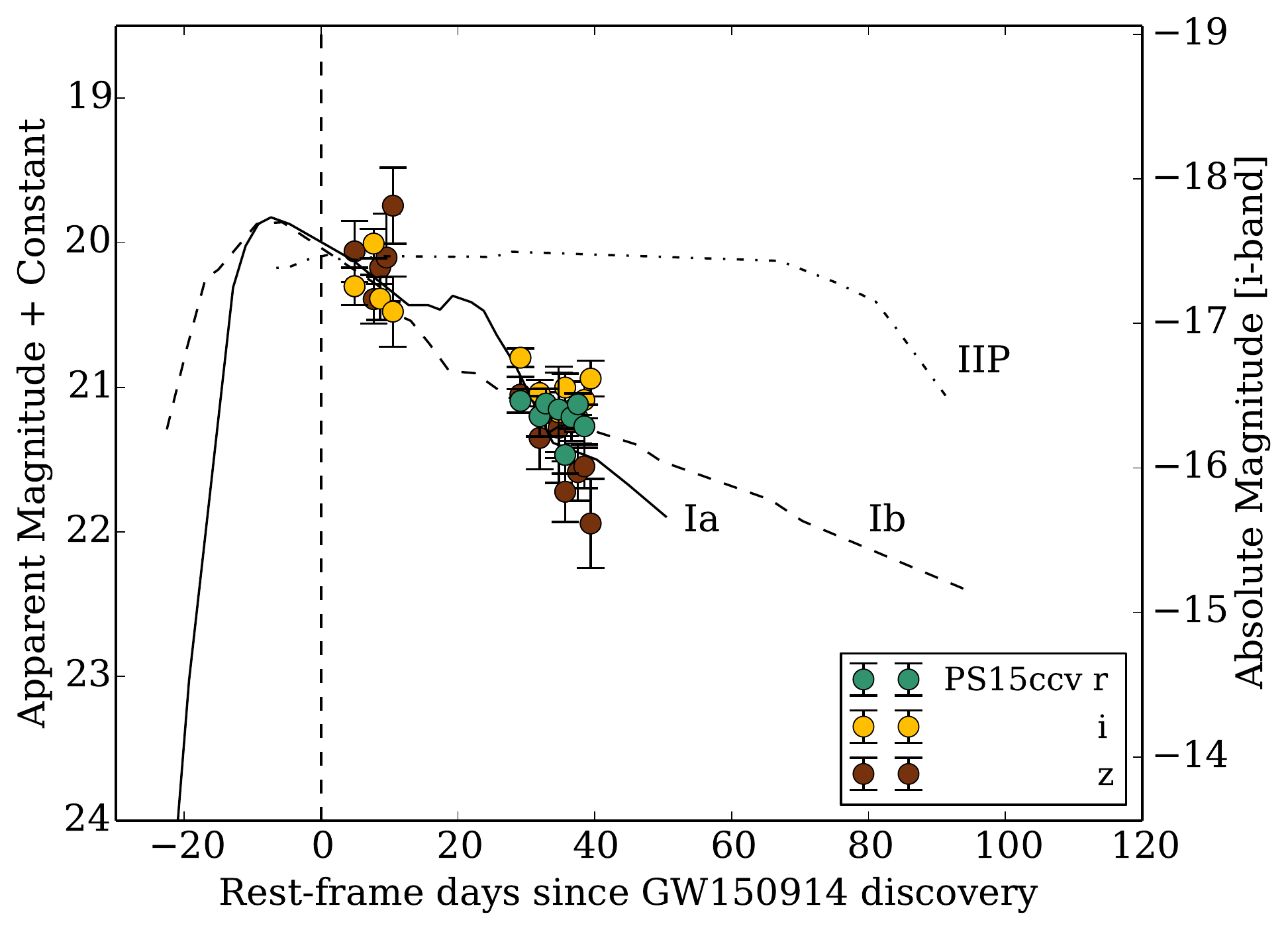}
\caption{Comparison of the measured lightcurve of PS15ccw and PS15ccv in the  $riz_{\rm P1}$ filters with three typical 
SNe at similar redshifts observed in the same filter systems.}
\label{fig:ps15ccw}
\end{figure}

The redshift of the closest galaxy  (SDSS J085730.63+043202.3)  is $z=0.072$ from SDSS DR12. 
PS15ccw is  6\farcs2 from the galaxy core which is a  projected physical distance of 8.2\,kpc. 
At  this  distance the absolute magnitude of the brightest PS1 point is 
$M_{\rm i}=-18.2$. We collected a lightcurve covering about 30 days in $gri_{\rm P1}$, 
and Fig.\,\ref{fig:ps15ccw} compares  lightcurves of typical 
supernovae. The three SN template lightcurves are an $i$-band
lightcurve of SN2005gb (type Ia, $z=0.09$) from the Sloan Digital Sky Survey-II Supernova (SN) survey
\citep{2008AJ....136.2306H} and two from the Pan-STARRS1 Medium Deep Survey. 
PS1-12bmp is a type II-P from the sample of \cite{2015ApJ...799..208S} and the \ips lightcurve
is shown here ($z=0.063$). PS1-13avb is a type Ib \citep{2015MNRAS.449..451W} at $z=0.0705$
and again the \ips lightcurve is shown. Each lightcurve has been scaled (arbitrarily) to the same peak magnitude 
of 18.8, but no $k-$corrections  have been applied since all the SNe are at similar redshifts. 
The time axis has been corrected for time dilation for each and is in the restframe. 
This  straightforward comparison indicates that the location, lightcurve and absolute magnitude of the transient
is compatible with it being a supernova in this galaxy and likely either a type Ia or type Ib/c. It is 
mostly likely a type Ia, since its peak absolute magnitude would quite bright for a type Ib/c. 

\subsubsection{PS15ccv}
This transient is located very close to the core (within 0\farcs6) of a spiral galaxy with 
a spectroscopic redshift of $z=0.071$ (SDSS J085523.02+044118.7), or a 
projected distance of 0.8\,kpc. The absolute magnitude of the brightest PS1 point is 
$M_{\rm i}=-17.9$. The transient is detected with a declining lightcurve 
in \rps\ips\zps over a period of 30 days. Similar to PS15ccw, a 
comparison to the same type Ia, Ib and II SNe in the 
same filters shows the lightcurve to be compatible with a SN
of type Ia or type Ibc. Again, 
the location, lightcurve and absolute magnitude of the transient
is compatible with it being a supernova in this galaxy and likely either a type Ia or type Ibc.  Almost certainly this implies that we discovered PS15ccw and 
PS15ccv 20-30 days after explosion, during their decline phase and hence they exploded 10-20
days before the GW detection. 
Another possibility is that they are both very old type II-P SNe which fall off the plateau between the first and 
second group of points.  Either way, we rule these out as related to GW150914. 

\subsubsection{QSO variability : PS15ckh, PS15ckk, PS15cek and PS15dfz}
PS15ckh is  the known QSO SDSS J092455.87+021924.9 which was studied in detail by 
\cite{2003AJ....126..666I} and shown to be a multiply lensed QSO. The high resolution images 
presented in 
\cite{2003AJ....126..666I}
resolve the source into 5 separate components, two of which are 
confirmed as the lensed QSO at a redshift of $z=1.524$.  In the epoch of the brightest
flux excess (on MJD=57310.61218), we measure $i_{\rm P1}$=19.2 in the difference image
at the position of component A (which Inada et al. measured at $i=18.87$). 
 We clearly resolve additional flux excess at a position
1\farcs9, directly south, at the  position of component B of 
\cite{2003AJ....126..666I}.  This indicates that both A and B have flux excesses in these epochs
at approximately the same ratio as measured in 
\cite{2003AJ....126..666I}. 
We do not detect a flux excess from component C in the difference images. 
Further analysis is warranted, but outside the scope of this paper. 

PS15ckk is detected as a bright transient in \ips, \zps, \yps $\sim 16 - 16.4$ in the difference images. 
However this is the known, catalogued variable AGN  NPM1G-09.0361 in  \cite{2001A&A...374...92V}.   
PS15cek is a known low redshift galaxy with an active nucleus also identified 
in  \cite{2001A&A...374...92V}  and known as 2MFGC 07447  (amongst other names)
in NED. The flux excess measured for PS15cek is $\ips\simeq17$ over 25 days. 
PS15dfz is the known object SDSS J095459.62+041408.3 which is a QSO 
at $z=0.844$ (with $i=19.28$) and is detected with a flux excess of \ips$=20.65\pm0.05$
on two epochs.

\subsubsection{PS15cvz}
This object is 3\farcs4 from the galaxy SDSS J100541.42+010530.0, which has a 
spectroscopic redshift of $z=0.069$  and is therefore 4.3\,kpc in projection. 
It was discovered at $\ips=19.6$ and detected twice on separate nights by PS1, 
the first being 2015 October 13 some 29 days after GW150914. It is also visible in a SNIFS 
image taken on 2015 October 28. The absolute magnitude of this first point is $M_i = -17.8$, 
and hence is compatible with being a supernova in this galaxy. 

\subsubsection{PS15cwm}
This is detected 5 times over a period of 10 days beginning on  57310.60, or 31 days after GW150914. 
The brightest magnitude reported is  $\ips=20.9$ and there is a series of non-detections 
stretching to about 20 days before the first PS1 point.  There is a galaxy with an early 
type spectrum at $z=0.224$  (SDSS J093001.39+065800.0) 
which is 13\farcs8 south east of the transient. This corresponds
to a distance of 48\,kpc, which is somewhat too large to be confident of association. 
A closer, but fainter galaxy  ($i=18.7$; SDSS J093001.16+065819.8) is 7\farcs4 distant (or 26\,kpc) 
and has a photometric redshift of $z=0.26\pm0.03$.  A redshift of $z=0.22\pm0.07$ 
implies $M_i  = -19.1^{-0.7}_{+0.9}$, which is within the plausible ranges of normal supernovae. 
It is likely that the transient is associated with one of these galaxies, or with an undetected 
dwarf at a similar redshift. The lightcurve evolution is compatible with a supernova. 

\subsubsection{PS15cvv}
This transient is detected multiple times in filter \ips between 57310.61 and 57320.62
with a fairly flat evolution at \ips=$20.20\pm0.06$. There is no host galaxy or 
star within 10 arcsec in either the PS1 3$\pi$ reference stack or SDSS DR12. 
A number of faint galaxies ($r\simeq 21-22$) are clustered around the transient at 
distances of 12 - 18 arcseconds and with uncertain photometric redshifts of 
$0.3-0.7$. It is likely this is a foreground supernova, and as it was 
detected first on 57310.61, or 31 days after  GW150914 (with non-detections on 57297.63) 
we have no evidence for it being a slowly rising transient.

\subsubsection{PS15cvx}
The transient is 2\farcs3 from the centre of the likely host galaxy  2dFGRS N152Z175	with  a spectroscopic redshift of  $z=0.152$. It was detected multiple times over 10 days beginning MJD=57308 and with a peak magnitude of 
\ips=20.1, it corresponds to $M_i =-19.2$. We have no evidence in the lightcurve for it being a slowly rising transient so its discovery date makes it incompatible with the time of GW150914.

\subsubsection{PS15cwa}
The transient is 3\farcs9 from an obvious galaxy in the PS1 images, and 
also catalogued as the UV source GALEXASC J101318.70-105447.8. No redshift information 
is available for either the host or the transient. It was detected  starting 57310.62, 
or 31 days after  GW150914 and faded slowly (0.6 mag) over 10 days. With the data available, this appears to be a normal supernova. 

\subsubsection{PS15cwk}
The transient is 15\farcs4 from the core of the 2MASS extended source 
2MASX J10135545-1252341 
which is a resolved SO/Sa galaxy in the PS1 multi-colour images and 
has a diameter of approximately 18\farcs0. It was detected  starting 57310.63, 
or 31 days after  GW150914 and remained flat for 10 days.   With the data available, this appears to be a normal supernova.

\subsubsection{PS15cvw}
PS15cvw is coincident with a 
$r=22.7$ object which SDSS DR12 classifies as a star. The 
star galaxy separation is not trustworthy at these flux levels and 
this is likely a faint host galaxy. Multiple detections beginning 57313.61, for 8 days indicate 
a slowly fading object that is compatible with being a supernova.

\subsubsection{PS15cmp}
This is only detected on one night (MJD=57314.6) but in all three filters that were taken
(\rps\ips\zps images) at \ips=21.68. It is not close to any catalogued source and the nearest 
 galaxy is SDSS J085423.78+035330.1 which is 31\farcs5 from the transient and at $z=0.097$, this would be 
55\,kpc. That is quite a large distance, although not unprecedented for SNe. We checked the minor planet
centre to see if it was a slow moving, known minor planet. The nearest object is 2016 AH73 ($V=22.5$), 
but at 2.7 arcmin away it is unlikely to be the same object. It is likely that this is a faint SN that has 
creeped above the detection limit on one night and remained below on the other images. 


\subsubsection{PS15crh}
PS15crh is coincident with a 
$r=21.9$ object which SDSS DR12 classifies as a star. The 
star galaxy separation is not trustworthy at these flux levels and 
this is likely a faint host galaxy and supernova. The transient was discovered on 
MJD=57314.58 (35 days after GW150914) at \rps=21.39, and detected at similar 
mags 5 days later (and also 50 days later in normal PSST operations).  The lightcurve data 
are compatible with it being a flat type II-P supernova.

\subsubsection{PS15cwh}
This faint transient is 13\farcs5 from the centre of the likely host galaxy  SDSS J085414.34+030504.0 with  a spectroscopic redshift of  $z=0.028$.   It was detected at $\ips=21.96\pm0.07$, between 57282.6 and 
57314.59 and hence would appear to be a long lived faint transient (with $M_{i} \simeq -13.4$), 7.5\,kpc
from the centre of the host which was detected three days after the GW150914. 
However in previous imaging of this field on 57181.26 by PSST 
\citep{2015IAUGA..2258303H,2015ATel.7153....1H}
we detected a source at \zps$=18.83\pm0.07$ 
and hence this is likely  to be a very old SN
which was $M_z = -16.6$ at 100 days before
GW150914 which is now in the tail phase. 
A bright stellar outburst in the host
\citep[e.g. an LBV type giant outburst]{2010MNRAS.408..181P} is possible, 
although  $M_z = -16.6$  would be quite
luminous for such a transient, 
or a foreground variable. 
With this historical detection, we rule out that this is an intrinsically faint transient
which occurred within 3 days of GW150914.


\subsubsection{PS15cri}
No host at all, but multiple detections in the \ips filter on 4 separate nights. This is likely a hostless 
supernova, \footnote{By hostless, we mean that either the host is too faint to be detected in our reference stacks 
or the SN has host galaxy which is quite far from the SN position and at a redshift that is
difficult to spatially associate it.}

\subsubsection{PS15cwe}
No host detection at all, but multiple detections on 4 separate nights in the \ips and \zps filters, Again it is  likely a hostless 
supernova. 

\subsubsection{PS15crk}
The transient is 6\farcs5 from an resolved galaxy in the PS1 images and the galaxy is 
catalogued as the UV source GALEXASC J103003.03-173138.7. With two detections on two 
separate nights it is likely a SN in this host. 

\subsubsection{PS15cwg}
The transient is 0\farcs7 from the core of a resolved galaxy in the PS1 images, and this galaxy
is also catalogued as the 2MASS extended source 2MASX J10191950-0916015.  With two detections on two 
separate nights it is likely a SN in this host. 

\subsubsection{PS15crj}
The transient is 1\farcs9 from a resolved galaxy in the PS1 images, and 
also catalogued as SDSS J094242.07+021811.1 with a photometric 
redshift of $z=0.41\pm0.09$. This is a likely supernova, but the photometric redshift is 
probably an overestimate of the true redshift  (see Section\,\ref{sec:discuss} for a discussion on 
the use of photometric redshifts) 

\subsubsection{PS15cwc}
This is a faint transient with multiple detections on 2 separate nights. It is coincident with a 
a faint, but clearly detected extended source in the PS1 \ips stacks.  With two detections on two 
separate nights it is likely a SN in this faint host.

\subsubsection{PS15cqx}
The transient is 2\farcs9 from the centre of the likely host galaxy  2MASX J10050385-0629439 with  a spectroscopic redshift of  $z=0.054481$. This is close to the bright core of the host galaxy, but it is detected on  at least 3 separate nights and appears real. With only foreground extinction applied it has $M_i = -16.6$, but could suffer from 
more extinction from within the host. It is consistent with being a SN. 

\subsubsection{PS15cqw}
This is a 2\farcs0 from a resolved galaxy in the PS1 images, and which os 
also catalogued as $r=19.3$ source SDSS J094506.56+011702.   The transient is one the fainter
detections at \ips=21.99, but is detected on three separate nights. The host has  a photometric 
redshift of $z=0.22\pm0.03$ in SDSS DR12. The object is consistent with being a SN. 

\subsubsection{PS15cwd}
Transient is detected in multiple nights and is 5\farcs5 from a resolved galaxy in the PS1 images, and  also catalogued as UV source GALEXASC J100806.68-142514.1. The object is consistent with being a SN. 

\subsubsection{PS15dgd}
This transient is offset by 5\farcs6 from the  core of the host 2MASX J10272567-1458190. It has 
no redshift information available. Detected on 3 separate nights, the object is consistent with being a SN. 

\subsubsection{PS15dgc} 
This is in the host galaxy 2MASX J10182042-1031244 0.11 (offset  by 6\farcs6 from the core) 
with a spectroscopic redshift of $z=0.055749$. Detected on four nights at \ips=20.4, it has
an absolute magnitude of $M_i = -16.6$, and is consistent with being a SN. 

\subsubsection{PS15dgb}
Offset from a resolved host galaxy in PS1 (uncatalogued to date and the PS1 survey name is 
 PSO J151.1810-15.0013) with no redshift information. Detected on 4 nights, it is consistent with being
a SN. 

\subsubsection{PS15dga}
This is 2\farcs2 from the core of the host galaxy  2MASX J10044252-0931139 0.037
with no redshift information.  Detected on 4 nights, it is consistent with being
a SN. 

\subsubsection{PS15dfy}
 This object is coincident with a faint host galaxy 
SDSS J095248.77+063804.2 at $r=21.45\pm0.08$ and has a photometric redshift of
$z=0.582\pm0.123$. It has also been recorded as a transient by CRTS and is known  
as CSS141127-095249+063804 and MLS130216-095249+063804 at varying magnitudes
between 19.5-20 \citep[e.g.][]{2013ATel.4872....1D}. It is likely a variable QSO. 

\subsubsection{PS15dfx}
Offset from clear host galaxy (uncatalogued to date)  detected in the PS1 image stack (object called 
PSO J147.7171-4.1633 in the PS1 3$\pi$ survey).  

\subsubsection{PS15dfw}
This is 0\farcs89 from the galaxy SDSS J094411.68+045452.8 with $r=20.42\pm0.09$  and a
photometric redshift $z=0.113\pm0.0814$. It is also detected some 45 days later in routine operations of 
PSST, at \wps=22.1. It is consistent with being a SN. 

\subsubsection{PS15dfv}
This is within 1\farcs0 from SDSS J094138.31-021021.7  at $r=20.18$  which is morphologically
classified as a star. It is listed in the MilliQuas catalogue as a QSO at $z=2.1$, hence this is 
likely QSO variability. 

\subsubsection{PS15dfu}
Offset from the galaxy SDSS J093424.45+064800.4 by 2\farcs7, which  has a 
spectroscopic redshift go $z=0.0866$. 

\subsubsection{PS15dft}
Coincident with the faint SDSS star $r=20.5$, and this is already known CV candidate 
ASASSN-15se 
\citep[for details of ASAS-SN see][]{2014ApJ...788...48S}\footnote{http://www.astronomy.ohio-state.edu/$\sim$assassin/transients.html}. 

\subsubsection{PS15dfs}
Coincident with SDSS J092137.60+120138.0 red galaxy, at 
$r=20.27$ and which has a photometric redshift of $z=0.424 \pm 0.079$

\subsubsection{PS15dfr}
Coincident with SDSS J091829.05+114010.6 ($r=22.29\pm0.14$) and which has a
photometric redshift of $z=0.549\pm0.117$. It is also the recurrent transient MLS140206-091829+114011 and possibly a stellar variable or a variable QSO.

\section{Quantifying the sensitivity of the search}
\label{sec:upperlim}

The histogram of measured redshifts is shown in Fig\,\ref{fig:zMi_hist}. This includes all transients
with either a spectroscopic redshift of the transient, or the host galaxy, or a photometric redshift of the 
candidate host. There are five photometric redshifts included for comparison. 
The histogram of peak absolute magnitudes (all in $\ips$, apart from PS15ckf which is in $\zps$) is also shown in Fig\,\ref{fig:zMi_hist}. This illustrates that our transient discoveries broadly sample the redshift regime $0.05 \lesssim  z \lesssim 0.15$, with absolute peak magnitudes 
$-19 \lesssim M_{\ips} \lesssim -16$. The latter are normal supernova magnitudes which illustrates that 
if an EM counterpart for GW150914 were to be accessible in our imaging it would need to 
reach supernova-like luminosities. 
If we assume that none of the detected transients are associated with GW150914, then it is useful
to set some upper limits to our search.

\begin{figure}
\includegraphics[width=8cm]{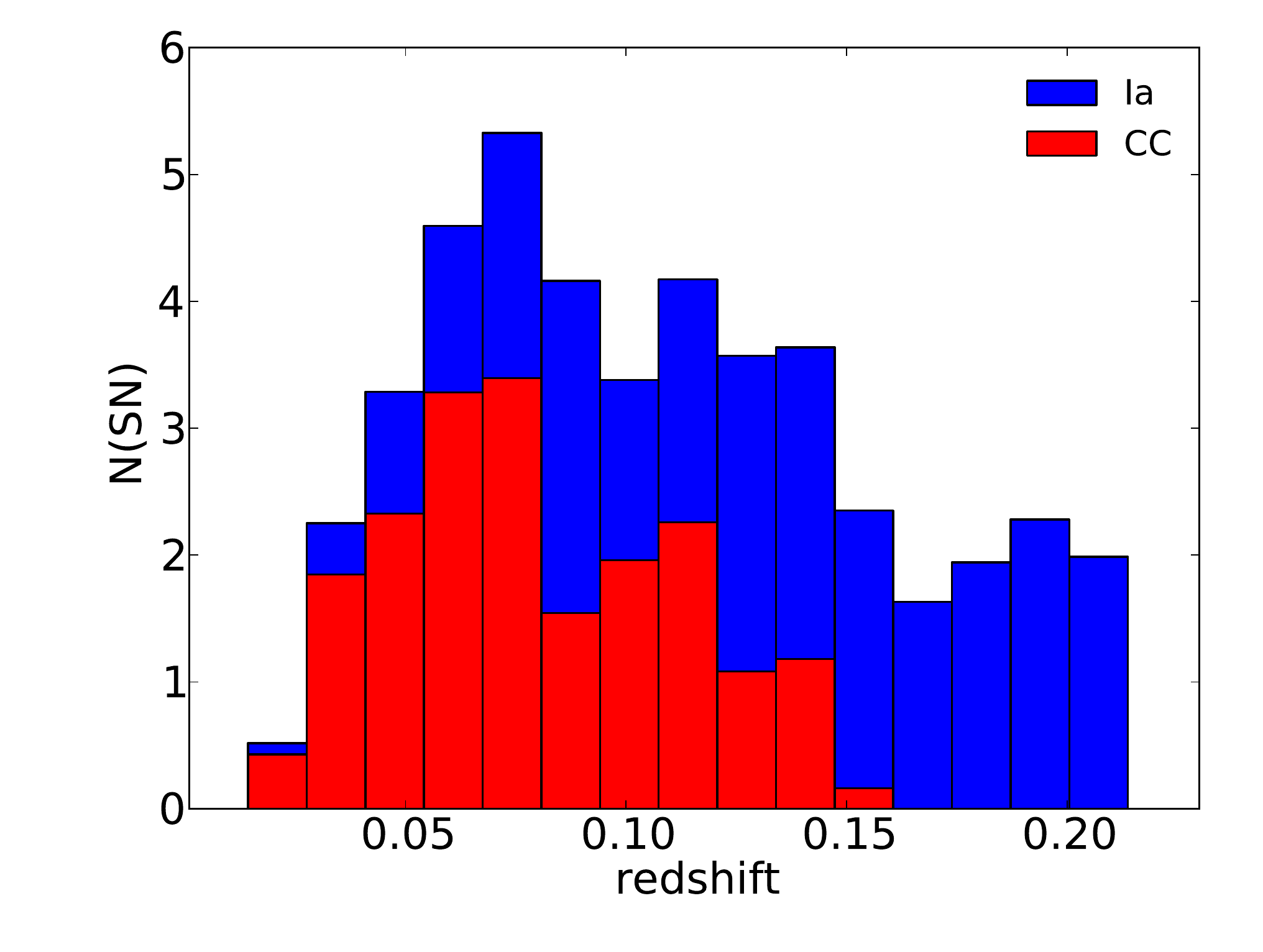}
\caption{Simulated numbers of core-collapse and type Ia SNe predicted to be detected within the sky survey area and their redshift distribution. See Section\,\ref{sec:rates} for details.}
\label{fig:rates}
\end{figure}

\subsection{Verifying the search efficiency by comparisons to SN rates}
\label{sec:rates}

To estimate the number of SNe that we expect to detect in our search we performed a simple survey simulation. In the simulation we input:
\begin{itemize}
\item the survey area, the time distribution of the observations, the detection efficiencies at the different epochs. The latter are shown in Fig.\,\ref{fig:maglim_mod1}
\item the light curves of template SNe are used to predict the observed light curves at different redshifts. For the computation we use the distance modulus derived from the redshift adopting standard cosmology and we include $k-$corrections for any given filter 
\citep[as in][]{2015A&A...584A..62C}
\item the rate per unit volume of the different types of SNe. As reference we use the local rate measurements of \cite{2011MNRAS.412.1441L} and the average rate evolution from the recent compilation of 
\cite{2015A&A...584A..62C}
\end{itemize}

The observed light curves along with the survey observing log allowed us to derive the control times for the different SN types for each considered redshift bins. The survey area and standard cosmology define the survey volume for each redshift bin. The expected number of SNe at each redshift is obtained simply by multiplying the SN rate by the control time and the survey volume and is shown in Fig.\,\ref{fig:rates}.
From this computation we predict a total number of ~50 SNe of which ~60 per cent should be of type Ia. If we limit to a redshift $z<0.1$ we expect 22 events with an equal number of type Ia and core collapse. 
Considering the uncertainty in SN rates and the sampling statistics the number of detected SNe and 
their redshift distribution is quite consistent with our predictions from measured cosmic SN rates.

To corroborate this with a simple calculation, the number of SNe within the volume defined by 
$z<0.1$ and our sky survey area can be predicted. The volumetric rates of 
 type Ia SNe within $z\lesssim0.15$ have been estimated as 
$2.9 \pm 0.6\times 10^{-5}$\,\mpcy (converted to $H_0=72$, which we use here) 
by  \cite{2010ApJ...713.1026D} and $2.5 \pm 0.5\times 10^{-5}$\,\mpcy\ 
by  \cite{2008A&A...479...49B}. Similarly 
the summary of 
\cite{2011ApJ...738..154H} of 
 the low redshift core-collapse SN surveys of 
\cite{1999A&A...351..459C,2008A&A...479...49B,2011MNRAS.412.1441L}
suggests a volumetric rate at $z\lesssim0.1$ of  $10 \pm 3\times 10^{-5}$\,\mpcy for 
core collapse (types II and Ibc). 
The number of type Ia SNe that these rates predict within the 
440 
square degrees, is the volumetric rate multiplied by the control time (CT), 
where CT is the time spent by each SN in a detectable magnitude range. 
For a type Ia SN at $z\simeq0.1$, the peak magnitude would be in the range 
\ips$ = 19.5^{+0.3}_{-0.5}$, and therefore it would be detectable above the 
\ips=21 limiting magnitude of Figure\,\ref{fig:maglim_mod1} if it exploded 
within the previous $50\pm15$\,days. Therefore the later deeper limits (which cover
the whole
440
square degree region) effectively encompass the earlier, shallower limits over 
a smaller portion of the field. With this control time of $45\pm15$\,days, 
the expected number of type Ia SNe in this volume defined by $z\lesssim0.1$  is  

\begin{equation}
n = V_{z<0.1}\frac{A_{\rm PS1}}{41253}\frac{45\pm15}{365}(2.9\pm0.6\times10^{-5})
\end{equation}

With a co-moving volume of $V_{z<0.1}=0.3$\,Gpc$^{3}$ (for $H_{0}=72$, $\Omega_{\rm M}=0.3$
and $\Omega_{\Lambda}=0.7$),  and a survey area of 
$A_{\rm PS1} = 440$
square degrees, 
we should detect approximately  $12^{+8}_{-5}$ type Ia in 
the PS1 footprints. We find 9 spectroscopically confirmed type Ia SNe within this redshift 
range, 8 spectroscopically confirmed type II or Ibc and 8 candidate SNe in galaxies with 
redshifts less than $z=0.1$. If these unconfirmed 8 are roughly split in equal proportions 
as the spectroscopically confirmed sample, then we  are likely to have discovered a sample 
of approximately 13 type Ia SNe. The unconfirmed sample (in galaxies with no host redshift) are
likely to be at higher redshift than $z\sim0.1$. Hence our recovered SN population 
is in good agreement to the expectations 
from the type Ia volumetric rates. 
%

\subsection{Interpreting the sensitivity limits in the context of GW150914}
\label{sec:upplim}

\begin{figure}
\includegraphics[width=8cm]{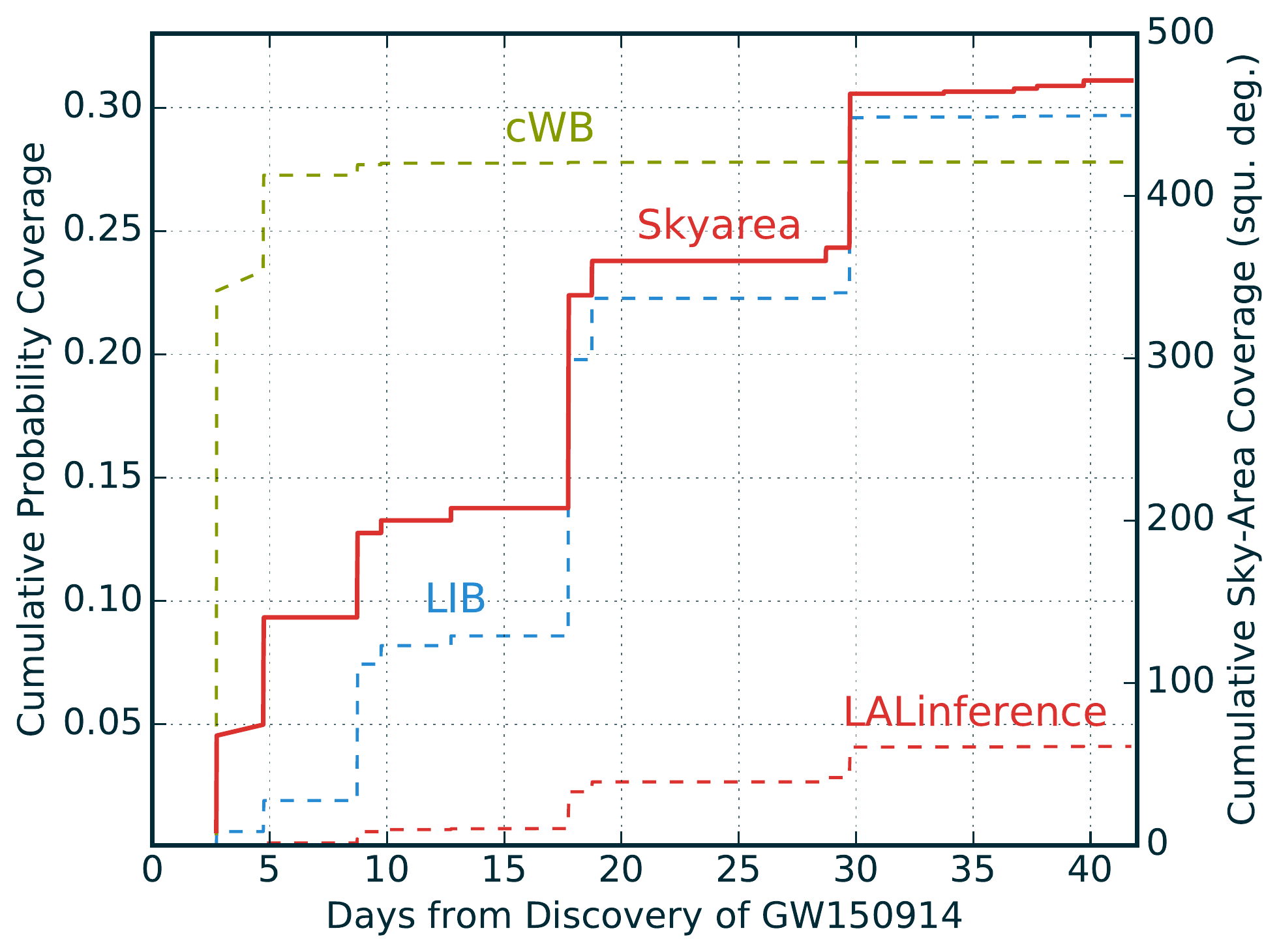}
\caption{Cummulative sky coverage mapped by Pan-STARRS1 as function of time since the GW150914 trigger is shown in solid red line. The cumulative probability of the LIGO sky maps covered are shown as broken lines which each referring to the particular sky probability map
released by LIGO.  The first two maps released early where the cWB and LIB, while the LALInference 
map was the most authoritative. }
\label{fig:skyarea}
\end{figure}

\begin{figure}
\includegraphics[width=9cm]{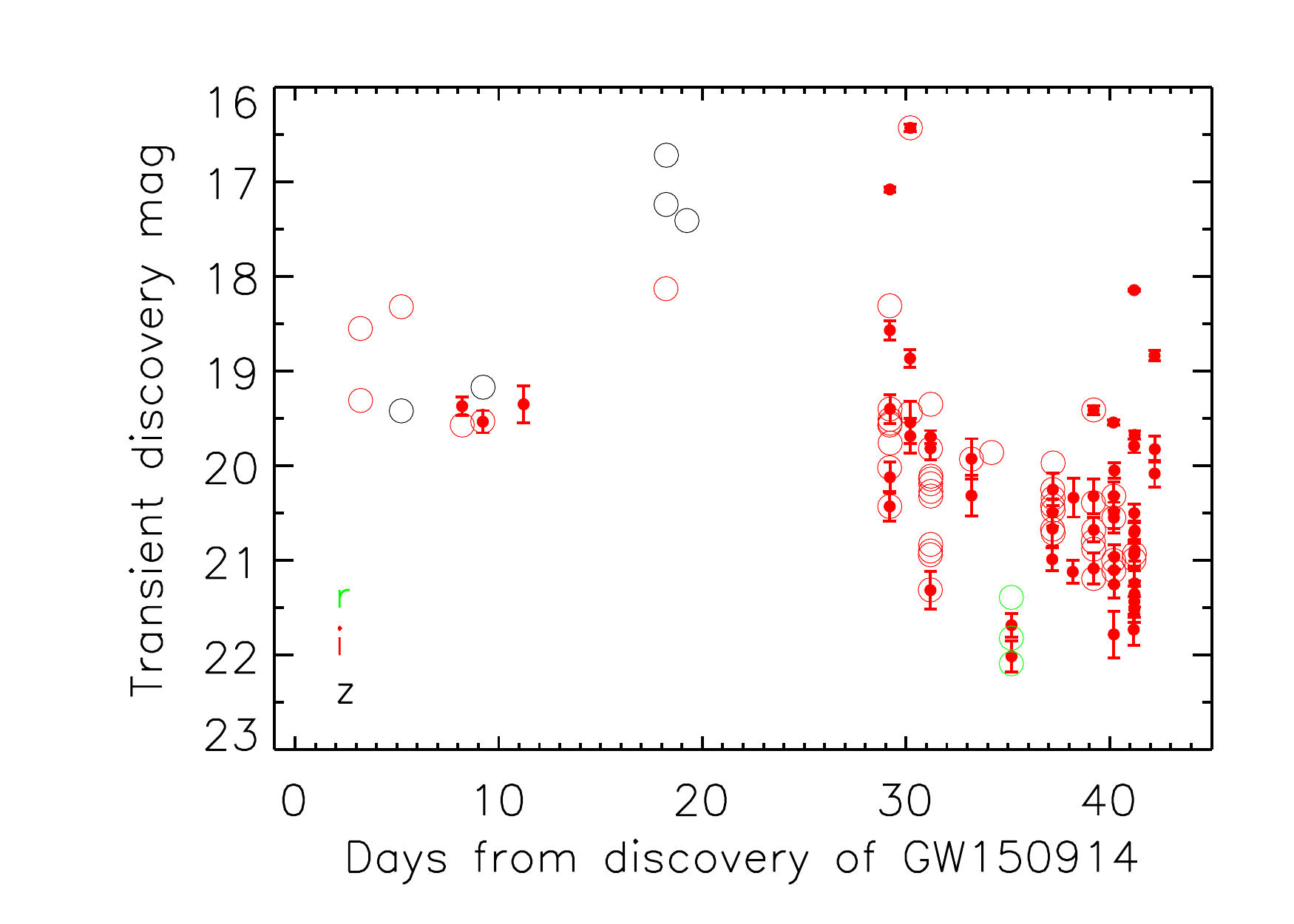}
\includegraphics[width=9cm]{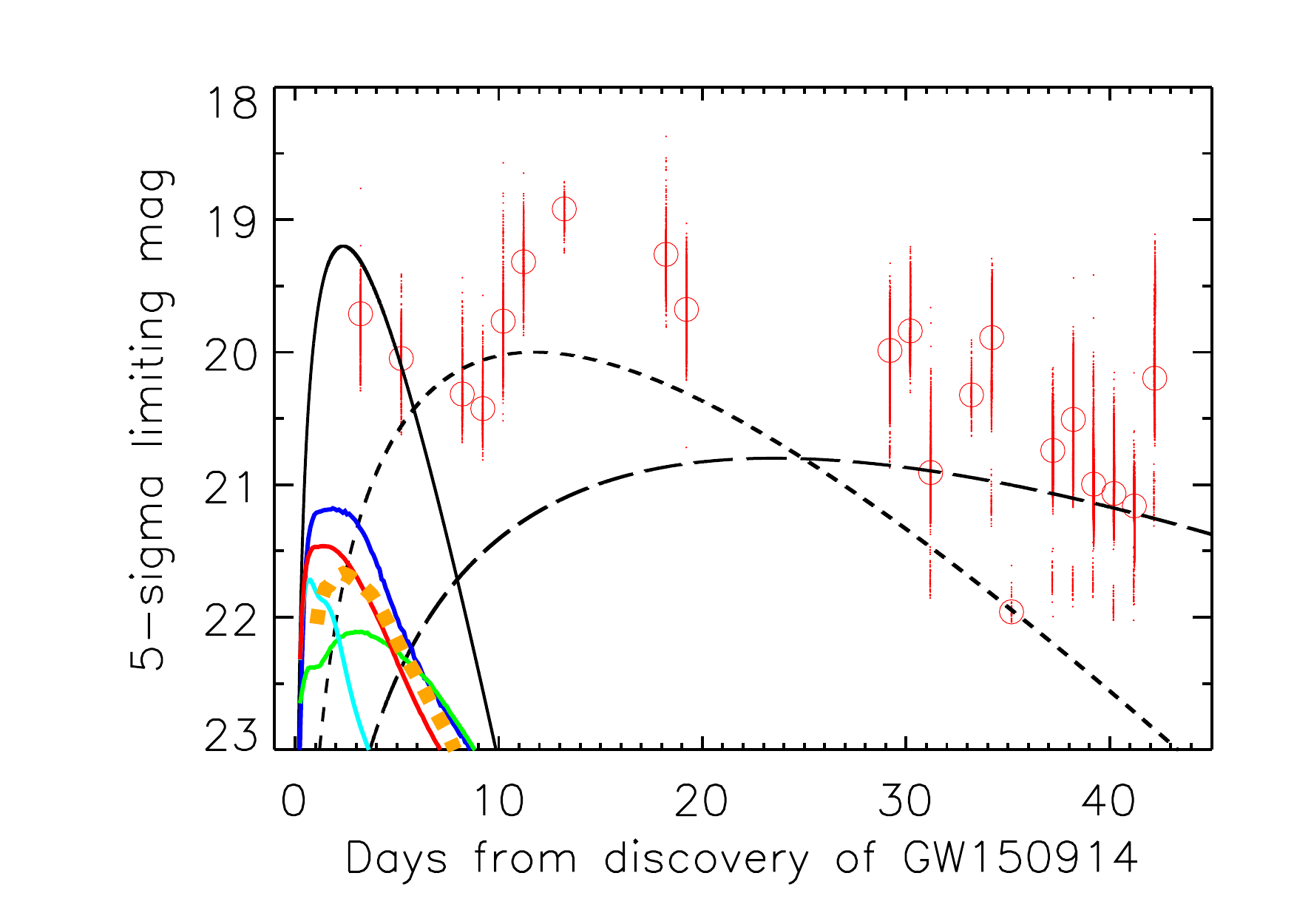}
\caption{{\bf Top Panel :} 
The magnitudes of transients discovered as a function of time from the discovery of 
GW150914. The solid red circles are the faintest \ips magnitude measured for each of the 56 transients. 
The open circles are the discovery magnitudes for each transient in the filter they were first detected (green is \rps, red is \ips and black is \zps). 
 {\bf Bottom panel:} The 5$\sigma$ detection limits for all \ips images determined  from fake source injection and recovery. Each individual GPC1 skycell is plotted as a point (there are approximately 51 skycells per PS1 pointing) with the median value on each night denoted by the open circles.
The simulated lightcurves of three transients with parameterised lightcurves and 
characteristic timescales as defined the 
text of 4d, 20d and 40d which have been scaled in flux to be detectable at signal levels similar to those for measured objects. The four coloured lightcurves (solid lines) in the bottom right 
are the model lightcurves for compact mergers from 
Kasen et al. (2015)
and 
Barnes \& Kasen (2013)
The orange dashed curve is the brightest BH-NS merger model
of Tanaka et al. (2014; model MS1Q3a75\_k1\_AB).
 The models  
and are discussed in the text in 
Section\,\ref{sec:upplim}.}
\label{fig:maglim_mod1}
\end{figure}

Our joint Pan-STARRS and PESSTO search cannot 
set significant and complete upper limits to the magnitude of any optical counterpart because we could not map out the 
southern sky localisation region. Furthermore, the obvious complication 
is illustrated in Figure\,\ref{fig:skymaps}, 
which shows that the northern sky localisation region was not mapped out uniformly in time. 
Ideally we would have surveyed the whole northern region within 24hrs  from the time of trigger and 
repeated it to give a flatter temporal distribution in our sensitivities. However as explained in 
Section\,\ref{sec:obs} this was not possible due to the Right Ascension range of the region. 
In addition, the sensitivity of the Pan-STARRS 
imaging of the error region is highly non-uniform.  As discussed in Section\,\ref{sec:obs}, the PS1 telescope
could only reach the north western area of the GW150914 probability map during the first 3-10 days
after the discovery and these fields were observed at typical airmasses of $1.8\lesssim \sec z \lesssim 2.8$ and close
to 18 degree twilight. 

The time sequence of mapping out the error region is shown in 
Figure\,\ref{fig:skymaps} and in 
Figure\,\ref{fig:skyarea}. 
The consequence of the field rising to accessible altitudes to allow 
longer observing sequences (in darker skies) is illustrated in    Figure\,\ref{fig:maglim_mod1} . These two panels illustrate the depths of the fields we reached, through 
the direct measurement of faint objects and calculations of the 5$\sigma$ limiting magnitudes. 
We
plot the \ips discovery magnitudes of each of the 56 transients in the $\rps\ips\zps$ filters 
and  the faintest magnitudes recorded. 
Then we show the estimated 5$\sigma$ limiting 
magnitude in each Pan-STARRS skycell. The 60 Pan-STARRS chips are processed individually
and warped onto sky coordinates in predefined skycells (approximately 51 per pointing). 
For each of these, we calculate the nominal 
5$\sigma$ limiting magnitude from photon statistics of the sky background and read noise. 
 A number of fake stars (typically 500) are then inserted into each of the skycells 
at magnitudes around the nominal  value and the 50 per cent  recovery efficiency is measured
(where an object must have a significance of 5$\sigma$ to be recovered). 
This is adopted as the  5$\sigma$ limiting magnitude. 
This quantifies the sensitivity of the full system from PS1 imaging through 
difference imaging, object detection and filtering as discussed in Section\,\ref{sec:obs}. 
The nominal calculation and the 50 per cent efficiency measurement are in good quantitative agreement. 
The fake stars are positioned randomly in the images, and are mostly on blank sky. 
Transients that are coincident with galaxies will typically have higher background contamination noise
than our randomly placed sample. This 
is hard to quantify since it will critically depend on the position within a galaxy, and the surface
brightness at that point. Candidates for compact binary mergers (e.g. NS-NS mergers) 
have been often found to lie well outside galaxies
  \citep{2014ARAA..52...43B}
and hence the limiting magnitudes we calculate 
would be appropriate. Nevertheless we caution that for objects inside galaxies, the limiting magnitudes
will be lower (as they would for any other imaging survey).
It illustrates the sky brightness and high airmass problems affecting the imaging in the first 10 days, 
then the poor weather on Haleakala in the next 10-15 days. The best observing 
conditions were between $27-45$ days after GW150914. During this period we typically 
reach 5$\sigma$ detection sensitivities of $\ips = 21.0\pm0.5$ in exposures which typically 
have exposure times 30-80 sec, airmass=1.5-2.7 and image quality of FWHM=1\farcs0 - 1\farcs4. 

There are now quantitative models for the lightcurves of merging neutron stars which include
heating from radioactive heavy elements and $^{56}$Ni 
combined with 
realistic opacities from r-processed elements (and ejecta masses), such as those 
recently calculated by 
\cite{2013ApJ...775...18B}, \cite{2013ApJ...774...25K}, \cite{2015MNRAS.450.1777K}
and \cite{2013ApJ...775..113T}.  These have been termed kilonovae, as they are fainter 
than supernovae and mostly have shorter timescales.

The compact binary that caused GW150914 
is estimated to be a a pair of binary black holes
and not merging neutron stars. However as there
are no quantitative predictions for what (if any) optical radiation merging black holes may produce, we show the various quantitative models for merging neutron stars. Even if they are not directly applicable here, they indicate the limits we would need to reach to place meaningful constraints. The models in Figure 7 are the disc wind outflows of compact object mergers (blue) of
\cite{2015MNRAS.450.1777K}  ; the 
r-process powered merger model which includes a 56Ni-dominated wind (red) of 
\cite{2013ApJ...775...18B} ;  a 
merger model with iron-group opacity with 
$M_{\rm ej} =0.01$\msun by the same authors (cyan); and a merger model for opacity dominated by r-process elements, with $M_{\rm ej} =0.1$\msun (green) also by the same authors. 
We have determined \ips synthetic magnitudes, based on the spectra in these papers and 
 scaled them to the closest plausible distance for 
GW150914 (190\,Mpc). 
The NS-NS merger models of 
\cite{2013ApJ...775..113T} are of 
similar peak luminosity and timescales and are useful 
alternatives.  In addition, we also show the brightest BH-NS 
merger models of \cite{2014ApJ...780...31T}
to illustrate the diversity of theorised 
objects (also scaled to 190\,Mpc). 
The comparison shows we
are not reaching the depth, at this distance to 
probe these types of merger model. However 
the depth of our survey fields at 30-40 days did
reach these sensitivities and if the field had been 
better placed in the sky we would have reached these
depths. This bodes well for future searches. 

\cite{2015MNRAS.446.1115M} have  further predicted that 
neutron powered pre-cursors of kilonovae may be detectable, 
through the decay of free neutrons. These are brighter, bluer 
and faster than the heavy element radioactive decay from 
the 
\cite{2013ApJ...774...25K} kilonova models. We do
not plot them here, as we are not probing
the short timescales predicted by those models (hours to 1-2 days).

For BH-BH mergers, the situation is much less clear and there are no published EM lightcurve
predictions that could be quantitatively compared with our data to at least 
determine how deep one might need to get in the UV/optical/NIR to detect a counterpart
\citep{2013IJMPD..2241011K}. 
\cite{2014ApJ...780...31T} have also  radioactively powered lightcurves for BH-NS 
mergers which they suggest could be comparable luminosity (or even higher) to 
NS-NS merger models. 
To quantify plausible transients we adopt a simple parameterised lightcurve for which we 
can vary the timescale and peak flux. We can then  use this
function to quantify the temporally varying limits we have. 
We use the following function

\begin{equation}
f(t) =\frac{e^2}{4} f_0\Bigg(\frac{t}{t_0}\Bigg)^{2}e^{-t/t_0}
\end{equation}

This function peaks at $t=2t_0$, has a peak value of $f_0$, and has a full-width-half-maximum of approximately FWHM=$3.38t_0$. 
Choosing flux values to simulate candidates peaking at AB magnitudes of $i=19, 19.5$ and 21.0
and values of 
$t_0 = 1.2, 5.9$ and 11.8d we illustrate the sensitivity of the Pan-STARRS search to 
transients with these three timescales in Fig\,\ref{fig:maglim_mod1}.  
The values of $t_0$ correspond to lightcurves with characteristic
timescales of $t_{\rm FWHM} = 4, 10, 40$ days. 
The simulated transients are 
set to explode at $t=0$ on the discovery epoch of GW150914.  The plot shows  we would be sensitive to transients with these timescales and peak  magnitudes of 4d and $i=19.2$, 20d and $i=20.0$, and 40d and $i=20.8$. 
If we assume a distance to GW150914 of 
$D_{L} = 400\pm 200$\,Mpc 
\citep[from][]{theprizepaper}
 then the following absolute magnitude limits are inferred 

\begin{enumerate}
\item {\bf 4d timescale :} with a sensitivity of $i=19.2$, we should have detected transients with a peak flux of $M_i \lesssim -18.8^{-0.9}_{+1.4}$ during the first five days after GW150914. Unfortunately, due to the RA constraints we were unable to map out the whole northern localisation region, as illustrated in Figure\,\ref{fig:skymaps}. It is quite possible 
that we missed a transient of this luminosity and timescale simply due to poor coverage in the early days. 
Future searches will hopefully have more accessible RA ranges to allow this interesting fast timescale to be 
probed. 
\item {\bf 20d timescale :} with a sensitivity of $i=20$,  we would be sensitive to transients with a peak flux of $M_i \lesssim -18.0^{-0.9}_{+1.4}$ during the first 3 to 17 days after GW150914. Again, as shown in Figure\,\ref{fig:skymaps}, only the northern most region was mapped in any detail up to 17 days and therefore 
these sensitivity limits again can not be applied over the whole region. 
\item {\bf 40d timescale :}  these are the most meaningful limits (\ips$<20.8$), since we were able to map the whole 
skymap region on the timescale of 30-40 days. We would have been sensitive to long lived transients with lightcurves as parameterised above with peak fluxes of  $M_i \lesssim -17.2^{-0.9}_{+1.4}$, over the whole 440 square degrees covered
(or 4.4 per cent of the probability contours).  Although one might expect most electromagnetic 
sources of GW 
to be fast decaying (as in the NS-NS and BH-NS mergers), massive asymmetric core-collapse 
may produce supernovae such as broad lined Ic events
\citep{2012ApJ...761...63P}. These have similar rise and decay timescales to the 40d parameterised
curve and are worth considering even though they are likely to be detectable by LIGO to much 
smaller distances (up to 50\,Mpc). 
\end{enumerate}

\begin{figure}
\centering
\includegraphics[scale=0.5]{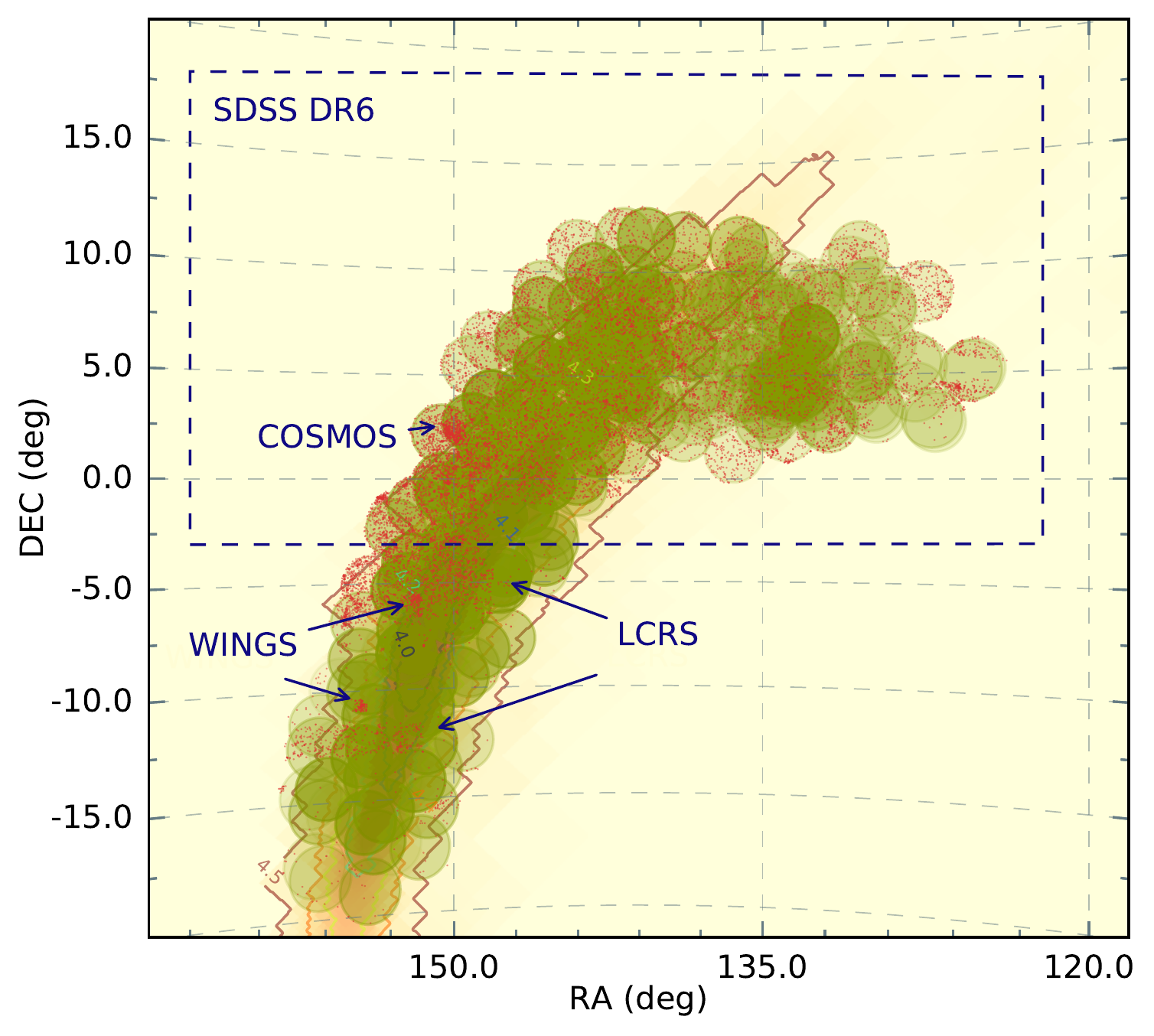}
\caption{
Catalogued galaxies in NED which  have a spectroscopic redshift $z\leq0.15$ and also lie within the PS1 survey area. The green circles are the same as in Fig\,\ref{fig:skymaps} and show the PS1 pointings. The red dots are all galaxies which have a catalogued spectroscopic redshift that is $z<0.15$. The sharp drop in the galaxy density below $\delta\simeq-2$ is due the boundary of the SDSS DR12 survey footprint. The smaller area surveys labelled are described in the text. 
completeness of galaxy catalogues outside the SDSS footprint.}
\label{fig:all-ned}
\end{figure}

\begin{figure}
\centering
\includegraphics[scale=0.4]{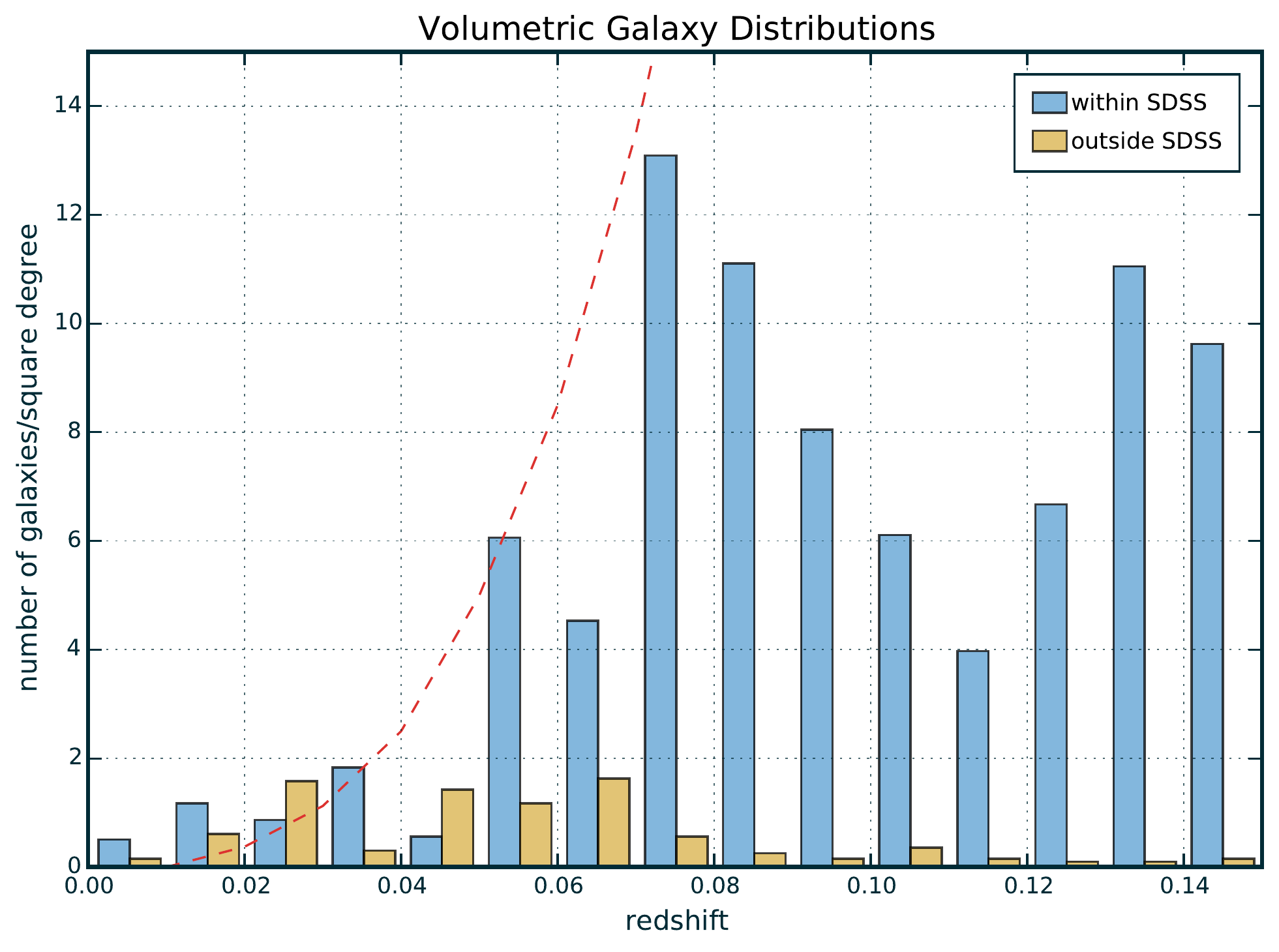}
\caption{Histogram of galaxy counts per square degree within the SDSS DR12 footprint (blue) and outside SDSS DR12. All galaxies included in this plot have a spectroscopic redshift. The dotted redline plots the volume of the Universe as a function of redshift. It is scaled arbitrarily to approximately match the galaxy counts in SDSS at $0.02< z < 0.03$, and illustrates that within SDSS the galaxy completeness falls off at $\sim0.07$. Outside SDSS, current catalogues are incomplete beyond $z\sim0.03$ or about 100\,Mpc. }
\label{fig:gal-hist}
\end{figure}

\begin{figure}
\centering
\includegraphics[scale=0.4]{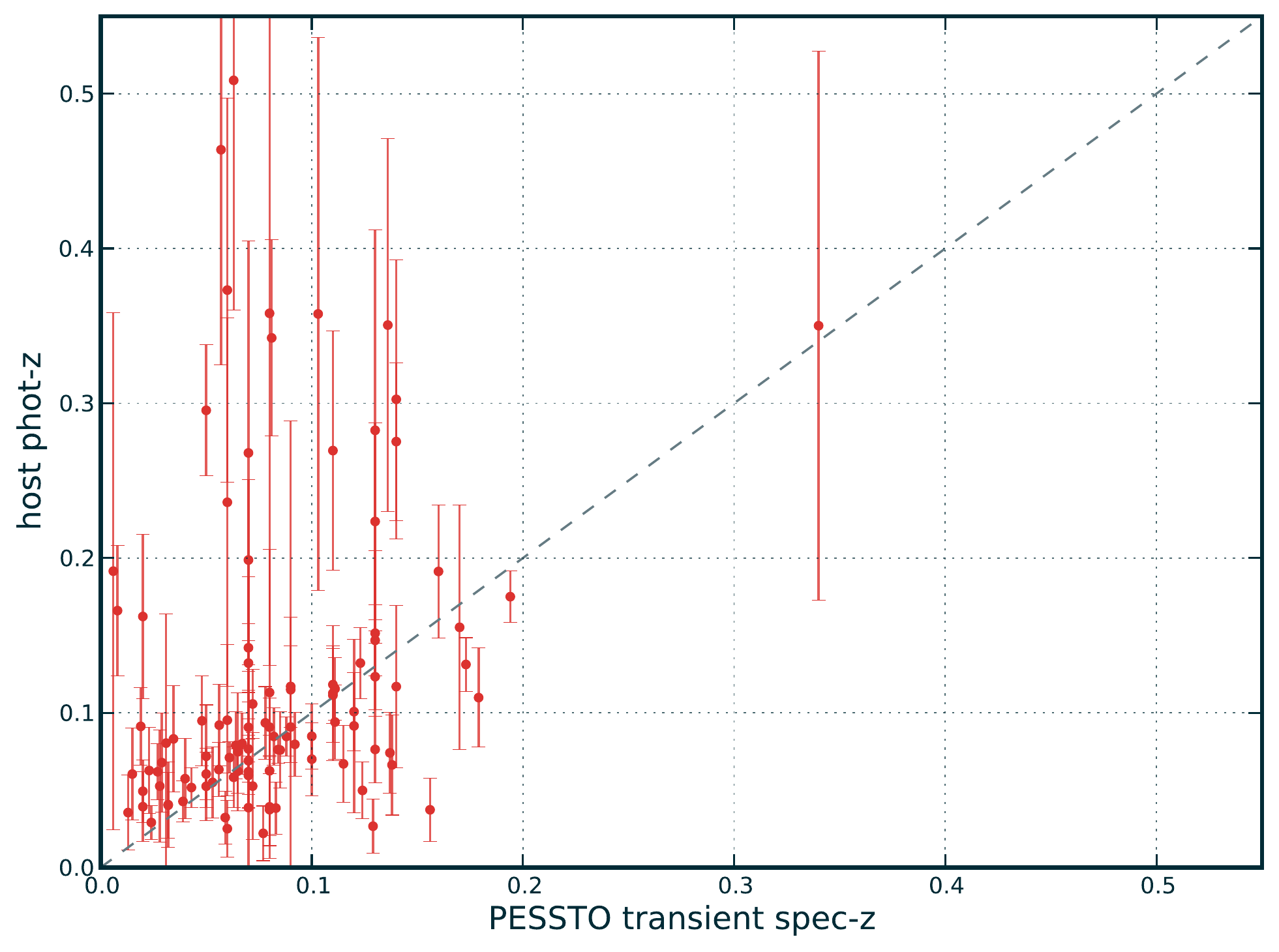}
\caption{The confirmed spectroscopic redshifts of transients versus the estimated photometric redshift from SDSS DR12. As the spectroscopic catalogues are incomplete, we demonstrate that using the photometric redshifts is not particularly useful to constrain the 
true redshifts of  transients at low redshifts. }
\label{fig:pessto-photz}
\end{figure}

\section{Discussion}
\label{sec:discuss}

\subsection{Our survey in context}
Our search of approximately 
440 
square degrees of the northern sky localisation region of GW150914 over a 41 day period from the  discovery yielded 56 reliable astrophysical candidates not including 
variable stars. This illustrates and quantifies the problem that will be
faced in locating the EM counterparts to GW sources in the future.  By reaching magnitudes of \ips$=21\pm0.5$, 
one will find of order 5-10 supernovae per fresh 100 square degrees  surveyed within a redshift of  $z\lesssim0.15$, which will be mostly old.  This means that any future search for the EM counterparts of GW sources 
requires significant spectroscopic resources to classify every transient in the field.  It seems clear that there
is no easy way to rapidly distinguish candidates and remove the interloping SNe 
other than spectroscopic confirmation or constraining the luminosities at other  wavebands
(gamma-ray, x-ray, infra-red or radio). 
Other early attempts at surveying the error box of GW150914 further illustrate the 
problem such as the recent work of 
\cite{2016ApJ...823L..33S,2016ApJ...823L..34A,2016arXiv160208764K} 
and the summary of the broad wavelength range follow-up campaign in \cite{2016arXiv160208492A}. 
The possibilities for the counterpart of GW150914 are as follows, and 
these could be applied to any counterpart search that returns a null result 

\begin{enumerate}
\item The EM counterpart was outside our survey region. This is probable, given the total probability we covered from the LIGO sky maps is only 4.2 per cent. The original sky map released implied we
were covering around 30 per cent with our pointings. 
In this case there is little we can add, as the  southern sky localisation region is significantly favoured. 
\item The EM counterpart was in our survey region, but fell below the limits. In this case the sensitivity limits from the model transients with three example timescales set useful targets (in terms of luminosity and timescale)
 to aim for in future searches
\item The EM counterpart to GW150914 was detected as one of the 56 transients but we don't recognise it 
as causally linked. This seems unlikely, but it is not ruled out. It maybe that one of the fainter transients 
discovered in the time window of $>$24\,days after GW150914 is associated and that without a confirming spectrum, 
or detailed lightcurve no useful discrimination from the SN population is possible. Future surveys of GW localisation regions 
must still search for known SNe that are rare by volume (or sky area) but are habitually  found in the GW 
regions.
\end{enumerate}

Given the above,  the only possibility for improvement (and increasing the probability of detection of
a GW counterpart) is to survey the regions rapidly and continuously  and be as spectroscopically 
complete as  as possible.   A reasonable question then is could survey strategies be adjusted to 
make use of redshift and flux information of the host galaxy population as suggested 
by, for example  \cite{2011CQGra..28h5016W} and more recently \cite{2015arXiv150803608G}. 

\subsection{Using galaxy catalogues with spectroscopic and photometric redshifts}

The use of galaxy catalogues to pick potential host galaxies within the sky localisation region 
has the advantage that larger aperture telescopes, with smaller FOV cameras, can focus on these
targets and produce significantly deeper images than 0.4-2m telescopes that aim to map the 
sky region. The bar to this has been the commonly known problem of the incompleteness of 
galaxy catalogues beyond distances of $\sim$100\,Mpc ($z\sim0.025$). Our search provides
a useful practical example to investigate if galaxy targeting, from catalogued sources would be useful
in the case of GW150914. In Figure\,\ref{fig:all-ned} 
we show the known galaxy catalogue within the sky localisation region for GW150914 and the inhomogeneity is immediately obvious.  The northern region is dominated
by galaxy counts from SDSS  and regions of high density are visible as the 
COSMOS \citep{2007ApJS..172....1S}
Las Campanas Redshift Survey \citep[LCRS;][]{1996ApJ...470..172S}
and 
WIde-field Nearby Galaxy-cluster Survey \citep[WINGS;][]{2006A&A...445..805F}. 
Such structure has been illustrated by \cite{2015arXiv150803608G} 
 in the ``Census of the Local Universe" (CLU) catalogue that they present, which is a union of 
existing catalogues. The CLU aims to catalogue all galaxies with $L > L^{\ast}_{B}$, where 
$L^{\ast}_{B}=(1.2\pm0.1)\times10^{10}h^{-2}$L$_{B,\odot}$  
(which corresponds to $M^{\ast}_B=-20.5$ for $h=0.7$). For comparison, the 
Milky Way galaxy is estimated at $M_B = -20.4$, therefore the CLU is aiming at galaxies
with masses larger than the Milky Way within a radius of about 200\,Mpc. While
\cite{2015arXiv150803608G} show that selecting these high mass galaxies from
the union of existing catalogues produces a reasonably encouraging large scale 
structure pattern (their Figure 1), we illustrate here that  for GW150914 such a 
galaxy targeted strategy would be rather incomplete. This stems mostly from the fact that 
the distance to GW150914 
$z\sim0.1$ (or 400\,Mpc) 
is much larger than expected for the first LIGO/Virgo  
bursts up to now ($z\lesssim0.05$, or 200\,Mpc). However even at 200\,Mpc, 
the CLU of \cite{2015arXiv150803608G}  drops to below 40 per cent completeness. 
The severe incompleteness of current galaxy catalogues, with spectroscopic 
redshifts, is highlighted in our Figure\,\ref{fig:all-ned} and Figure\,\ref{fig:gal-hist}. 
For this we selected all galaxies within NED with a spectroscopic redshift within the 
PS1 footprints (Figure\,\ref{fig:all-ned}), then we selected a region within  
the SDSS DR12 footprint and outside the SDSS area. The number of galaxies (with no 
luminosity cut-off) per square degree is shown in Figure\,\ref{fig:gal-hist}.
A simple calculation of co-moving volume is plotted in red, scaled arbitrarily to 
the galaxy counts at $z\sim0.02$. This illustrates that the galaxy counts in the 
SDSS area do not fall off until $z\simeq0.07$ (300\,Mpc). However outside 
SDSS, the galaxy catalogues are clearly incomplete by some margin beyond 
$z\simeq0.03$ (120\,Mpc). The conclusion from this is that given the unexpectedly
large distance estimate to GW150914, a targeted galaxy search would not have been a viable 
strategy in this case. 

As the Pan-STARRS1 survey will provide a \grizy\ multi-colour survey of the whole sky 
above $\delta\simeq-30^{\rm \circ}$ 
\citep[][Chambers et al. 2016, in prep]{2013MNRAS.435.1825M}
and photometric redshifts of all galaxies 
\citep{2012ApJ...746..128S}, 
we should consider the possibility of using photometric redshifts of transient host galaxies 
to guide candidate selection. Photometric redshifts of galaxy samples within the 
range 
$ z < 0.5$ 
have encouragingly small RMS scatters of typically 
2.4 per cent
\citep{2012ApJ...746..128S}. 
In the second data release (SSDR2), the PESSTO survey \citep{2015A&A...579A..40S} produced a catalogue of 
all transients classified by the survey and the cross-matched host galaxies
\footnote{Publicly available from http://www.eso.org/qi/}. This catalogue 
including the transient object redshift from the PESSTO spectral classifications, 
the host galaxy spectroscopic redshift (where it existed in NED) and the host galaxy 
photometric redshift (where it existed in SDSS DR12).  This sample of 106 low redshift transients
gives a useful sample of 
transients for which a confirmed spectroscopic redshift exist and a photometric 
host galaxy redshift estimate can be compared. This is plotted in Figure\,\ref{fig:pessto-photz}
and illustrates that at low redshift, where GW counterparts are expected to be ($z\lesssim0.15$), 
the use of photometric redshift information is of limited value. Around 20 per cent of transients
have host photometric redshifts that differ by more than a factor 2 from their true value. 
Perhaps the most useful constraint that could be made is that a selection of host $z_{\rm phot}\leq0.15$
(83 objects), produces a pure sample of transients with $z_{\rm spec}\leq0.15$. Only 
3 objects lie beyond $z_{\rm spec}>0.15$, giving a purity of 96 per cent. One 
might be tempted to use this as a selection for low redshift transients, however the incompleteness 
(about 20 per cent) caused by the objects with high host $z_{\rm phot}$ values precludes this 
as a useful selection criteria.  It is also dubious whether the $z_{\rm phot}$ selection is actually
driving the pure sample of $z_{\rm spec}\leq0.15$ objects, since this  redshift limit is 
driven by the limiting magnitudes of the imaging surveys for PESSTO and the sensitivity 
limit of spectroscopic classification (both around $r\sim20.5$ for EFOSC2 and the feeder surveys
of LSQ, PS1, OGLE etc as described in Smartt et al. 2015).  The poor recovery of true
redshifts from host $z_{\rm phot}$ illustrates that the relatively high values for 5 of the transients
found in the GW150914 region (illustrated in Figure\,\ref{fig:zMi_hist} and discussed in Section\,\ref{sec:trans})
are likely to be examples of systematic over-estimation as visible in the 20 per cent of significant outliers
in Figure\,\ref{fig:pessto-photz}. 

In conclusion, we find that attempting to implement a refined strategy to search for the counterpart of 
GW150914  (or any GW source at distances greater than $\sim$120\,Mpc) by using galaxy catalogues would
not have been a useful exercise. This targeted 
galaxy method would certainly we useful 
at distances below 100\,Mpc. However the most plausible skymap and distance  information needs to be released early and be reliable for this to be useful and neither occurred for GW150914. 
If GW sources turn out to have a higher redshift distribution than we expected (e.g. $z > 0.05$, rather than below $z<0.02$) then  
the spectroscopic galaxy catalogues  currently 
available are incomplete by large margins (particularly outside the SDSS DR12 footprint). 
It appears that the most useful strategy is tiling out the entire localisation region and being 
as spectroscopically complete as possible. Further thought should be given to 
interpreting the probability maps in a manner that leads to investment of exposure time as 
a function of sky probability. 

\subsection{Multi-messenger searches and temporal coincidences}
This early multi-messenger search should be seen in 
context with other wide-field follow-up from 
high energy triggers with poor sky localisation.   \cite{2015ApJ...811...52A} 
triggered a range of wide-field optical facilities to try and identify coincident 
sources with detections of high energy neutrinos from IceCube. They searched the error circle region of 0.9 square degrees with the PTF, Swift and ROTSE and found one transient, PTF12csy. However
searching through archival data from the Pan-STARRS1 
3$\pi$ survey showed that PTF12csy was visible about 158
days before the neutrino detection and unlikely to
be causally related. 
\cite{2015ApJ...806...52S} searched for the optical counterparts to high energy  Fermi GRBs, which are detected in the GBM with poor
spatial localisation (of the order 10s to 100 square degrees).
In this case the PTF camera covered between 30-147
square degrees for 35 separate Fermi GRBs. In each 
of these cases around $\sim$10 optical transients
are found that would warrant follow-up. These are
mostly SNe or AGN, and 
\cite{2015ApJ...806...52S} showed that photometric 
monitoring along with rapid spectral typing 
could uncover the elusive afterglows.   A key component to 
both of these is identifying transients which 
are both spatially and temporally coincident with the high energy
trigger. The work here, searching 440 square degrees (finding 56 transients) goes a step beyond in terms of sky area and 
contamination by already existing transients. 

\begin{figure}
\centering
\includegraphics[width=8cm]{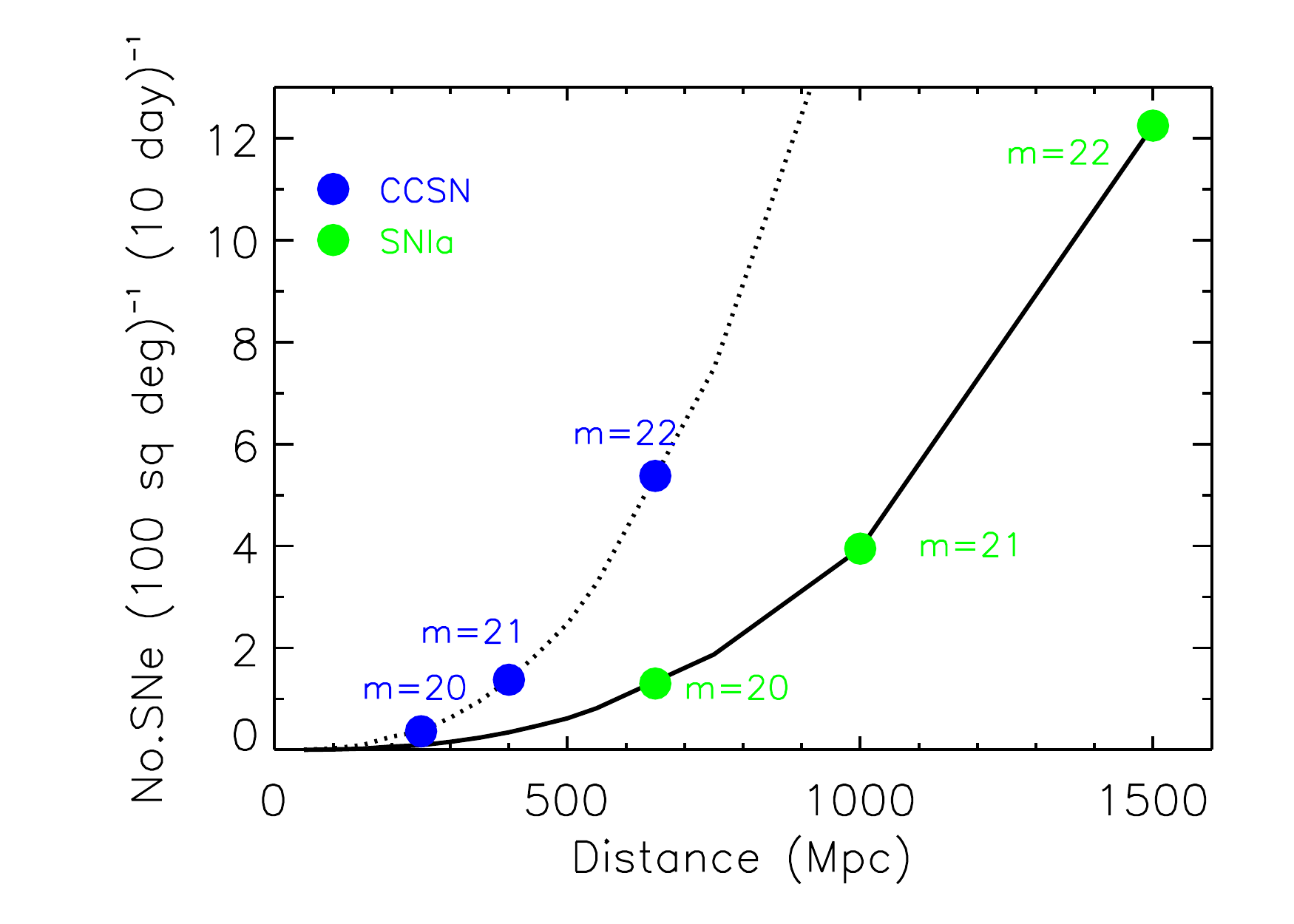}
\caption{The predicted number of SNe per 100 square degrees per 10 day period, determined from the cosmic SN rates (Section\,\ref{sec:rates}). The core-collapse rates are the dotted  line and the Ia rates are solid line.  The solid symbols illustrate the peak magnitude of CCNSe ($M = -17$) and SNIa ($M = -19$)  AB mags at the distances, illustrating the numbers expected for difference survey depths. For example, a survey which reaches $m=22$ can except to find around 6 CCSN which would have exploded in a 10 day window around the GW.  }
\label{fig:sncontam}
\end{figure}

A key part of removing the contaminating SN population is to use the date and time
of the GW source to reject transients which are not temporally coincident. 
As discussed and illustrated in this paper, a survey of $\sim$hundreds of square degrees
will find mostly old SNe that have exploded in the weeks previously. Effectively we are
sampling the history of explosions over a past period which is defined by the length of 
time they are visible for. A more interesting question is how many new SNe per day per 
100 sq degrees are expected to be detected. In Fig.\,\ref{fig:sncontam} we show these numbers which are calculated from the cosmic SN rates (discussed in Section\,\ref{sec:rates}). 
Co-moving volumes were calculated with the cosmology used in this paper to 
determine the rates, and luminosity distance is plotted to be useable for magnitude
estimates. 
A survey reaching $m=21$ will typically find 1 new CCSN and 4 new SN Ia per 100 square degrees per 10 days. Which means if we can date the epoch of the explosions with an uncertainty of 10 days, then the numbers in this figure are a reasonable estimate of the 
rate of contaminating sources. Dating requires a combination of spectra and lightcurves
and is generally easier for type Ia SNe than CCNSe since the later show more diversity. 
This prediction is in reasonable agreement with what we have seen in this search. 
For $m=20.5$, we would expect about 12 new SNe in the 440 square degrees in 
a 10 day window and we find about 6 that (within the uncertainties) are plausibly within the
window, plus there are many more candidates for which we have no dates. 
This does not require that the SNe are found very close to explosion, only that 
with information around peak that they can be dated. Of course lightcurves which reach
closer to the explosion time are very useful to aid dating. For example 
PTF have demonstrated the science return from discoveries within 24hrs of a 
deep non-detection 
 \citep[e.g.][]{2013ApJ...775L...7C} and the ATLAS survey 
\citep{2011PASP..123...58T} will soon be capable of surveying the entire northern sky multiple times per
night to $m_{\rm AB} \simeq 20$. 
For possible GW sources, we expect to be looking for temporal and spatial coincidence of 
unusual transients, rare types of SNe or those that 
might be linked to long GRBs (e.g. broad lined Ic SNe) or short GRBs (kilonovae). 
The early searches may be forced to consider temporal coincidences and the 
numbers in Fig.\,\ref{fig:sncontam} serve as a guide. The numbers scale linearly with the
uncertainty in the explosion epoch.

\section{Conclusions}

For the first detection of GW waves from the LIGO experiment, we have searched for an optical counterpart to the source. We used the sky probability maps provided by LIGO to 
focus our search on 442 square degrees of sky with the Pan-STARRS1 telescope. 
We discovered 56 astrophysical transients over a period of 41 days and through 
a combination of spectra mostly from PESSTO and SNIFS, host galaxy redshifts and 
photometric monitoring, we quantified these objects. All appear to be fairly normal 
SNe and AGN variability and none is obviously linked to GW150914. 

The distance estimated by LIGO to GW150914 of 
$D_{L} = 410^{+160}_{-180}$\,Mpc, means that relatively faint kilonova type lightcurves
would not be detectable by our images which reach \ips$\simeq 20.0$ over
the kilonova-like $\sim10$ day timescales. The fact that GW150914
is likely to be a BH-BH coalescence means we do not have quantitative 
models to compare with our limits.  We used analytic parameterised lightcurves
with different timescales to illustrate the capability of our survey. Had we 
covered a significant fraction of the probability, we could reach 
the sensitivity limits of  $M_i \leq -17.2^{-0.9}_{+1.4}$ at the distance of GW150914.
We treat this early search as a lessons learned 
exercise and our main findings are : 

\begin{itemize}
\item Access to the most reliable skymaps as early as possible is key to 
focusing future searches. The final probability of the most reliable skymap that we covered was 4.2 per cent. This was significantly less than the initial cumulative probability from the first skymap released (which was about 30 per cent). It is critical to invest telescope exposure times on the highest probability regions as early as possible. 
\item Spectroscopic classification of sources is essential to determine redshift and 
date the explosion epoch. We provide estimates of the number of contaminating SNe per
sky area per 10 days as a guide to the rate of unrelated sources in the GW sky maps. 
\item We illustrate the capability of the Pan-STARRS1 telescope to survey hundreds of square degrees of sky to AB mags of 19-21.5 rapidly and to produce transients daily. PESSTO is a powerful classification survey that can play an important role in the classification of these 
optical transients. 
\item Future searches will benefit from an even quicker response time to map out the sky localisation region within 1 day. We have demonstrated that this is possible in this paper and 
quantified the contaminating sources arising. The goal of our, and future surveys will be to map the high probability to region to $m_{\rm AB}\simeq22-23$\,mag within 1-2 days. 
\end{itemize}

\bigskip
\noindent
{\bf ACKNOWLEDGMENTS}
Pan-STARRS is supported by the University of Hawaii and the National Aeronautics and Space Administration's Planetary Defense Office  under Grant No. NNX14AM74G. The Pan-STARRS-LIGO effort is in collaboration with the LIGO Consortium and supported by Queen's University Belfast.  The Pan-STARRS1 Sky Surveys have been made possible through contributions by the Institute for Astronomy, the University of Hawaii, the Pan-STARRS Project Office, the Max Planck Society and its participating institutes, the Max Planck Institute for Astronomy, Heidelberg and the Max Planck Institute for Extraterrestrial Physics, Garching, The Johns Hopkins University, Durham University, the University of Edinburgh, the Queen's University Belfast, the Harvard-Smithsonian Center for Astrophysics, the Las Cumbres Observatory Global Telescope Network Incorporated, the National Central University of Taiwan, the Space Telescope Science Institute, and the National Aeronautics and Space Administration under Grant No. NNX08AR22G issued through the Planetary Science Division of the NASA Science Mission Directorate, the National Science Foundation Grant No. AST-1238877, the University of Maryland, Eotvos Lorand University (ELTE), and the Los Alamos National Laboratory. This work is based (in part) on observations collected at the European Organisation for Astronomical Research in the Southern Hemisphere, Chile as part of PESSTO, (the Public ESO Spectroscopic Survey for Transient Objects Survey) ESO programs 188.D-3003,  191.D-0935. 
Some of the data presented herein were obtained at the Palomar Observatory, California Institute of Technology.
SJS acknowledges funding from the European Research Council under the European Union's Seventh Framework Programme (FP7/2007-2013)/ERC Grant agreement n$^{\rm o}$ [291222] and  STFC grants ST/I001123/1 and ST/L000709/1. MF is  supported by the European Union FP7 programme through ERC grant number 320360.  KM acknowledges support from the STFC through an Ernest Rutherford Fellowship
FOE acknowledges support from FONDECYT through postdoctoral grant 3140326.

This research has made use of the NASA/IPAC Extragalactic Database (NED) which is operated by the Jet Propulsion Laboratory, California Institute of Technology, under contract with the National Aeronautics and Space Administration and data products from the Two Micron All Sky Survey, which is a joint project of the University of Massachusetts and the Infrared Processing and Analysis Center/California Institute of Technology, funded by the National Aeronautics and Space Administration and the National Science Foundation.

\bibliographystyle{mn2e}

\bibliography{lib.bib}

\begin{thebibliography}{136}
\expandafter\ifx\csname natexlab\endcsname\relax\def\natexlab#1{#1}\fi

\bibitem[{{Aartsen} {et~al}\mbox{.}(2015){Aartsen}, {Abraham}, {Ackermann},
  {Adams}, {Aguilar}, {Ahlers}, {Ahrens}, {Altmann}, {Anderson}, {Archinger},
  \& et~al.}]{2015ApJ...811...52A}
{Aartsen} M.~G. {et~al.}, 2015, \apj, 811, 52

\bibitem[{{Aasi} {et~al}\mbox{.}(2014){Aasi}, {Abadie}, {Abbott}, {Abbott},
  {Abbott}, {Abernathy}, {Accadia}, {Acernese}, {Adams}, {Adams}, \&
  et~al.}]{2014ApJS..211....7A}
{Aasi} J. {et~al.}, 2014, \apjs, 211, 7

\bibitem[{{Aasi} {et~al}\mbox{.}(2013){Aasi}, {Abadie}, {Abbott}, {Abbott},
  {Abbott}, {Abernathy}, {Accadia}, {Acernese}, {Adams}, {Adams}, \&
  et~al.}]{2013PhRvD..87b2002A}
{Aasi} J. {et~al.}, 2013, \prd, 87, 022002

\bibitem[{{Abadie} {et~al}\mbox{.}(2012){Abadie}, {Abbott}, {Abbott}, {Abbott},
  {Abernathy}, {Accadia}, {Acernese}, {Adams}, {Adhikari}, {Affeldt}, \&
  et~al.}]{2012PhRvD..85h2002A}
{Abadie} J. {et~al.}, 2012, \prd, 85, 082002

\bibitem[{{Abadie} {et~al}\mbox{.}(2010){Abadie}, {Abbott}, {Abbott},
  {Abernathy}, {Accadia}, {Acernese}, {Adams}, {Adhikari}, {Ajith}, {Allen}, \&
  et~al.}]{2010CQGra..27q3001A}
{Abadie} J. {et~al.}, 2010, Classical and Quantum Gravity, 27, 173001

\bibitem[{{Abazajian} {et~al}\mbox{.}(2009){Abazajian}, {Adelman-McCarthy},
  {Ag{\"u}eros}, {Allam}, {Allende Prieto}, {An}, {Anderson}, {Anderson},
  {Annis}, {Bahcall}, \& et~al.}]{2009ApJS..182..543A}
{Abazajian} K.~N. {et~al.}, 2009, \apjs, 182, 543

\bibitem[{{Abbott} {et~al}\mbox{.}(2016{\natexlab{a}}){Abbott}, {Abbott},
  {Abbott}, {Abernathy}, {Acernese}, {Ackley}, {Adams}, {Adams}, {Addesso},
  {Adhikari}, \& et~al.}]{2016arXiv160208492A}
{Abbott} B.~P. {et~al.}, 2016{\natexlab{a}}, ArXiv e-prints

\bibitem[{{Abbott} {et~al}\mbox{.}(2016{\natexlab{b}}){Abbott}, {Abbott},
  {Abbott}, {Abernathy}, {Acernese}, {Ackley}, {Adams}, {Adams}, {Addesso},
  {Adhikari}, \& et~al.}]{theprizepaper}
{Abbott} B.~P. {et~al.}, 2016{\natexlab{b}}, Physical Review Letters, 116,
  061102

\bibitem[{{Abbott} {et~al}\mbox{.}(2016{\natexlab{c}}){Abbott}, {Abbott},
  {Abbott}, {Abernathy}, {Acernese}, {Ackley}, {Adams}, {Adams}, {Addesso},
  {Adhikari}, \& et~al.}]{2016arXiv160203842A}
{Abbott} B.~P. {et~al.}, 2016{\natexlab{c}}, ArXiv e-prints

\bibitem[{{Acernese} {et~al}\mbox{.}(2015){Acernese}, {Agathos}, {Agatsuma},
  {Aisa}, {Allemandou}, {Allocca}, {Amarni}, {Astone}, {Balestri}, {Ballardin},
  \& et~al.}]{2015CQGra..32b4001A}
{Acernese} F. {et~al.}, 2015, Classical and Quantum Gravity, 32, 024001

\bibitem[{{Aldering} {et~al}\mbox{.}(2006){Aldering}, {Antilogus}, {Bailey},
  {Baltay}, {Bauer}, {Blanc}, {Bongard}, {Copin}, {Gangler}, {Gilles},
  {Kessler}, {Kocevski}, {Lee}, {Loken}, {Nugent}, {Pain}, {P{\'e}contal},
  {Pereira}, {Perlmutter}, {Rabinowitz}, {Rigaudier}, {Scalzo}, {Smadja},
  {Thomas}, {Wang}, {Weaver}, \& {Nearby Supernova
  Factory}}]{2006ApJ...650..510A}
{Aldering} G. {et~al.}, 2006, \apj, 650, 510

\bibitem[{{Anderson} {et~al}\mbox{.}(2014){Anderson},
  {Gonz{\'a}lez-Gait{\'a}n}, {Hamuy}, {Guti{\'e}rrez}, {Stritzinger}, {Olivares
  E.}, {Phillips}, {Schulze}, {Antezana}, {Bolt}, {Campillay}, {Castell{\'o}n},
  {Contreras}, {de Jaeger}, {Folatelli}, {F{\"o}rster}, {Freedman},
  {Gonz{\'a}lez}, {Hsiao}, {Krzemi{\'n}ski}, {Krisciunas}, {Maza}, {McCarthy},
  {Morrell}, {Persson}, {Roth}, {Salgado}, {Suntzeff}, \&
  {Thomas-Osip}}]{2014ApJ...786...67A}
{Anderson} J.~P. {et~al.}, 2014, \apj, 786, 67

\bibitem[{{Annis} {et~al}\mbox{.}(2016){Annis}, {Soares-Santos}, {Berger},
  {Brout}, {Chen}, {Chornock}, {Cowperthwaite}, {Diehl}, {Doctor},
  {Drlica-Wagner}, {Drout}, {Farr}, {Finley}, {Flaugher}, {Foley}, {Frieman},
  {Gruendl}, {Herner}, {Holz}, {Kessler}, {Lin}, {Marriner}, {Neilsen}, {Rest},
  {Sako}, {Smith}, {Smith}, {Sobreira}, {Walker}, {Yanny}, {Abbott}, {Abdalla},
  {Allam}, {Benoit-L{\'e}vy}, {Bernstein}, {Bertin}, {Buckley-Geer}, {Burke},
  {Capozzi}, {Carnero Rosell}, {Carrasco Kind}, {Carretero}, {Castander},
  {Cenko}, {Crocce}, {Cunha}, {D'Andrea}, {da Costa}, {Desai}, {Dietrich},
  {Eifler}, {Evrard}, {Fernandez}, {Fischer}, {Fong}, {Fosalba}, {Fox},
  {Fryer}, {Garcia-Bellido}, {Gaztanaga}, {Gerdes}, {Goldstein}, {Gruen},
  {Gutierrez}, {Honscheid}, {James}, {Karliner}, {Kasen}, {Kent}, {Kuehn},
  {Kuropatkin}, {Lahav}, {Li}, {Lima}, {Maia}, {Martini}, {Metzger}, {Miller},
  {Miquel}, {Mohr}, {Nichol}, {Nord}, {Ogando}, {Peoples}, {Petravic},
  {Plazas}, {Quataert}, {Romer}, {Roodman}, {Rykoff}, {Sanchez}, {Santiago},
  {Scarpine}, {Schindler}, {Schubnell}, {Sevilla-Noarbe}, {Sheldon}, {Smith},
  {Stebbins}, {Swanson}, {Tarle}, {Thaler}, {Thomas}, {Tucker}, {Vikram},
  {Wechsler}, {Weller}, \& {Wester}}]{2016ApJ...823L..34A}
{Annis} J. {et~al.}, 2016, \apjl, 823, L34

\bibitem[{{Bacon} {et~al}\mbox{.}(2001){Bacon}, {Copin}, {Monnet}, {Miller},
  {Allington-Smith}, {Bureau}, {Carollo}, {Davies}, {Emsellem}, {Kuntschner},
  {Peletier}, {Verolme}, \& {de Zeeuw}}]{2001MNRAS.326...23B}
{Bacon} R. {et~al.}, 2001, \mnras, 326, 23

\bibitem[{{Baltay} {et~al}\mbox{.}(2013){Baltay}, {Rabinowitz}, {Hadjiyska},
  {Walker}, {Nugent}, {Coppi}, {Ellman}, {Feindt}, {McKinnon}, {Horowitz}, \&
  {Effron}}]{2013PASP..125..683B}
{Baltay} C. {et~al.}, 2013, \pasp, 125, 683

\bibitem[{{Barnes} \& {Kasen}(2013)}]{2013ApJ...775...18B}
{Barnes} J., {Kasen} D., 2013, \apj, 775, 18

\bibitem[{{Berger}(2014)}]{2014ARAA..52...43B}
{Berger} E., 2014, \araa, 52, 43

\bibitem[{{Berger}, {Fong} \& {Chornock}(2013){Berger}, {Fong}, \&
  {Chornock}}]{2013ApJ...774L..23B}
{Berger} E., {Fong} W., {Chornock} R., 2013, \apjl, 774, L23

\bibitem[{{Berry} {et~al}\mbox{.}(2015){Berry}, {Mandel}, {Middleton},
  {Singer}, {Urban}, {Vecchio}, {Vitale}, {Cannon}, {Farr}, {Farr}, {Graff},
  {Hanna}, {Haster}, {Mohapatra}, {Pankow}, {Price}, {Sidery}, \&
  {Veitch}}]{2015ApJ...804..114B}
{Berry} C.~P.~L. {et~al.}, 2015, \apj, 804, 114

\bibitem[{{Blondin} \& {Tonry}(2007)}]{2007ApJ...666.1024B}
{Blondin} S., {Tonry} J.~L., 2007, \apj, 666, 1024

\bibitem[{{Botticella} {et~al}\mbox{.}(2008){Botticella}, {Riello},
  {Cappellaro}, {Benetti}, {Altavilla}, {Pastorello}, {Turatto}, {Greggio},
  {Patat}, {Valenti}, {Zampieri}, {Harutyunyan}, {Pignata}, \&
  {Taubenberger}}]{2008A&A...479...49B}
{Botticella} M.~T. {et~al.}, 2008, \aap, 479, 49

\bibitem[{{Botticella} {et~al}\mbox{.}(2010){Botticella}, {Trundle},
  {Pastorello}, {Rodney}, {Rest}, {Gezari}, {Smartt}, {Narayan}, {Huber},
  {Tonry}, {Young}, {Smith}, {Bresolin}, {Valenti}, {Kotak}, {Mattila},
  {Kankare}, {Wood-Vasey}, {Riess}, {Neill}, {Forster}, {Martin}, {Stubbs},
  {Burgett}, {Chambers}, {Dombeck}, {Flewelling}, {Grav}, {Heasley}, {Hodapp},
  {Kaiser}, {Kudritzki}, {Luppino}, {Lupton}, {Magnier}, {Monet}, {Morgan},
  {Onaka}, {Price}, {Rhoads}, {Siegmund}, {Sweeney}, {Wainscoat}, {Waters},
  {Waterson}, \& {Wynn-Williams}}]{2010ApJ...717L..52B}
{Botticella} M.~T. {et~al.}, 2010, \apjl, 717, L52

\bibitem[{{Brink} {et~al}\mbox{.}(2013){Brink}, {Richards}, {Poznanski},
  {Bloom}, {Rice}, {Negahban}, \& {Wainwright}}]{2013MNRAS.435.1047B}
{Brink} H., {Richards} J.~W., {Poznanski} D., {Bloom} J.~S., {Rice} J.,
  {Negahban} S., {Wainwright} M., 2013, \mnras, 435, 1047

\bibitem[{{Cao} {et~al}\mbox{.}(2013){Cao}, {Kasliwal}, {Arcavi}, {Horesh},
  {Hancock}, {Valenti}, {Cenko}, {Kulkarni}, {Gal-Yam}, {Gorbikov}, {Ofek},
  {Sand}, {Yaron}, {Graham}, {Silverman}, {Wheeler}, {Marion}, {Walker},
  {Mazzali}, {Howell}, {Li}, {Kong}, {Bloom}, {Nugent}, {Surace}, {Masci},
  {Carpenter}, {Degenaar}, \& {Gelino}}]{2013ApJ...775L...7C}
{Cao} Y. {et~al.}, 2013, \apjl, 775, L7

\bibitem[{{Cappellaro} {et~al}\mbox{.}(2015){Cappellaro}, {Botticella},
  {Pignata}, {Grado}, {Greggio}, {Limatola}, {Vaccari}, {Baruffolo}, {Benetti},
  {Bufano}, {Capaccioli}, {Cascone}, {Covone}, {De Cicco}, {Falocco}, {Della
  Valle}, {Jarvis}, {Marchetti}, {Napolitano}, {Paolillo}, {Pastorello},
  {Radovich}, {Schipani}, {Spiro}, {Tomasella}, \&
  {Turatto}}]{2015A&A...584A..62C}
{Cappellaro} E. {et~al.}, 2015, \aap, 584, A62

\bibitem[{{Cappellaro}, {Evans} \& {Turatto}(1999){Cappellaro}, {Evans}, \&
  {Turatto}}]{1999A&A...351..459C}
{Cappellaro} E., {Evans} R., {Turatto} M., 1999, \aap, 351, 459

\bibitem[{{Cenko} {et~al}\mbox{.}(2013){Cenko}, {Kulkarni}, {Horesh}, {Corsi},
  {Fox}, {Carpenter}, {Frail}, {Nugent}, {Perley}, {Gruber}, {Gal-Yam},
  {Groot}, {Hallinan}, {Ofek}, {Rau}, {MacLeod}, {Miller}, {Bloom},
  {Filippenko}, {Kasliwal}, {Law}, {Morgan}, {Polishook}, {Poznanski},
  {Quimby}, {Sesar}, {Shen}, {Silverman}, \& {Sternberg}}]{2013ApJ...769..130C}
{Cenko} S.~B. {et~al.}, 2013, \apj, 769, 130

\bibitem[{{Chen} \& {Holz}(2015)}]{2015arXiv150900055C}
{Chen} H.-Y., {Holz} D.~E., 2015, ArXiv e-prints

\bibitem[{{Chornock} {et~al}\mbox{.}(2014){Chornock}, {Berger}, {Gezari},
  {Zauderer}, {Rest}, {Chomiuk}, {Kamble}, {Soderberg}, {Czekala}, {Dittmann},
  {Drout}, {Foley}, {Fong}, {Huber}, {Kirshner}, {Lawrence}, {Lunnan},
  {Marion}, {Narayan}, {Riess}, {Roth}, {Sanders}, {Scolnic}, {Smartt},
  {Smith}, {Stubbs}, {Tonry}, {Burgett}, {Chambers}, {Flewelling}, {Hodapp},
  {Kaiser}, {Magnier}, {Martin}, {Neill}, {Price}, \&
  {Wainscoat}}]{2014ApJ...780...44C}
{Chornock} R. {et~al.}, 2014, \apj, 780, 44

\bibitem[{{Clocchiatti} {et~al}\mbox{.}(1996){Clocchiatti}, {Benetti},
  {Wheeler}, {Wren}, {Boisseau}, {Cappellaro}, {Turatto}, {Patat}, {Swartz},
  {Harkness}, {Brotherton}, {Wills}, {Hemenway}, {Cornell}, {Frueh}, \&
  {Kaiser}}]{1996AJ....111.1286C}
{Clocchiatti} A. {et~al.}, 1996, \aj, 111, 1286

\bibitem[{{Cornish} \& {Littenberg}(2015)}]{2015CQGra..32m5012C}
{Cornish} N.~J., {Littenberg} T.~B., 2015, Classical and Quantum Gravity, 32,
  135012

\bibitem[{{Cowperthwaite} \& {Berger}(2015)}]{2015ApJ...814...25C}
{Cowperthwaite} P.~S., {Berger} E., 2015, \apj, 814, 25

\bibitem[{{Denneau} {et~al}\mbox{.}(2013){Denneau}, {Jedicke}, {Grav},
  {Granvik}, {Kubica}, {Milani}, {Vere{\v s}}, {Wainscoat}, {Chang},
  {Pierfederici}, {Kaiser}, {Chambers}, {Heasley}, {Magnier}, {Price}, {Myers},
  {Kleyna}, {Hsieh}, {Farnocchia}, {Waters}, {Sweeney}, {Green}, {Bolin},
  {Burgett}, {Morgan}, {Tonry}, {Hodapp}, {Chastel}, {Chesley}, {Fitzsimmons},
  {Holman}, {Spahr}, {Tholen}, {Williams}, {Abe}, {Armstrong}, {Bressi},
  {Holmes}, {Lister}, {McMillan}, {Micheli}, {Ryan}, {Ryan}, \&
  {Scotti}}]{2013PASP..125..357D}
{Denneau} L. {et~al.}, 2013, \pasp, 125, 357

\bibitem[{{Dilday} {et~al}\mbox{.}(2010){Dilday}, {Smith}, {Bassett}, {Becker},
  {Bender}, {Castander}, {Cinabro}, {Filippenko}, {Frieman}, {Galbany},
  {Garnavich}, {Goobar}, {Hopp}, {Ihara}, {Jha}, {Kessler}, {Lampeitl},
  {Marriner}, {Miquel}, {Moll{\'a}}, {Nichol}, {Nordin}, {Riess}, {Sako},
  {Schneider}, {Sollerman}, {Wheeler}, {{\"O}stman}, {Bizyaev}, {Brewington},
  {Malanushenko}, {Malanushenko}, {Oravetz}, {Pan}, {Simmons}, \&
  {Snedden}}]{2010ApJ...713.1026D}
{Dilday} B. {et~al.}, 2010, \apj, 713, 1026

\bibitem[{{Drake} {et~al}\mbox{.}(2013){Drake}, {Djorgovski}, {Mahabal},
  {Graham}, {Williams}, {Prieto}, {Catelan}, {Christensen}, {Larson}, \&
  {Beshore}}]{2013ATel.4872....1D}
{Drake} A.~J. {et~al.}, 2013, The Astronomer's Telegram, 4872, 1

\bibitem[{{Drake} {et~al}\mbox{.}(2010){Drake}, {Djorgovski}, {Prieto},
  {Mahabal}, {Balam}, {Williams}, {Graham}, {Catelan}, {Beshore}, \&
  {Larson}}]{2010ApJ...718L.127D}
{Drake} A.~J. {et~al.}, 2010, \apjl, 718, L127

\bibitem[{{Drout} {et~al}\mbox{.}(2014){Drout}, {Chornock}, {Soderberg},
  {Sanders}, {McKinnon}, {Rest}, {Foley}, {Milisavljevic}, {Margutti},
  {Berger}, {Calkins}, {Fong}, {Gezari}, {Huber}, {Kankare}, {Kirshner},
  {Leibler}, {Lunnan}, {Mattila}, {Marion}, {Narayan}, {Riess}, {Roth},
  {Scolnic}, {Smartt}, {Tonry}, {Burgett}, {Chambers}, {Hodapp}, {Jedicke},
  {Kaiser}, {Magnier}, {Metcalfe}, {Morgan}, {Price}, \&
  {Waters}}]{2014ApJ...794...23D}
{Drout} M.~R. {et~al.}, 2014, \apj, 794, 23

\bibitem[{{Essick} {et~al}\mbox{.}(2015){Essick}, {Vitale}, {Katsavounidis},
  {Vedovato}, \& {Klimenko}}]{2015ApJ...800...81E}
{Essick} R., {Vitale} S., {Katsavounidis} E., {Vedovato} G., {Klimenko} S.,
  2015, \apj, 800, 81

\bibitem[{{Fasano} {et~al}\mbox{.}(2006){Fasano}, {Marmo}, {Varela},
  {D'Onofrio}, {Poggianti}, {Moles}, {Pignatelli}, {Bettoni}, {Kj{\ae}rgaard},
  {Rizzi}, {Couch}, \& {Dressler}}]{2006A&A...445..805F}
{Fasano} G. {et~al.}, 2006, \aap, 445, 805

\bibitem[{{Filippenko}(1997)}]{1997ARA&A..35..309F}
{Filippenko} A.~V., 1997, \araa, 35, 309

\bibitem[{{Finkbeiner} {et~al}\mbox{.}(2015){Finkbeiner}, {Schlafly},
  {Schlegel}, {Padmanabhan}, {Juric}, {Burgett}, {Chambers}, {Denneau},
  {Draper}, {Flewelling}, {Hodapp}, {Kaiser}, {Magnier}, {Metcalfe}, {Morgan},
  {Price}, {Stubbs}, \& {Tonry}}]{2015arXiv151201214F}
{Finkbeiner} D.~P. {et~al.}, 2015, ArXiv e-prints

\bibitem[{{Flesch}(2015)}]{2015PASA...32...10F}
{Flesch} E.~W., 2015, \pasa, 32, 10

\bibitem[{{Fraser} {et~al}\mbox{.}(2013){Fraser}, {Magee}, {Kotak}, {Smartt},
  {Smith}, {Polshaw}, {Drake}, {Boles}, {Lee}, {Burgett}, {Chambers}, {Draper},
  {Flewelling}, {Hodapp}, {Kaiser}, {Kudritzki}, {Magnier}, {Price}, {Tonry},
  {Wainscoat}, \& {Waters}}]{2013ApJ...779L...8F}
{Fraser} M. {et~al.}, 2013, \apjl, 779, L8

\bibitem[{{Galbany} {et~al}\mbox{.}(2016){Galbany}, {Hamuy}, {Phillips},
  {Suntzeff}, {Maza}, {de Jaeger}, {Moraga}, {Gonz{\'a}lez-Gait{\'a}n},
  {Krisciunas}, {Morrell}, {Thomas-Osip}, {Krzeminski}, {Gonz{\'a}lez},
  {Antezana}, {Wishnjewski}, {McCarthy}, {Anderson}, {Guti{\'e}rrez},
  {Stritzinger}, {Folatelli}, {Anguita}, {Galaz}, {Green}, {Impey}, {Kim},
  {Kirhakos}, {Malkan}, {Mulchaey}, {Phillips}, {Pizzella}, {Prosser},
  {Schmidt}, {Schommer}, {Sherry}, {Strolger}, {Wells}, \&
  {Williger}}]{2016AJ....151...33G}
{Galbany} L. {et~al.}, 2016, \aj, 151, 33

\bibitem[{{Gall} {et~al}\mbox{.}(2015){Gall}, {Polshaw}, {Kotak}, {Jerkstrand},
  {Leibundgut}, {Rabinowitz}, {Sollerman}, {Sullivan}, {Smartt}, {Anderson},
  {Benetti}, {Baltay}, {Feindt}, {Fraser}, {Gonz{\'a}lez-Gait{\'a}n},
  {Inserra}, {Maguire}, {McKinnon}, {Valenti}, \&
  {Young}}]{2015A&A...582A...3G}
{Gall} E.~E.~E. {et~al.}, 2015, \aap, 582, A3

\bibitem[{{Ganeshalingam}, {Li} \& {Filippenko}(2011){Ganeshalingam}, {Li}, \&
  {Filippenko}}]{2011MNRAS.416.2607G}
{Ganeshalingam} M., {Li} W., {Filippenko} A.~V., 2011, \mnras, 416, 2607

\bibitem[{{Gehrels} {et~al}\mbox{.}(2015){Gehrels}, {Cannizzo}, {Kanner},
  {Kasliwal}, {Nissanke}, \& {Singer}}]{2015arXiv150803608G}
{Gehrels} N., {Cannizzo} J.~K., {Kanner} J., {Kasliwal} M.~M., {Nissanke} S.,
  {Singer} L.~P., 2015, ArXiv e-prints

\bibitem[{{Gezari} {et~al}\mbox{.}(2015){Gezari}, {Jones}, {Sanders},
  {Soderberg}, {Hung}, {Heinis}, {Smartt}, {Rest}, {Scolnic}, {Chornock},
  {Berger}, {Foley}, {Huber}, {Price}, {Stubbs}, {Riess}, {Kirshner}, {Smith},
  {Wood-Vasey}, {Schiminovich}, {Martin}, {Burgett}, {Chambers}, {Flewelling},
  {Kaiser}, {Tonry}, \& {Wainscoat}}]{2015ApJ...804...28G}
{Gezari} S. {et~al.}, 2015, \apj, 804, 28

\bibitem[{{Goldstein} {et~al}\mbox{.}(2015){Goldstein}, {D'Andrea}, {Fischer},
  {Foley}, {Gupta}, {Kessler}, {Kim}, {Nichol}, {Nugent}, {Papadopoulos},
  {Sako}, {Smith}, {Sullivan}, {Thomas}, {Wester}, {Wolf}, {Abdalla},
  {Banerji}, {Benoit-Levy}, {Bertin}, {Brooks}, {Carnero Rosell}, {Castander},
  {da Costa}, {Covarrubias}, {DePoy}, {Desai}, {Diehl}, {Doel}, {Eifler},
  {Fausti Neto}, {Finley}, {Flaugher}, {Fosalba}, {Frieman}, {Gerdes}, {Gruen},
  {Gruendl}, {James}, {Kuehn}, {Kuropatkin}, {Lahav}, {Li}, {Maia}, {Makler},
  {March}, {Marshall}, {Martini}, {Merritt}, {Miquel}, {Nord}, {Ogando},
  {Plazas}, {Romer}, {Roodman}, {Sanchez}, {Scarpine}, {Schubnell},
  {Sevilla-Noarbe}, {Smith}, {Soares-Santos}, {Sobreira}, {Suchyta}, {Swanson},
  {Tarle}, {Thaler}, \& {Walker}}]{2015AJ....150...82G}
{Goldstein} D.~A. {et~al.}, 2015, \aj, 150, 82

\bibitem[{{Gonz{\'a}lez-Gait{\'a}n}
  {et~al}\mbox{.}(2015){Gonz{\'a}lez-Gait{\'a}n}, {Tominaga}, {Molina},
  {Galbany}, {Bufano}, {Anderson}, {Gutierrez}, {F{\"o}rster}, {Pignata},
  {Bersten}, {Howell}, {Sullivan}, {Carlberg}, {de Jaeger}, {Hamuy},
  {Baklanov}, \& {Blinnikov}}]{2015MNRAS.451.2212G}
{Gonz{\'a}lez-Gait{\'a}n} S. {et~al.}, 2015, \mnras, 451, 2212

\bibitem[{{Graham} \& {Schady}(2015)}]{2015arXiv151101466G}
{Graham} J.~F., {Schady} P., 2015, ArXiv e-prints

\bibitem[{{Guetta} \& {Della Valle}(2007)}]{2007ApJ...657L..73G}
{Guetta} D., {Della Valle} M., 2007, \apjl, 657, L73

\bibitem[{{Hanna}, {Mandel} \& {Vousden}(2014){Hanna}, {Mandel}, \&
  {Vousden}}]{2014ApJ...784....8H}
{Hanna} C., {Mandel} I., {Vousden} W., 2014, \apj, 784, 8

\bibitem[{{Harutyunyan} {et~al}\mbox{.}(2008){Harutyunyan}, {Pfahler},
  {Pastorello}, {Taubenberger}, {Turatto}, {Cappellaro}, {Benetti},
  {Elias-Rosa}, {Navasardyan}, {Valenti}, {Stanishev}, {Patat}, {Riello},
  {Pignata}, \& {Hillebrandt}}]{2008A&A...488..383H}
{Harutyunyan} A.~H. {et~al.}, 2008, \aap, 488, 383

\bibitem[{{Holtzman} {et~al}\mbox{.}(2008){Holtzman}, {Marriner}, {Kessler},
  {Sako}, {Dilday}, {Frieman}, {Schneider}, {Bassett}, {Becker}, {Cinabro},
  {DeJongh}, {Depoy}, {Doi}, {Garnavich}, {Hogan}, {Jha}, {Konishi},
  {Lampeitl}, {Marshall}, {McGinnis}, {Miknaitis}, {Nichol}, {Prieto}, {Riess},
  {Richmond}, {Romani}, {Smith}, {Takanashi}, {Tokita}, {van der Heyden},
  {Yasuda}, \& {Zheng}}]{2008AJ....136.2306H}
{Holtzman} J.~A. {et~al.}, 2008, \aj, 136, 2306

\bibitem[{{Horiuchi} {et~al}\mbox{.}(2011){Horiuchi}, {Beacom}, {Kochanek},
  {Prieto}, {Stanek}, \& {Thompson}}]{2011ApJ...738..154H}
{Horiuchi} S., {Beacom} J.~F., {Kochanek} C.~S., {Prieto} J.~L., {Stanek}
  K.~Z., {Thompson} T.~A., 2011, \apj, 738, 154

\bibitem[{{Huber} {et~al}\mbox{.}(2015{\natexlab{a}}){Huber}, {Carter
  Chambers}, {Flewelling}, {Smartt}, {Smith}, \&
  {Wright}}]{2015IAUGA..2258303H}
{Huber} M., {Carter Chambers} K., {Flewelling} H., {Smartt} S.~J., {Smith} K.,
  {Wright} D., 2015{\natexlab{a}}, IAU General Assembly, 22, 58303

\bibitem[{{Huber} {et~al}\mbox{.}(2015{\natexlab{b}}){Huber}, {Chambers},
  {Flewelling}, {Willman}, {Primak}, {Schultz}, {Gibson}, {Magnier}, {Waters},
  {Tonry}, {Wainscoat}, {Smith}, {Wright}, {Smartt}, {Foley}, {Jha}, {Rest}, \&
  {Scolnic}}]{2015ATel.7153....1H}
{Huber} M. {et~al.}, 2015{\natexlab{b}}, The Astronomer's Telegram, 7153, 1

\bibitem[{{Inada} {et~al}\mbox{.}(2003){Inada}, {Becker}, {Burles},
  {Castander}, {Eisenstein}, {Hall}, {Johnston}, {Pindor}, {Richards},
  {Schechter}, {Sekiguchi}, {White}, {Brinkmann}, {Frieman}, {Kleinman},
  {Krzesi{\'n}ski}, {Long}, {Neilsen}, {Newman}, {Nitta}, {Schneider},
  {Snedden}, \& {York}}]{2003AJ....126..666I}
{Inada} N. {et~al.}, 2003, \aj, 126, 666

\bibitem[{{Inserra} {et~al}\mbox{.}(2013{\natexlab{a}}){Inserra}, {Pastorello},
  {Turatto}, {Pumo}, {Benetti}, {Cappellaro}, {Botticella}, {Bufano},
  {Elias-Rosa}, {Harutyunyan}, {Taubenberger}, {Valenti}, \&
  {Zampieri}}]{2013A&A...555A.142I}
{Inserra} C. {et~al.}, 2013{\natexlab{a}}, \aap, 555, A142

\bibitem[{{Inserra} {et~al}\mbox{.}(2013{\natexlab{b}}){Inserra}, {Smartt},
  {Jerkstrand}, {Valenti}, {Fraser}, {Wright}, {Smith}, {Chen}, {Kotak},
  {Pastorello}, {Nicholl}, {Bresolin}, {Kudritzki}, {Benetti}, {Botticella},
  {Burgett}, {Chambers}, {Ergon}, {Flewelling}, {Fynbo}, {Geier}, {Hodapp},
  {Howell}, {Huber}, {Kaiser}, {Leloudas}, {Magill}, {Magnier}, {McCrum},
  {Metcalfe}, {Price}, {Rest}, {Sollerman}, {Sweeney}, {Taddia},
  {Taubenberger}, {Tonry}, {Wainscoat}, {Waters}, \&
  {Young}}]{2013ApJ...770..128I}
{Inserra} C. {et~al.}, 2013{\natexlab{b}}, \apj, 770, 128

\bibitem[{{Jin} {et~al}\mbox{.}(2015){Jin}, {Li}, {Cano}, {Covino}, {Fan}, \&
  {Wei}}]{2015ApJ...811L..22J}
{Jin} Z.-P., {Li} X., {Cano} Z., {Covino} S., {Fan} Y.-Z., {Wei} D.-M., 2015,
  \apjl, 811, L22

\bibitem[{{Kaiser} {et~al}\mbox{.}(2010){Kaiser}, {Burgett}, {Chambers},
  {Denneau}, {Heasley}, {Jedicke}, {Magnier}, {Morgan}, {Onaka}, \&
  {Tonry}}]{2010SPIE.7733E..12K}
{Kaiser} N. {et~al.}, 2010, in Society of Photo-Optical Instrumentation
  Engineers (SPIE) Conference Series, Vol. 7733, Society of Photo-Optical
  Instrumentation Engineers (SPIE) Conference Series

\bibitem[{{Kamble} \& {Kaplan}(2013)}]{2013IJMPD..2241011K}
{Kamble} A., {Kaplan} D.~L.~A., 2013, International Journal of Modern Physics
  D, 22, 41011

\bibitem[{{Kankare} {et~al}\mbox{.}(2015){Kankare}, {Kotak}, {Pastorello},
  {Fraser}, {Mattila}, {Smartt}, {Bruce}, {Chambers}, {Elias-Rosa},
  {Flewelling}, {Fremling}, {Harmanen}, {Huber}, {Jerkstrand}, {Kangas},
  {Kuncarayakti}, {Magee}, {Magnier}, {Polshaw}, {Smith}, {Sollerman}, \&
  {Tomasella}}]{2015A&A...581L...4K}
{Kankare} E. {et~al.}, 2015, \aap, 581, L4

\bibitem[{{Kasen}, {Badnell} \& {Barnes}(2013){Kasen}, {Badnell}, \&
  {Barnes}}]{2013ApJ...774...25K}
{Kasen} D., {Badnell} N.~R., {Barnes} J., 2013, \apj, 774, 25

\bibitem[{{Kasen}, {Fern{\'a}ndez} \& {Metzger}(2015){Kasen}, {Fern{\'a}ndez},
  \& {Metzger}}]{2015MNRAS.450.1777K}
{Kasen} D., {Fern{\'a}ndez} R., {Metzger} B.~D., 2015, \mnras, 450, 1777

\bibitem[{{Kasliwal} {et~al}\mbox{.}(2016){Kasliwal}, {Cenko}, {Singer},
  {Corsi}, {Cao}, {Barlow}, {Bhalerao}, {Bellm}, {Cook}, {Duggan}, {Ferretti},
  {Frail}, {Horesh}, {Kendrick}, {Kulkarni}, {Lunnan}, {Palliyaguru}, {Laher},
  {Masci}, {Manulis}, {Miller}, {Nugent}, {Perley}, {Prince}, {Quimby}, {Rana},
  {Rebbapragada}, {Sesar}, {Singhal}, {Surace}, \& {Van
  Sistine}}]{2016arXiv160208764K}
{Kasliwal} M.~M. {et~al.}, 2016, ArXiv e-prints

\bibitem[{{Kasliwal} \& {Nissanke}(2014)}]{2014ApJ...789L...5K}
{Kasliwal} M.~M., {Nissanke} S., 2014, \apjl, 789, L5

\bibitem[{{Kelly} \& {Kirshner}(2012)}]{2012ApJ...759..107K}
{Kelly} P.~L., {Kirshner} R.~P., 2012, \apj, 759, 107

\bibitem[{{Klimenko} {et~al}\mbox{.}(2016){Klimenko}, {Vedovato}, {Drago},
  {Salemi}, {Tiwari}, {Prodi}, {Lazzaro}, {Ackley}, {Tiwari}, {Da Silva}, \&
  {Mitselmakher}}]{2016PhRvD..93d2004K}
{Klimenko} S. {et~al.}, 2016, \prd, 93, 042004

\bibitem[{{Koppenhoefer} {et~al}\mbox{.}(2009){Koppenhoefer}, {Afonso},
  {Saglia}, \& {Henning}}]{2009A&A...494..707K}
{Koppenhoefer} J., {Afonso} C., {Saglia} R.~P., {Henning} T., 2009, \aap, 494,
  707

\bibitem[{{Lantz} {et~al}\mbox{.}(2004){Lantz}, {Aldering}, {Antilogus},
  {Bonnaud}, {Capoani}, {Castera}, {Copin}, {Dubet}, {Gangler}, {Henault},
  {Lemonnier}, {Pain}, {Pecontal}, {Pecontal}, \&
  {Smadja}}]{2004SPIE.5249..146L}
{Lantz} B. {et~al.}, 2004, in Society of Photo-Optical Instrumentation
  Engineers (SPIE) Conference Series, Vol. 5249, Optical Design and
  Engineering, {Mazuray} L., {Rogers} P.~J., {Wartmann} R., eds., pp. 146--155

\bibitem[{{Le Guillou} {et~al}\mbox{.}(2015){Le Guillou}, {Mitra}, {Baumont},
  {Chotard}, {Leget}, {Fraser}, {Galbany}, {Dennefeld}, {Inserra}, {Maguire},
  {Smartt}, {Smith}, {Sullivan}, {Valenti}, {Yaron}, {Young}, {Manulis},
  {Baltay}, {Ellman}, {Hadjiyska}, {McKinnon}, {Rabinowitz}, {Rostami},
  {Feindt}, {Kowalski}, {Nugent}, {Wright}, {Chambers}, {Flewelling}, {Huber},
  {Magnier}, {Tonry}, {Waters}, \& {Wainscoat}}]{2015ATel.7102....1G}
{Le Guillou} L.~L. {et~al.}, 2015, The Astronomer's Telegram, 7102, 1

\bibitem[{{Lee} {et~al}\mbox{.}(2014){Lee}, {Koppenhoefer}, {Seitz}, {Bender},
  {Riffeser}, {Kodric}, {Hopp}, {Snigula}, {G{\"o}ssl}, {Kudritzki}, {Burgett},
  {Chambers}, {Hodapp}, {Kaiser}, \& {Waters}}]{2014ApJ...797...22L}
{Lee} C.-H. {et~al.}, 2014, \apj, 797, 22

\bibitem[{{Li} {et~al}\mbox{.}(2011){Li}, {Leaman}, {Chornock}, {Filippenko},
  {Poznanski}, {Ganeshalingam}, {Wang}, {Modjaz}, {Jha}, {Foley}, \&
  {Smith}}]{2011MNRAS.412.1441L}
{Li} W. {et~al.}, 2011, \mnras, 412, 1441

\bibitem[{{Lynch} {et~al}\mbox{.}(2015){Lynch}, {Vitale}, {Essick},
  {Katsavounidis}, \& {Robinet}}]{2015arXiv151105955L}
{Lynch} R., {Vitale} S., {Essick} R., {Katsavounidis} E., {Robinet} F., 2015,
  ArXiv e-prints

\bibitem[{{Magnier} {et~al}\mbox{.}(2013){Magnier}, {Schlafly}, {Finkbeiner},
  {Juric}, {Tonry}, {Burgett}, {Chambers}, {Flewelling}, {Kaiser}, {Kudritzki},
  {Morgan}, {Price}, {Sweeney}, \& {Stubbs}}]{2013ApJS..205...20M}
{Magnier} E.~A. {et~al.}, 2013, \apjs, 205, 20

\bibitem[{{McCrum} {et~al}\mbox{.}(2014){McCrum}, {Smartt}, {Kotak}, {Rest},
  {Jerkstrand}, {Inserra}, {Rodney}, {Chen}, {Howell}, {Huber}, {Pastorello},
  {Tonry}, {Bresolin}, {Kudritzki}, {Chornock}, {Berger}, {Smith},
  {Botticella}, {Foley}, {Fraser}, {Milisavljevic}, {Nicholl}, {Riess},
  {Stubbs}, {Valenti}, {Wood-Vasey}, {Wright}, {Young}, {Drout}, {Czekala},
  {Burgett}, {Chambers}, {Draper}, {Flewelling}, {Hodapp}, {Kaiser}, {Magnier},
  {Metcalfe}, {Price}, {Sweeney}, \& {Wainscoat}}]{2014MNRAS.437..656M}
{McCrum} M. {et~al.}, 2014, \mnras, 437, 656

\bibitem[{{McCrum} {et~al}\mbox{.}(2015){McCrum}, {Smartt}, {Rest}, {Smith},
  {Kotak}, {Rodney}, {Young}, {Chornock}, {Berger}, {Foley}, {Fraser},
  {Wright}, {Scolnic}, {Tonry}, {Urata}, {Huang}, {Pastorello}, {Botticella},
  {Valenti}, {Mattila}, {Kankare}, {Farrow}, {Huber}, {Stubbs}, {Kirshner},
  {Bresolin}, {Burgett}, {Chambers}, {Draper}, {Flewelling}, {Jedicke},
  {Kaiser}, {Magnier}, {Metcalfe}, {Morgan}, {Price}, {Sweeney}, {Wainscoat},
  \& {Waters}}]{2015MNRAS.448.1206M}
{McCrum} M. {et~al.}, 2015, \mnras, 448, 1206

\bibitem[{{Metcalfe} {et~al}\mbox{.}(2013){Metcalfe}, {Farrow}, {Cole},
  {Draper}, {Norberg}, {Burgett}, {Chambers}, {Denneau}, {Flewelling},
  {Kaiser}, {Kudritzki}, {Magnier}, {Morgan}, {Price}, {Sweeney}, {Tonry},
  {Wainscoat}, \& {Waters}}]{2013MNRAS.435.1825M}
{Metcalfe} N. {et~al.}, 2013, \mnras, 435, 1825

\bibitem[{{Metzger} {et~al}\mbox{.}(2015){Metzger}, {Bauswein}, {Goriely}, \&
  {Kasen}}]{2015MNRAS.446.1115M}
{Metzger} B.~D., {Bauswein} A., {Goriely} S., {Kasen} D., 2015, \mnras, 446,
  1115

\bibitem[{{Metzger} \& {Berger}(2012)}]{2012ApJ...746...48M}
{Metzger} B.~D., {Berger} E., 2012, \apj, 746, 48

\bibitem[{{Metzger} {et~al}\mbox{.}(2010){Metzger}, {Mart{\'{\i}}nez-Pinedo},
  {Darbha}, {Quataert}, {Arcones}, {Kasen}, {Thomas}, {Nugent}, {Panov}, \&
  {Zinner}}]{2010MNRAS.406.2650M}
{Metzger} B.~D. {et~al.}, 2010, \mnras, 406, 2650

\bibitem[{{Narayan}, {Paczynski} \& {Piran}(1992){Narayan}, {Paczynski}, \&
  {Piran}}]{1992ApJ...395L..83N}
{Narayan} R., {Paczynski} B., {Piran} T., 1992, \apjl, 395, L83

\bibitem[{{Ngiam} {et~al}\mbox{.}(2011){Ngiam}, {Chen}, {Bhaskar}, Koh, \&
  Ng}]{Ngiam11}
{Ngiam} J., {Chen} Z., {Bhaskar} S.~A., Koh P.~W., Ng A.~Y., 2011, in Advances
  in Neural Information Processing Systems 24, Shawe-Taylor J., Zemel R.,
  Bartlett P., Pereira F., Weinberger K., eds., Curran Associates, Inc., pp.
  1125--1133

\bibitem[{{Nicholl} {et~al}\mbox{.}(2013){Nicholl}, {Smartt}, {Jerkstrand},
  {Inserra}, {McCrum}, {Kotak}, {Fraser}, {Wright}, {Chen}, {Smith}, {Young},
  {Sim}, {Valenti}, {Howell}, {Bresolin}, {Kudritzki}, {Tonry}, {Huber},
  {Rest}, {Pastorello}, {Tomasella}, {Cappellaro}, {Benetti}, {Mattila},
  {Kankare}, {Kangas}, {Leloudas}, {Sollerman}, {Taddia}, {Berger}, {Chornock},
  {Narayan}, {Stubbs}, {Foley}, {Lunnan}, {Soderberg}, {Sanders},
  {Milisavljevic}, {Margutti}, {Kirshner}, {Elias-Rosa}, {Morales-Garoffolo},
  {Taubenberger}, {Botticella}, {Gezari}, {Urata}, {Rodney}, {Riess},
  {Scolnic}, {Wood-Vasey}, {Burgett}, {Chambers}, {Flewelling}, {Magnier},
  {Kaiser}, {Metcalfe}, {Morgan}, {Price}, {Sweeney}, \&
  {Waters}}]{2013Natur.502..346N}
{Nicholl} M. {et~al.}, 2013, \nat, 502, 346

\bibitem[{{Nicholl} {et~al}\mbox{.}(2015){Nicholl}, {Smartt}, {Jerkstrand},
  {Inserra}, {Sim}, {Chen}, {Benetti}, {Fraser}, {Gal-Yam}, {Kankare},
  {Maguire}, {Smith}, {Sullivan}, {Valenti}, {Young}, {Baltay}, {Bauer},
  {Baumont}, {Bersier}, {Botticella}, {Childress}, {Dennefeld}, {Della Valle},
  {Elias-Rosa}, {Feindt}, {Galbany}, {Hadjiyska}, {Le Guillou}, {Leloudas},
  {Mazzali}, {McKinnon}, {Polshaw}, {Rabinowitz}, {Rostami}, {Scalzo},
  {Schmidt}, {Schulze}, {Sollerman}, {Taddia}, \& {Yuan}}]{2015MNRAS.452.3869N}
{Nicholl} M. {et~al.}, 2015, \mnras, 452, 3869

\bibitem[{{Nissanke}, {Kasliwal} \& {Georgieva}(2013){Nissanke}, {Kasliwal}, \&
  {Georgieva}}]{2013ApJ...767..124N}
{Nissanke} S., {Kasliwal} M., {Georgieva} A., 2013, \apj, 767, 124

\bibitem[{{Ofek} {et~al}\mbox{.}(2007){Ofek}, {Cameron}, {Kasliwal}, {Gal-Yam},
  {Rau}, {Kulkarni}, {Frail}, {Chandra}, {Cenko}, {Soderberg}, \&
  {Immler}}]{2007ApJ...659L..13O}
{Ofek} E.~O. {et~al.}, 2007, \apjl, 659, L13

\bibitem[{{Pastorello} {et~al}\mbox{.}(2010){Pastorello}, {Botticella},
  {Trundle}, {Taubenberger}, {Mattila}, {Kankare}, {Elias-Rosa}, {Benetti},
  {Duszanowicz}, {Hermansson}, {Beckman}, {Bufano}, {Fraser}, {Harutyunyan},
  {Navasardyan}, {Smartt}, {van Dyk}, {Vink}, \&
  {Wagner}}]{2010MNRAS.408..181P}
{Pastorello} A. {et~al.}, 2010, \mnras, 408, 181

\bibitem[{{Phillips} {et~al}\mbox{.}(1992){Phillips}, {Wells}, {Suntzeff},
  {Hamuy}, {Leibundgut}, {Kirshner}, \& {Foltz}}]{1992AJ....103.1632P}
{Phillips} M.~M., {Wells} L.~A., {Suntzeff} N.~B., {Hamuy} M., {Leibundgut} B.,
  {Kirshner} R.~P., {Foltz} C.~B., 1992, \aj, 103, 1632

\bibitem[{{Piro} \& {Thrane}(2012)}]{2012ApJ...761...63P}
{Piro} A.~L., {Thrane} E., 2012, \apj, 761, 63

\bibitem[{{Polshaw} {et~al}\mbox{.}(2015){Polshaw}, {Kotak}, {Chambers},
  {Smartt}, {Taubenberger}, {Kromer}, {Gall}, {Hillebrandt}, {Huber}, {Smith},
  \& {Wainscoat}}]{2015A&A...580L..15P}
{Polshaw} J. {et~al.}, 2015, \aap, 580, L15

\bibitem[{{Prentice} {et~al}\mbox{.}(2016){Prentice}, {Mazzali}, {Pian},
  {Gal-Yam}, {Kulkarni}, {Rubin}, {Corsi}, {Fremling}, {Sollerman}, {Yaron},
  {Arcavi}, {Zheng}, {Kasliwal}, {Filippenko}, {Cenko}, {Cao}, \&
  {Nugent}}]{2016arXiv160201736P}
{Prentice} S.~J. {et~al.}, 2016, ArXiv e-prints

\bibitem[{{Quimby} {et~al}\mbox{.}(2011){Quimby}, {Kulkarni}, {Kasliwal},
  {Gal-Yam}, {Arcavi}, {Sullivan}, {Nugent}, {Thomas}, {Howell}, {Nakar},
  {Bildsten}, {Theissen}, {Law}, {Dekany}, {Rahmer}, {Hale}, {Smith}, {Ofek},
  {Zolkower}, {Velur}, {Walters}, {Henning}, {Bui}, {McKenna}, {Poznanski},
  {Cenko}, \& {Levitan}}]{2011Natur.474..487Q}
{Quimby} R.~M. {et~al.}, 2011, \nat, 474, 487

\bibitem[{{Quimby} {et~al}\mbox{.}(2013){Quimby}, {Yuan}, {Akerlof}, \&
  {Wheeler}}]{2013MNRAS.431..912Q}
{Quimby} R.~M., {Yuan} F., {Akerlof} C., {Wheeler} J.~C., 2013, \mnras, 431,
  912

\bibitem[{{Rest} {et~al}\mbox{.}(2014){Rest}, {Scolnic}, {Foley}, {Huber},
  {Chornock}, {Narayan}, {Tonry}, {Berger}, {Soderberg}, {Stubbs}, {Riess},
  {Kirshner}, {Smartt}, {Schlafly}, {Rodney}, {Botticella}, {Brout}, {Challis},
  {Czekala}, {Drout}, {Hudson}, {Kotak}, {Leibler}, {Lunnan}, {Marion},
  {McCrum}, {Milisavljevic}, {Pastorello}, {Sanders}, {Smith}, {Stafford},
  {Thilker}, {Valenti}, {Wood-Vasey}, {Zheng}, {Burgett}, {Chambers},
  {Denneau}, {Draper}, {Flewelling}, {Hodapp}, {Kaiser}, {Kudritzki},
  {Magnier}, {Metcalfe}, {Price}, {Sweeney}, {Wainscoat}, \&
  {Waters}}]{2014ApJ...795...44R}
{Rest} A. {et~al.}, 2014, \apj, 795, 44

\bibitem[{{Rubin} {et~al}\mbox{.}(2015){Rubin}, {Gal-Yam}, {De Cia}, {Horesh},
  {Khazov}, {Ofek}, {Kulkarni}, {Arcavi}, {Manulis}, {Yaron}, {Vreeswijk},
  {Kasliwal}, {Ben-Ami}, {Perley}, {Cao}, {Cenko}, {Rebbapragada},
  {Wo{\'z}niak}, {Filippenko}, {Clubb}, {Nugent}, {Pan}, {Badenes}, {Howell},
  {Valenti}, {Sand}, {Sollerman}, {Johansson}, {Leonard}, {Horst}, {Armen},
  {Fedrow}, {Quimby}, {Mazzali}, {Pian}, {Sternberg}, {Matheson}, {Sullivan},
  {Maguire}, \& {Lazarevic}}]{2015arXiv151200733R}
{Rubin} A. {et~al.}, 2015, ArXiv e-prints

\bibitem[{{Saglia} {et~al}\mbox{.}(2012){Saglia}, {Tonry}, {Bender}, {Greisel},
  {Seitz}, {Senger}, {Snigula}, {Phleps}, {Wilman}, {Bailer-Jones}, {Klement},
  {Rix}, {Smith}, {Green}, {Burgett}, {Chambers}, {Heasley}, {Kaiser},
  {Magnier}, {Morgan}, {Price}, {Stubbs}, \& {Wainscoat}}]{2012ApJ...746..128S}
{Saglia} R.~P. {et~al.}, 2012, \apj, 746, 128

\bibitem[{{Sanders} {et~al}\mbox{.}(2015){Sanders}, {Soderberg}, {Gezari},
  {Betancourt}, {Chornock}, {Berger}, {Foley}, {Challis}, {Drout}, {Kirshner},
  {Lunnan}, {Marion}, {Margutti}, {McKinnon}, {Milisavljevic}, {Narayan},
  {Rest}, {Kankare}, {Mattila}, {Smartt}, {Huber}, {Burgett}, {Draper},
  {Hodapp}, {Kaiser}, {Kudritzki}, {Magnier}, {Metcalfe}, {Morgan}, {Price},
  {Tonry}, {Wainscoat}, \& {Waters}}]{2015ApJ...799..208S}
{Sanders} N.~E. {et~al.}, 2015, \apj, 799, 208

\bibitem[{{Schlafly} \& {Finkbeiner}(2011)}]{2011ApJ...737..103S}
{Schlafly} E.~F., {Finkbeiner} D.~P., 2011, \apj, 737, 103

\bibitem[{{Schlafly} {et~al}\mbox{.}(2012){Schlafly}, {Finkbeiner},
  {Juri{\'c}}, {Magnier}, {Burgett}, {Chambers}, {Grav}, {Hodapp}, {Kaiser},
  {Kudritzki}, {Martin}, {Morgan}, {Price}, {Rix}, {Stubbs}, {Tonry}, \&
  {Wainscoat}}]{2012ApJ...756..158S}
{Schlafly} E.~F. {et~al.}, 2012, \apj, 756, 158

\bibitem[{{Scoville} {et~al}\mbox{.}(2007){Scoville}, {Aussel}, {Brusa},
  {Capak}, {Carollo}, {Elvis}, {Giavalisco}, {Guzzo}, {Hasinger}, {Impey},
  {Kneib}, {LeFevre}, {Lilly}, {Mobasher}, {Renzini}, {Rich}, {Sanders},
  {Schinnerer}, {Schminovich}, {Shopbell}, {Taniguchi}, \&
  {Tyson}}]{2007ApJS..172....1S}
{Scoville} N. {et~al.}, 2007, \apjs, 172, 1

\bibitem[{{Shappee} {et~al}\mbox{.}(2014){Shappee}, {Prieto}, {Grupe},
  {Kochanek}, {Stanek}, {De Rosa}, {Mathur}, {Zu}, {Peterson}, {Pogge},
  {Komossa}, {Im}, {Jencson}, {Holoien}, {Basu}, {Beacom}, {Szczygie{\l}},
  {Brimacombe}, {Adams}, {Campillay}, {Choi}, {Contreras}, {Dietrich},
  {Dubberley}, {Elphick}, {Foale}, {Giustini}, {Gonzalez}, {Hawkins}, {Howell},
  {Hsiao}, {Koss}, {Leighly}, {Morrell}, {Mudd}, {Mullins}, {Nugent},
  {Parrent}, {Phillips}, {Pojmanski}, {Rosing}, {Ross}, {Sand}, {Terndrup},
  {Valenti}, {Walker}, \& {Yoon}}]{2014ApJ...788...48S}
{Shappee} B.~J. {et~al.}, 2014, \apj, 788, 48

\bibitem[{{Shectman} {et~al}\mbox{.}(1996){Shectman}, {Landy}, {Oemler},
  {Tucker}, {Lin}, {Kirshner}, \& {Schechter}}]{1996ApJ...470..172S}
{Shectman} S.~A., {Landy} S.~D., {Oemler} A., {Tucker} D.~L., {Lin} H.,
  {Kirshner} R.~P., {Schechter} P.~L., 1996, \apj, 470, 172

\bibitem[{{Singer} {et~al}\mbox{.}(2015){Singer}, {Kasliwal}, {Cenko},
  {Perley}, {Anderson}, {Anupama}, {Arcavi}, {Bhalerao}, {Bue}, {Cao},
  {Connaughton}, {Corsi}, {Cucchiara}, {Fender}, {Fox}, {Gehrels}, {Goldstein},
  {Gorosabel}, {Horesh}, {Hurley}, {Johansson}, {Kann}, {Kouveliotou}, {Huang},
  {Kulkarni}, {Masci}, {Nugent}, {Rau}, {Rebbapragada}, {Staley}, {Svinkin},
  {Th{\"o}ne}, {de Ugarte Postigo}, {Urata}, \&
  {Weinstein}}]{2015ApJ...806...52S}
{Singer} L.~P. {et~al.}, 2015, \apj, 806, 52

\bibitem[{{Singer} \& {Price}(2016)}]{2016PhRvD..93b4013S}
{Singer} L.~P., {Price} L.~R., 2016, \prd, 93, 024013

\bibitem[{{Singer} {et~al}\mbox{.}(2014){Singer}, {Price}, {Farr}, {Urban},
  {Pankow}, {Vitale}, {Veitch}, {Farr}, {Hanna}, {Cannon}, {Downes}, {Graff},
  {Haster}, {Mandel}, {Sidery}, \& {Vecchio}}]{2014ApJ...795..105S}
{Singer} L.~P. {et~al.}, 2014, \apj, 795, 105

\bibitem[{{Smartt} {et~al}\mbox{.}(2009){Smartt}, {Eldridge}, {Crockett}, \&
  {Maund}}]{2009MNRAS.395.1409S}
{Smartt} S.~J., {Eldridge} J.~J., {Crockett} R.~M., {Maund} J.~R., 2009,
  \mnras, 395, 1409

\bibitem[{{Smartt} {et~al}\mbox{.}(2015){Smartt}, {Valenti}, {Fraser},
  {Inserra}, {Young}, {Sullivan}, {Pastorello}, {Benetti}, {Gal-Yam}, {Knapic},
  {Molinaro}, {Smareglia}, {Smith}, {Taubenberger}, {Yaron}, {Anderson},
  {Ashall}, {Balland}, {Baltay}, {Barbarino}, {Bauer}, {Baumont}, {Bersier},
  {Blagorodnova}, {Bongard}, {Botticella}, {Bufano}, {Bulla}, {Cappellaro},
  {Campbell}, {Cellier-Holzem}, {Chen}, {Childress}, {Clocchiatti},
  {Contreras}, {Dall'Ora}, {Danziger}, {de Jaeger}, {De Cia}, {Della Valle},
  {Dennefeld}, {Elias-Rosa}, {Elman}, {Feindt}, {Fleury}, {Gall},
  {Gonzalez-Gaitan}, {Galbany}, {Morales Garoffolo}, {Greggio}, {Guillou},
  {Hachinger}, {Hadjiyska}, {Hage}, {Hillebrandt}, {Hodgkin}, {Hsiao}, {James},
  {Jerkstrand}, {Kangas}, {Kankare}, {Kotak}, {Kromer}, {Kuncarayakti},
  {Leloudas}, {Lundqvist}, {Lyman}, {Hook}, {Maguire}, {Manulis}, {Margheim},
  {Mattila}, {Maund}, {Mazzali}, {McCrum}, {McKinnon}, {Moreno-Raya},
  {Nicholl}, {Nugent}, {Pain}, {Pignata}, {Phillips}, {Polshaw}, {Pumo},
  {Rabinowitz}, {Reilly}, {Romero-Ca{\~n}izales}, {Scalzo}, {Schmidt},
  {Schulze}, {Sim}, {Sollerman}, {Taddia}, {Tartaglia}, {Terreran},
  {Tomasella}, {Turatto}, {Walker}, {Walton}, {Wyrzykowski}, {Yuan}, \&
  {Zampieri}}]{2015A&A...579A..40S}
{Smartt} S.~J. {et~al.}, 2015, \aap, 579, A40

\bibitem[{{Smith} {et~al}\mbox{.}(2008){Smith}, {Chornock}, {Li},
  {Ganeshalingam}, {Silverman}, {Foley}, {Filippenko}, \&
  {Barth}}]{2008ApJ...686..467S}
{Smith} N., {Chornock} R., {Li} W., {Ganeshalingam} M., {Silverman} J.~M.,
  {Foley} R.~J., {Filippenko} A.~V., {Barth} A.~J., 2008, \apj, 686, 467

\bibitem[{{Smith} {et~al}\mbox{.}(2007){Smith}, {Li}, {Foley}, {Wheeler},
  {Pooley}, {Chornock}, {Filippenko}, {Silverman}, {Quimby}, {Bloom}, \&
  {Hansen}}]{2007ApJ...666.1116S}
{Smith} N. {et~al.}, 2007, \apj, 666, 1116

\bibitem[{{Soares-Santos} {et~al}\mbox{.}(2016){Soares-Santos}, {Kessler},
  {Berger}, {Annis}, {Brout}, {Buckley-Geer}, {Chen}, {Cowperthwaite}, {Diehl},
  {Doctor}, {Drlica-Wagner}, {Farr}, {Finley}, {Flaugher}, {Foley}, {Frieman},
  {Gruendl}, {Herner}, {Holz}, {Lin}, {Marriner}, {Neilsen}, {Rest}, {Sako},
  {Scolnic}, {Sobreira}, {Walker}, {Wester}, {Yanny}, {Abbott}, {Abdalla},
  {Allam}, {Armstrong}, {Banerji}, {Benoit-L{\'e}vy}, {Bernstein}, {Bertin},
  {Brown}, {Burke}, {Capozzi}, {Carnero Rosell}, {Carrasco Kind}, {Carretero},
  {Castander}, {Cenko}, {Chornock}, {Crocce}, {D'Andrea}, {da Costa}, {Desai},
  {Dietrich}, {Drout}, {Eifler}, {Estrada}, {Evrard}, {Fairhurst}, {Fernandez},
  {Fischer}, {Fong}, {Fosalba}, {Fox}, {Fryer}, {Garcia-Bellido}, {Gaztanaga},
  {Gerdes}, {Goldstein}, {Gruen}, {Gutierrez}, {Honscheid}, {James},
  {Karliner}, {Kasen}, {Kent}, {Kuropatkin}, {Kuehn}, {Lahav}, {Li}, {Lima},
  {Maia}, {Margutti}, {Martini}, {Matheson}, {McMahon}, {Metzger}, {Miller},
  {Miquel}, {Mohr}, {Nichol}, {Nord}, {Ogando}, {Peoples}, {Plazas},
  {Quataert}, {Romer}, {Roodman}, {Rykoff}, {Sanchez}, {Scarpine}, {Schindler},
  {Schubnell}, {Sevilla-Noarbe}, {Sheldon}, {Smith}, {Smith}, {Smith},
  {Stebbins}, {Sutton}, {Swanson}, {Tarle}, {Thaler}, {Thomas}, {Tucker},
  {Vikram}, {Wechsler}, \& {Weller}}]{2016ApJ...823L..33S}
{Soares-Santos} M. {et~al.}, 2016, \apjl, 823, L33

\bibitem[{{Steele}, {Copperwheat} \& {Piascik}(2015){Steele}, {Copperwheat}, \&
  {Piascik}}]{GCN18371}
{Steele} I.~S., {Copperwheat} C., {Piascik} A., 2015, GRB Coordinates Network,
  18371

\bibitem[{{Taddia} {et~al}\mbox{.}(2015){Taddia}, {Sollerman}, {Leloudas},
  {Stritzinger}, {Valenti}, {Galbany}, {Kessler}, {Schneider}, \&
  {Wheeler}}]{2015A&A...574A..60T}
{Taddia} F. {et~al.}, 2015, \aap, 574, A60

\bibitem[{{Tanaka} \& {Hotokezaka}(2013)}]{2013ApJ...775..113T}
{Tanaka} M., {Hotokezaka} K., 2013, \apj, 775, 113

\bibitem[{{Tanaka} {et~al}\mbox{.}(2014){Tanaka}, {Hotokezaka}, {Kyutoku},
  {Wanajo}, {Kiuchi}, {Sekiguchi}, \& {Shibata}}]{2014ApJ...780...31T}
{Tanaka} M., {Hotokezaka} K., {Kyutoku} K., {Wanajo} S., {Kiuchi} K.,
  {Sekiguchi} Y., {Shibata} M., 2014, \apj, 780, 31

\bibitem[{{Tanvir} {et~al}\mbox{.}(2013){Tanvir}, {Levan}, {Fruchter},
  {Hjorth}, {Hounsell}, {Wiersema}, \& {Tunnicliffe}}]{2013Natur.500..547T}
{Tanvir} N.~R., {Levan} A.~J., {Fruchter} A.~S., {Hjorth} J., {Hounsell} R.~A.,
  {Wiersema} K., {Tunnicliffe} R.~L., 2013, \nat, 500, 547

\bibitem[{{The LIGO Scientific Collaboration} {et~al}\mbox{.}(2015){The LIGO
  Scientific Collaboration}, {Aasi}, {Abbott}, {Abbott}, {Abbott}, {Abernathy},
  {Ackley}, {Adams}, {Adams}, {Addesso}, \& et~al.}]{2015CQGra..32g4001T}
{The LIGO Scientific Collaboration} {et~al.}, 2015, Classical and Quantum
  Gravity, 32, 074001

\bibitem[{{The LIGO Scientific Collaboration} \& {The Virgo
  Collaboration}(2012)}]{2012arXiv1203.2674T}
{The LIGO Scientific Collaboration}, {The Virgo Collaboration}, 2012, ArXiv
  e-prints

\bibitem[{{The LIGO Scientific Collaboration} \& {the Virgo
  Collaboration}(2015)}]{GCN18858}
{The LIGO Scientific Collaboration}, {the Virgo Collaboration}, 2015, GRB
  Coordinates Network, 18858

\bibitem[{{The LIGO Scientific Collaboration} \& {the Virgo
  Collaboration}(2016)}]{2016arXiv160203840T}
{The LIGO Scientific Collaboration}, {the Virgo Collaboration}, 2016, ArXiv
  e-prints

\bibitem[{{The LIGO Scientific Collaboration} \& {the Virgo
  Collaborations}(2015)}]{GCN18330}
{The LIGO Scientific Collaboration}, {the Virgo Collaborations}, 2015, GRB
  Coordinates Network, 18330

\bibitem[{{Tonry}(2011)}]{2011PASP..123...58T}
{Tonry} J.~L., 2011, \pasp, 123, 58

\bibitem[{{Tonry} {et~al}\mbox{.}(2012{\natexlab{a}}){Tonry}, {Stubbs},
  {Kilic}, {Flewelling}, {Deacon}, {Chornock}, {Berger}, {Burgett}, {Chambers},
  {Kaiser}, {Kudritzki}, {Hodapp}, {Magnier}, {Morgan}, {Price}, \&
  {Wainscoat}}]{2012ApJ...745...42T}
{Tonry} J.~L. {et~al.}, 2012{\natexlab{a}}, \apj, 745, 42

\bibitem[{{Tonry} {et~al}\mbox{.}(2012{\natexlab{b}}){Tonry}, {Stubbs},
  {Lykke}, {Doherty}, {Shivvers}, {Burgett}, {Chambers}, {Hodapp}, {Kaiser},
  {Kudritzki}, {Magnier}, {Morgan}, {Price}, \&
  {Wainscoat}}]{2012ApJ...750...99T}
{Tonry} J.~L. {et~al.}, 2012{\natexlab{b}}, \apj, 750, 99

\bibitem[{{Utrobin}, {Chugai} \& {Botticella}(2010){Utrobin}, {Chugai}, \&
  {Botticella}}]{2010ApJ...723L..89U}
{Utrobin} V.~P., {Chugai} N.~N., {Botticella} M.~T., 2010, \apjl, 723, L89

\bibitem[{{Valenti} {et~al}\mbox{.}(2011){Valenti}, {Fraser}, {Benetti},
  {Pignata}, {Sollerman}, {Inserra}, {Cappellaro}, {Pastorello}, {Smartt},
  {Ergon}, {Botticella}, {Brimacombe}, {Bufano}, {Crockett}, {Eder}, {Fugazza},
  {Haislip}, {Hamuy}, {Harutyunyan}, {Ivarsen}, {Kankare}, {Kotak}, {Lacluyze},
  {Magill}, {Mattila}, {Maza}, {Mazzali}, {Reichart}, {Taubenberger},
  {Turatto}, \& {Zampieri}}]{2011MNRAS.416.3138V}
{Valenti} S. {et~al.}, 2011, \mnras, 416, 3138

\bibitem[{{Veres} {et~al}\mbox{.}(2015){Veres}, {Jedicke}, {Fitzsimmons},
  {Denneau}, {Granvik}, {Bolin}, {Chastel}, {Wainscoat}, {Burgett}, {Chambers},
  {Flewelling}, {Kaiser}, {Magnier}, {Morgan}, {Price}, {Tonry}, \&
  {Waters}}]{2015Icar..261...34V}
{Veres} P. {et~al.}, 2015, \icarus, 261, 34

\bibitem[{{V{\'e}ron-Cetty} \& {V{\'e}ron}(2001)}]{2001A&A...374...92V}
{V{\'e}ron-Cetty} M.-P., {V{\'e}ron} P., 2001, \aap, 374, 92

\bibitem[{{White}, {Daw} \& {Dhillon}(2011){White}, {Daw}, \&
  {Dhillon}}]{2011CQGra..28h5016W}
{White} D.~J., {Daw} E.~J., {Dhillon} V.~S., 2011, Classical and Quantum
  Gravity, 28, 085016

\bibitem[{{Woosley} \& {Bloom}(2006)}]{2006ARA&A..44..507W}
{Woosley} S.~E., {Bloom} J.~S., 2006, \araa, 44, 507

\bibitem[{{Wright}(2015)}]{Wright15}
{Wright} D., 2015, PhD thesis, Department of Physics and Astronomy, Queen's
  University Belfast.

\bibitem[{{Wright} {et~al}\mbox{.}(2015){Wright}, {Smartt}, {Smith}, {Miller},
  {Kotak}, {Rest}, {Burgett}, {Chambers}, {Flewelling}, {Hodapp}, {Huber},
  {Jedicke}, {Kaiser}, {Metcalfe}, {Price}, {Tonry}, {Wainscoat}, \&
  {Waters}}]{2015MNRAS.449..451W}
{Wright} D.~E. {et~al.}, 2015, \mnras, 449, 451

\bibitem[{{Yang} {et~al}\mbox{.}(2015){Yang}, {Jin}, {Li}, {Covino}, {Zheng},
  {Hotokezaka}, {Fan}, {Piran}, \& {Wei}}]{2015NatCo...6E7323Y}
{Yang} B. {et~al.}, 2015, Nature Communications, 6, 7323

\end{thebibliography}

\newpage

\appendix

\section{Details of Pan-STARSS1 fields}

Table\,\ref{tab:a1} gives a summary of the Pan-STARRS1 observations. 
Table\,\ref{tab:a2} gives the RA of the proximity to twilight for each filter used to survey the
sky probability maps. This lists the maximum RA accessible in each filter for the date given
when the sub us 16 degrees below the horizon (for \gps\rps\ips), 14 degrees below (for \zps)
and 10 degrees below (for \yps).  This is indicated in the plots in Figure 1 as vertical lines for the \ips filter. 

\begin{table*}
\label{tab:a1}
\caption{Summary of Pan-STARRS1 observations}
\begin{tabular}{lllll}\hline\hline
Date  & MJD & Filters & Exposure Times (sec) & Number of PS1 Exposures \\\hline
20150917 & 57282 & \ips\zps\yps & 45,60 & 34 \\
20150919 & 57284 & \ips\zps\yps & 20,23,35 & 59 \\
20150922 & 57287 & \ips\zps\yps & 40,45,68 & 45 \\
20150923 & 57288 & \ips\zps\yps & 25,30,37 & 49 \\
20150924 & 57289 & \ips\zps\yps & 30,45 & 45 \\
20150925 & 57290 & \ips\zps\yps & 30,45 & 44 \\
20150927 & 57292 & \ips & 35 & 4 \\
20151002 & 57297 & \ips\zps\yps & 25,35 & 57 \\
20151003 & 57298 & \ips\zps\yps & 25,40 & 65 \\
20151013 & 57308 & \ips\zps\yps & 25,35 & 88 \\
20151014 & 57309 & \ips\zps\yps & 30,60 & 77 \\
20151015 & 57310 & \ips\rps\zps\yps & 30,60,200 & 72 \\
20151017 & 57312 & \ips & 60 & 5 \\
20151018 & 57313 & \ips\rps\zps\yps & 30,80,200 & 70 \\
20151019 & 57314 & \ips\rps\zps\yps & 35,200 & 22 \\
20151021 & 57316 & \ips\rps\zps\yps & 30,45,50,200 & 98 \\
20151022 & 57317 & \ips\rps\zps\yps & 30,45,60,200 & 79 \\
20151023 & 57318 & \ips\rps\zps\yps & 30,60,65,200 & 87 \\
20151024 & 57319 & \ips\rps\zps\yps & 30,45,200 & 102 \\
20151025 & 57320 & \ips\rps\zps\yps & 30,60,65,200 & 90 \\
20151026 & 57321 & \ips\rps\zps & 35,50,60,200 & 64 \\\hline
\end{tabular}
\end{table*}

\begin{table*}\caption{Proximity to twilght of Pan-STARRS1 fields}
\label{tab:a2}
\begin{tabular}{llllllll}\hline\hline
Date     &    Plot       &       \multicolumn{3}{c}{LST of twilight}          &      \multicolumn{3}{c}{RA of an HA = 4.5 hrs at twilight} \\
             &               &    16 deg &  14 deg &  10 deg   &                               \\
             &                &  \ips-band  & \zps-band  & \yps-band    &  \ips-band  &  \zps-band &  \yps-band \\\hline

17 Sept &  first 3 days  &   04:27  &  04:36  &  04:53   &    08:57  &   09:06   & 09:23   \\
              &                    &          &         &          &    134.25 &   136.50  & 140.75  \\
\\
27 Sept  & 3-10 days     &   05:09  &  05:18  &  05:35   &    09:39  &   09:48   & 10:05   \\
              &                    &          &         &          &   144.75  &  147.00   & 151.25    \\                                                    
\\
4 Oct   &  10-17 days    &   05:39  &  05:47  &  06:05   &    10:09  &   10:17   & 10:35 \\
            &                      &          &         &          &    152.25 &   154.25  & 158.75 \\
\\
11 Oct  &  17-24 days    &   06:08  &  06:17  &  06:34   &    10:38  &   10:47   & 11:04 \\
            &                       &          &         &          &    159.50 &   161.75  & 166.00 \\
\\
18 Oct  &  24-31 days    &   06:38  &  06:47  &  07:04   &    11:08  &   11:17   & 11:34 \\
            &                    &          &         &          &    167.0  &   169.25  & 173.50 \\
\\
25 Oct  &  $>$ 31 days     &   07:08  &  07:17  &  07:34   &    11:38  &   11:47   & 12:04\\
            &                    &          &         &          &    174.5  &   176.75  &  181.00 \\\hline
\end{tabular}
\end{table*}
\end{document}